\documentclass[%
reprint,
jpa,
superscriptaddress,
amsmath,amssymb,
]{revtex4-2}

\usepackage{graphicx}% Include figure files
\usepackage{dcolumn}% Align table columns on decimal point
\usepackage{bm}% bold math
\usepackage[dvipsnames]{xcolor}
\usepackage{changes}

\usepackage{xr-hyper}
\usepackage{hyperref}

\begin{document}

	\title{Predictability Analysis and Prediction of Discrete Weather and Financial Time-Series Data with a Hamiltonian-Based Filter-Projection Approach}
	
	\author{Henrik Kiefer}
		\affiliation{%
		Freie  Universität  Berlin,  Department  of  Physics,  14195  Berlin,  Germany}%
	%\altaffiliation{henrik.kiefer@fu-berlin.de}%
	\author{Denis Furtel}
		\affiliation{%
		Freie  Universität  Berlin,  Department  of  Physics,  14195  Berlin,  Germany}%
	%\altaffiliation{denis.furtel@gmail.com}%
	\author{Cihan Ayaz}
		\affiliation{%
		Freie  Universität  Berlin,  Department  of  Physics,  14195  Berlin,  Germany}%
	%\altaffiliation{cihanayaz@fu-berlin.de}%
	\author{Anton Klimek}
		\affiliation{%
		Freie  Universität  Berlin,  Department  of  Physics,  14195  Berlin,  Germany}%
	%\altaffiliation{kanton@physik.fu-berlin.de}%
	\author{Jan O. Daldrop}
		\affiliation{%
		Freie  Universität  Berlin,  Department  of  Physics,  14195  Berlin,  Germany}%
	%\altaffiliation{daldrop@fu-berlin.de}%
	\author{Roland R. Netz}
	\altaffiliation{rnetz@physik.fu-berlin.de}%
	\affiliation{%
		Freie  Universität  Berlin,  Department  of  Physics,  14195  Berlin,  Germany}%
			\affiliation{%
				Centre for Condensed Matter Theory, Department of Physics, Indian Institute of Science, Bangalore 560012, India}%

	\date{\today}	
	\begin{abstract}
		\noindent 
The generalized Langevin equation (GLE), derived by projection from a general many-body Hamiltonian, exactly describes the dynamics of an arbitrary coarse-grained variable in a complex environment. However, analysis and prediction of real-world data with the GLE is hampered by slow transient or seasonal data components and time-discretization effects. Machine-learning (ML) techniques work but are computer-resource demanding and difficult to interpret. We show that by convolution filtering, time-series data decompose into fast, transient and seasonal components that each obey Hamiltonian dynamics and, thus, can be separately analyzed by projection techniques. We introduce methods to extract all GLE parameters from highly discretized time-series data and to forecast future data including the environmental stochasticity. For daily-resolved weather data, our analysis reveals non-Markovian memory that decays over a few days. Our prediction accuracy is comparable to commercial (weather.com) and ML long short-term memory (LSTM) methods at a reduced computational cost by a factor of $10^2-10^3$ compared to  LSTM. For financial data, memory is very short-ranged and the dynamics effectively is Markovian, in agreement with the efficient-market hypothesis; consequently, models simpler than the GLE are sufficient. Our GLE framework is an efficient and interpretable method for the analysis and prediction of complex time-series data. 
	
	\end{abstract}
	\maketitle

	\newpage

A major purpose  of  science is the analysis, understanding  and prediction  of time-series data.
With the non-stop generation of  data from computer simulations and experiments 
and the continuous growth in computing power and storage capacity,
the need to make sense of time-series data is steadily  increasing \cite{cryer1986time}.
 Traditionally,   
 structurally simple differential equations
 were derived or posited
 that can be straightforwardly interpreted,
 decomposed  and combined to build dynamical models of increasing complexity. 
For example, in meteorology  or climatology, where  prediction accuracy is  vital for disaster management and  habitability planning,
traditional  models rely on deterministic  physical equations \cite{lorenc1986analysis};
in recent years stochastic models  have attracted increasing interest  \cite{franzke2015stochastic, watkins2021generalized}. 	
Likewise, in  economy,
there is huge demand for building insightful models for  the dynamics  of  financial data such as stock market indices  and other financial assets. 
In contrast to classical economics 
  \cite{ord2017principles, petropoulos2022forecasting},
   fluctuations are at the core of  recent econophysics  models 
   \cite{gopikrishnan1999scaling,  takayasu_empirical_2002,bouchaud2003theory}.\\
\indent As an alternative to analysis and prediction based on simple equations, 
 machine-learning (ML) methods have been developed 
 \cite{hochreiter1997long, zhang2003time, tseng2002combining, pathak2018model, raissi2018hidden, han2018solving}.
Despite their great success and universal applicability to all  kinds of data, they are also criticized:
Firstly, determination of the  exorbitant number of parameters requires excessive computations for training  \cite{al2015efficient};
secondly, ML  is essentially a black box and  the lack of interpretability  of the learned rules can be unsatisfactory for natural-science
applications and also critical for high-stake societal and political decisions  \cite{papernot2017practical, rudin2019stop, rodrigues2021deeper}. \\
\indent In fact, the equation that exactly describes the dynamics of an arbitrary observable over time, be it the temperature in Berlin or the 
Dollar-Yen exchange rate, is the generalized Langevin equation (GLE) \cite{mori1965transport,zwanzig1961memory}, 
which is an  integro-differential equation of motion  for a time-dependent phase-space observable $A(t)$. 
 Remarkably, the GLE is derived from the Hamiltonian for an arbitrarily complex many-body system
that does not need to be known in explicit form and  in principle can encompass the entire universe,
in which limit  the Hamiltonian  becomes  autonomous, i.e. time-independent \cite{schmitt_analyzing_2006},
and standard GLE formulations apply \cite{netz2024derivation}.
While we use  the GLE derived from classical Hamilton dynamics, a quantum version exists and can be used if the 
observable exhibits quantum properties \cite{nakajima1958quantum}, which, however, is not the case for the systems considered here.
The GLE  generalizes  the  Newtonian equation of motion by explicitly accounting for 
the interaction of the observable $A(t)$ with its responsive environment in terms of dissipative friction  and a time-dependent force
that is typically interpreted stochastically. Thus, the GLE correctly accounts for the loss of information when projecting the high-dimensional phase-space dynamics
onto the low-dimensional observable dynamics and, hence, constitutes an exact coarse-graining method.
 The GLE is non-Markovian and depends on the entire history of  $A(t)$,
quantified by the memory function $\Gamma(t)$.\\
\indent Several methods to extract all GLE parameters from time-series data exist \cite{straub1987calculation,horenko_data-based_2007, daldrop_butane_2018, grogan_data-driven_2020,Lei2024}.
 The GLE has been shown to correctly describe the dynamics of diverse physico-chemical systems, such as  bond-length vibrations, dihedral rotations and protein folding
\cite{satija2019generalized,ayaz2021non, brunig2022time,dalton2022protein,dalton2024role,Huang2024}
 and also more complex  systems, such as the motion of living organisms and financial and meteorological data
 \cite{schmitt_analyzing_2006, mitterwallner2020non, hassanibesheli2020reconstructing,klimek2023data}. 
In fact, the GLE has also been suggested  for time-series prediction,  albeit in approximate  form  \cite{chorin_optimal_2000}. 
Time-series are always discretized, 
which poses severe problems for the applicability of the GLE, which is a time-continuous integro-differential equation \cite{niemann2008usage, tepper2024accurate}. 
Thus, while the  GLE is universally applicable to arbitrary   systems and   in principle provides a suitable  framework for  time-series
modeling, it was actually never used to forecast real-world complex data. This is due to four  complications:
i) The multi-scale dynamic coupling between short-time stochastics and long-time transient or seasonal effects in the data, ii) the presence of non-Gaussian correlations in many 
non-trivial systems, iii) time-discretization effects, and iv) the  presence of slowly-relaxing non-equilibrium effects that render many tools of statistical mechanics useless. We solve all these problems by our combined filtering-projection approach.

\subsection*{Filtering, Generalized Langevin Equation and Prediction}

	\begin{figure*}[hbt!]
		\centering
		\includegraphics[width=\textwidth]{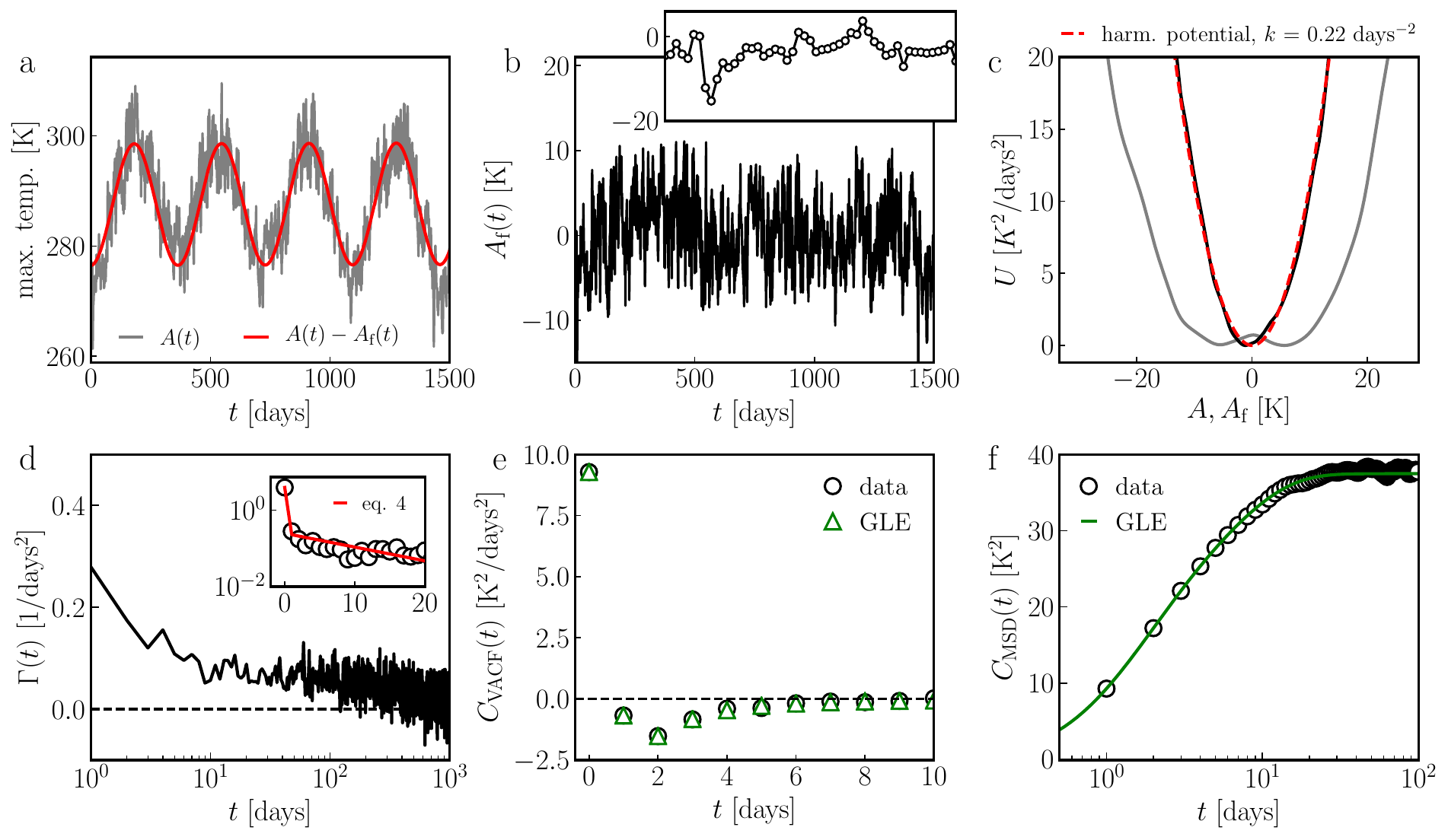}
		\caption{ {\bf\sffamily GLE analysis of weather data}. {\bf\sffamily a}: Daily resolved maximal air temperature $A(t)$  in Berlin-Tegel between January 15, 2006 and February 22, 2010 (gray) and  sum of seasonal and trend components  $A(t)-A_\text{f}(t)$ (red), see Methods \ref{sec:filtering} for details.  {\bf\sffamily b}: Filtered temperature component $A_\text{f}(t)$ that consists of the quickly varying temperature components
			and is described by the GLE.  The inset shows the first 50 days of the trajectory in the main plot.
			{\bf\sffamily c}: Effective potential $U$ of the unfiltered (gray) and filtered temperature data (black), 
			compared  with the harmonic potential $U=k A_\text{f}^2/2$ (red) according to the Mori-GLE in equation~(\ref{eq:mori_GLE}). 
			{\bf\sffamily d}:  Memory kernel $\Gamma(t)$ extracted from $A_\text{f}(t)$ using the Volterra method (see Supplementary Information \ref{app:volterra}). 
			The red line in the inset is a fit according to equation~(\ref{eq:meteo_kernel}). 
			{\bf\sffamily e}: Discrete velocity autocorrelation function $C_{\text{VACF}}(t)$ of  the filtered temperature data $A_\text{f}(t)$ (black circles) 
			compared with the GLE prediction based on a fit of
			the memory kernel in equation~(\ref{eq:meteo_kernel}) (green triangles, see Supplementary Information \ref{app:mitterwallner}). The GLE parameters are listed in Table~\ref{tab:fits_kernel_tmax}. {\bf\sffamily f}: Mean-squared displacement $C_{\text{MSD}}(t)$ of the filtered data $A_\text{f}(t)$  (black circles) together with the GLE prediction according to 
			equation~(\ref{eq:vacf2}) in Supplementary Information \ref{app:msd_harm} using the fitted memory kernel in equation~(\ref{eq:meteo_kernel}) (green line).}
		
		\label{fig:maxTemp1}
	\end{figure*}

In time-series analysis the data is often  decomposed into  trends, seasonal (i.e. periodic) contributions and a remaining part  that is typically neglected or  treated stochastically \cite{zhang2003time, petropoulos2022forecasting}.
This  ad hoc decomposition renders the equation governing the dynamics of the  remaining  part ill-defined. 
In this work, we  treat the phase-space function $A(t)$ representing the  time-series data by  
convolution filtering  according to 
\begin{equation}
	\label{eq:filtered_var}
A_\text{f}(t) = \int_{-\infty}^\infty  dt' A(t-t') f(t').
\end{equation}
The filter function $f(t)$ is general and, in our work, contains of a combination of low- and band-pass filters. 
  The decomposition of the observable  $A(t) =A_\text{f}(t) + A_\text{lp}(t) + A_\text{bp}(t)$ into the  filtered component $A_\text{f}(t)$,
 the slow low-pass-filtered  trend $A_\text{lp}(t)$ and the periodic band-pass-filtered part $A_\text{bp}(t)$ is exact.  Importantly, 
 $A_\text{f}(t)$, $A_\text{lp}(t)$ and  $A_\text{bp}(t)$ each obey Hamiltonian dynamics in phase space and, thus, 
 can be treated with the full arsenal of statistical-mechanics methods,
 see  Methods section \ref{sec:filtering} for details \cite{netz24filtering}.
The  slow and oscillating components  $A_\text{lp}(t)$ and  $A_\text{bp}(t)$ can be treated by deterministic models,
which amounts to fitting with periodic functions.
 In contrast,  the rapidly  moving filtered observable $A_\text{f}(t)$  is described by the Mori-GLE  \cite{mori1965transport}
 \begin{eqnarray}
		\label{eq:mori_GLE}
		\ddot{A}_\text{f}(t) &= -k  (A_\text{f}(t) -\langle A_\text{f} \rangle) + F_\text{R}(t) \nonumber - \int_0^t\mathrm{d}s\,\Gamma(t-s) \dot{A}_\text{f}(s) ,  \\ 
	\end{eqnarray}
where  $\ddot{A}_\text{f}(t)$ and  $ \dot{A}_\text{f}(t)$ denote second and first time derivatives of ${A}_\text{f}(t)$, 
  $k$ denotes the stiffness of an effective  harmonic potential, 
$\Gamma(t)$ accounts for non-Markovian friction   and $F_\text{R}(t)$ for environmental forces.
Note that $A_\text{f}(t)$ and $F_\text{R}(t)$ are phase-space functions while $k$ and $\Gamma(t)$ are not.
 The form of the GLE is actually not unique but depends  on the specific  projection used \cite{zwanzig1961memory, ayazhybrid2022, vroylandt2022derivation},
yet, the GLE equation~(\ref{eq:mori_GLE})  is exact even for non-linear systems, i.e. possibly present non-Gaussian effects  are accounted for by $F_\text{R}(t)$.
All parameters of the   GLE in equation~(\ref{eq:mori_GLE}) can be extracted from  time-series data;
in particular  $k = \langle \dot{A}_\text{f}^2 \rangle/ \langle A_\text{f}^2 \rangle$, $\Gamma(t)$ follows from a Volterra 
 equation derived from equation~(\ref{eq:mori_GLE})
 and  $F_\text{R}(t)$ is approximated  as a Gaussian random process  with second moment  \cite{mori1965transport}
	\begin{eqnarray}
		\label{eq:FDT2}
		\langle F_\text{R}(t) F_\text{R}(t^\prime)\rangle =  B  \Gamma(|t-t^\prime|),
	\end{eqnarray}
with $B=  \langle \dot{A}_\text{f}^2 \rangle $ (see  Methods section \ref{sec:Mori} for details).
This  becomes accurate for high-pass filtered observables, since they exhibit 
negligible non-Gaussian effects \cite{ayazhybrid2022, vroylandt2022derivation}, as demonstrated below. 
As usual, phase-space averages  $\langle \cdot \rangle$ are replaced by  time averages
and  $\langle A_\text{f} \rangle$ vanishes after high-pass filtering. Non-equilibrium effects need not be considered since the Hamiltonian is time-independent \cite{netz2024derivation}
and slow transient effects are removed by filtering \cite{netz24filtering}.
\\\indent To include the environmental influence on the prediction, we calculate  the future force $F_\text{R}(t)$ based on its past realization. 
In previous GLE formulations this was neglected \cite{chorin_optimal_2000}, which is approximate even if one is only interested in the mean future behavior of $A(t)$;
in  autoregressive moving average (ARMA) or autoregressive integrated moving average
(ARIMA) methods \cite{petropoulos2022forecasting} 
uncorrelated Gaussian noise is assumed, i.e. the right-hand side of equation~(\ref{eq:FDT2}) 
is approximated by a delta function. 
\\\indent In a nutshell, our GLE-based  prediction of filtered observables
consists of the following steps:
First, from the   power-spectrum of the original time-series $A(t)$,  the frequency ranges  of periodic  and slow components are inferred, which 
are  used to set  low- and band-pass  filters to obtain  the filtered time-series $A_\text{f}(t)$.
From $A_\text{f}(t)$,  the coefficient $k$, the memory kernel $\Gamma(t)$  and the  force trajectory $F_\text{R}(t)$ are determined,
fully accounting for  data discreteness in time,
which reveal  non-Markovian  effects and  whether the data contains predictable features. 
Then, $\Gamma(t)$ and the past $F_\text{R}(t)$ are  used to construct the future $F_\text{R}(t)$ as a conditional Gaussian process
according to equation~(\ref{eq:FDT2}). From the future  $F_\text{R}(t)$, the future $A_\text{f}(t)$ follows from equation~(\ref{eq:mori_GLE}).
We use step-by-step prediction, similar  to  recent ML-based  weather prediction   \cite{lam2023learning} (details are given in Methods section \ref{sec:GLE}).
\\\indent In the following, we compare our forecasts of weather  and financial data  with 
ML LSTM  predictions  \cite{hochreiter1997long} and the simpler yet robust Facebook Prophet algorithm \cite{taylor_forecasting_2018}. 
LSTM networks are recurrent neural networks that are designed for time-series data prediction,  
Prophet is suited for data  dominated by oscillations and trends (further details are given in Methods sections \ref{sec:LSTM}	and \ref{sec:prophet}).
\\\indent Our results indicate that  GLE    forecasts compe in accuracy with LSTM and Prophet at a fraction of computational cost with the main  benefit over ML  methods that  all model parameters are interpretable 
and that   the data  predictability can be deduced model-independently.

	\subsection*{ GLE Analysis and Forecasting of Weather Data}
	\label{sec:appl_meteo}

The daily maximal temperature  trajectory $A(t)$  in Berlin-Tegel in Fig.~\ref{fig:maxTemp1}a (gray line) 
exhibits pronounced seasonal variation, typical for continental European climate, see Methods  \ref{sec:data}. 
The corresponding effective potential in (c), $U(A) = -\langle\dot{A}^2\rangle \log \rho( A)$ calculated from  the probability distribution $\rho(A)$  (gray line),
exhibits two pronounced minima reflecting the mean winter and summer temperatures; thus, $A(t)$ is significantly non-Gaussian. 
The filter complement  $A(t)-A_\text{f}(t)$ (red line in Fig.~\ref{fig:maxTemp1}a) reflects the seasonal and slow transient contributions; the filtered observable  $A_\text{f}(t)$ 
(see Methods \ref{sec:filtering}) in (b) shows pronounced and fast  fluctuations over a temperature range of $\pm$10 K and is modeled
by  the Mori-GLE in equation~(\ref{eq:mori_GLE}). 
The effective potential of $A_\text{f}(t)$ in Fig.~\ref{fig:maxTemp1}c (black line) is perfectly  harmonic,  
$U(A_\text{f} )=k A_\text{f}^2/2$ (red-dashed line), with  $k = \langle \dot{A}_\text{f}^2 \rangle/\langle A_\text{f}^2 \rangle = 0.22$ days$^{-2}$ determined from  the $A_\text{f}(t)$ trajectory,
which suggests  that $A_\text{f}(t)$ is effectively  a Gaussian stochastic variable and that equation (\ref{eq:FDT2}) is an accurate approximation. 
The friction  memory  kernel $\Gamma(t)$ of $A_\text{f}(t)$  in Fig.~\ref{fig:maxTemp1}d
extracted using the Volterra scheme,  explained in Supplementary Information \ref{app:volterra},
 decays over a few days. The inset reveals
 a significant  $\delta$-contribution at $t=0$  followed by an exponential  decay, which is well described by  a fit (red line) according to 
	\begin{equation}
		\label{eq:meteo_kernel}
		\Gamma(t) = 2 a \delta(t) + \frac{b}{\tau} e^{-t/\tau}.
	\end{equation}
 We find the parameters $a = 1.85$ days$^{-1}$, $b=2.53$ days$^{-1}$ and a memory time of $\tau = 12.07$ days,
which clearly reveals the non-Markovian nature of the daily temperature variation
(see Supplementary Information \ref{app:approx_gle} and Supplementary Information  \ref{app:filtering_examples} for further discussion). 
 The $\delta$-component of  the discrete data in Fig.~\ref{fig:maxTemp1}d (inset) follows from the continuum model equation~(\ref{eq:meteo_kernel}) as 
   $2a/\Delta t$, where $\Delta t=$1 day  is the data  time resolution.

The memory kernel extracted from the discretized Volterra equation depends significantly on the  time discretization  step $\Delta t$ \cite{mitterwallner2020non,klimek2023data, tepper2024accurate},
mostly due to approximating derivatives  by finite-time differences. 
To estimate  GLE parameters $k$, $B$, and $\Gamma(t)$ representative for  the continuum limit, we 
fit the discretized velocity autocorrelation function  (VACF) $C_{\text{VACF}}(i\Delta t)$ calculated from the GLE using a memory kernel $\Gamma(t)$ in the form of equation~(\ref{eq:meteo_kernel}) 
to the data \cite{mitterwallner2020non,klimek2023data},  see Supplementary Information \ref{app:mitterwallner} for details
(by doing so, we approximate  the maximal daily temperature as a  discretized temperature trajectory).  
The perfect agreement between the VACF predicted from the GLE (green symbols, equation~(\ref{eq:vacf3}) in Supplementary Information \ref{app:mitterwallner})
 and calculated from the actual filtered  data $A_\text{f}(t)$ (black symbols) in Fig.~\ref{fig:maxTemp1}e
 shows that the GLE accurately describes the temporal  temperature correlations.
 Similarly, the GLE prediction for the  mean-squared displacement (MSD) of the filtered temperature in Fig.~\ref{fig:maxTemp1}f (green line) agrees very well with the 
actual data (black symbols). 
 The GLE parameters from the Volterra extraction and from the fit of the discretized VACF 
in Table~\ref{tab:fits_kernel_tmax} differ substantially, indicating the significance of discretization effects. Thus, we  use the GLE parameters  
that account for discretization effects   for the following prediction.\\
\indent Three  different  predictability mechanisms  exist, whose relevance is quantified by the characteristic  times scales defined and given in Table~\ref{tab:fits_kernel_tmax}:
i)   The persistence time $\tau_{\rm per}$ is  substantially shorter than $\Delta t$, so prediction by simply extrapolating 
the current temperature change is not possible. 
ii) The relaxation time $\tau_{\rm rel}$ is  substantially longer than $\Delta t$, so prediction by considering the relaxation 
back to $A_\text{f}=0$   is possible  but only if the current $A_\text{f}$ deviates considerably (i.e.  beyond  $\langle A_{\rm f}^2 \rangle ^{1/2}$)  from zero.
iii) The memory time $\tau$ is also  longer than $\Delta t$, which means that the random force $F_R(t)$ 
is predictable,  according to  equation~(\ref{eq:FDT2}).
We conclude that  predictability of  weather  should be possible and is due to a combination of relaxation
 and memory effects but not due to data persistence  (see Supplementary Information \ref{app:predictability} for details). 
For completeness we note that 
 the persistence time $\tau_{\rm per}$ is  considerably shorter than the relaxation time  $\tau_{\rm rel}$,
  thus, the dynamics of $A_\text{f}(t)$ is overdamped and no oscillations in $A_\text{f}(t)$ are expected (in agreement with the data in Fig.~\ref{fig:maxTemp1}b). 

 \begin{table}[!htbp]
		\centering
		\caption{Estimated values of the  memory-kernel and GLE parameters  $a$, $b$, $\tau$,
$k$ and $B$ in equations~(\ref{eq:mori_GLE}) - (\ref{eq:meteo_kernel})  for the daily maximal temperature data shown in Fig.~\ref{fig:maxTemp1}a  
using the Volterra method in Supplementary Information \ref{app:volterra} and the discrete estimation method in Supplementary Information \ref{app:mitterwallner}. 
The persistence time  $\tau_\text{per}$ quantifies the time over which  the dynamics of   $A_\text{f}(t)$ is ballistic,
 the relaxation time  $\tau_{\rm rel}$  the time over which $A_\text{f}(t)$ is correlated. 
 $ \langle A_\text{f}^2 \rangle^{1/2}$ is the standard deviation of $A_\text{f}(t)$.
 The fitting parameters of all weather  data analyzed in this work are summarized in Supplementary Information \ref{app:fit_kernels_params}.}
\begin{ruledtabular}
	\begin{tabular}{l c c } 
				
				  & Volterra & Discrete Estimation \\ 
				
				\hline
				
				$a$ [days$^{-1}$] & 1.85 & 4.31  \\
				$b$ [days$^{-1}$] & 2.53 & 2.07 \\
				$\tau$ [days] & 12.07 & 3.04   \\
				$k$ [days$^{-2}$] & 0.22 & 1.57  \\
				$B$ [K$^{2}$/days$^{2}$] & 9.28 & 29.46  \\
				$\Delta t $ [days] & 1 & 1  \\
				\hline
				$\tau_{\rm per}=1/(a+b)$  [days]&0.23 & 0.16 \\
				$\tau_{\rm rel}=(a+b)/k$  [days]&19.91 & 4.06 \\
				$ \langle A_\text{f}^2 \rangle^{1/2}= (B/k)^{1/2} $ [K] & 6.49 & 4.33\\ 
			\end{tabular}
		\end{ruledtabular}
		\label{tab:fits_kernel_tmax}
\end{table}

	\begin{figure*}[hbt!]
		\centering

		\includegraphics[width=\textwidth]{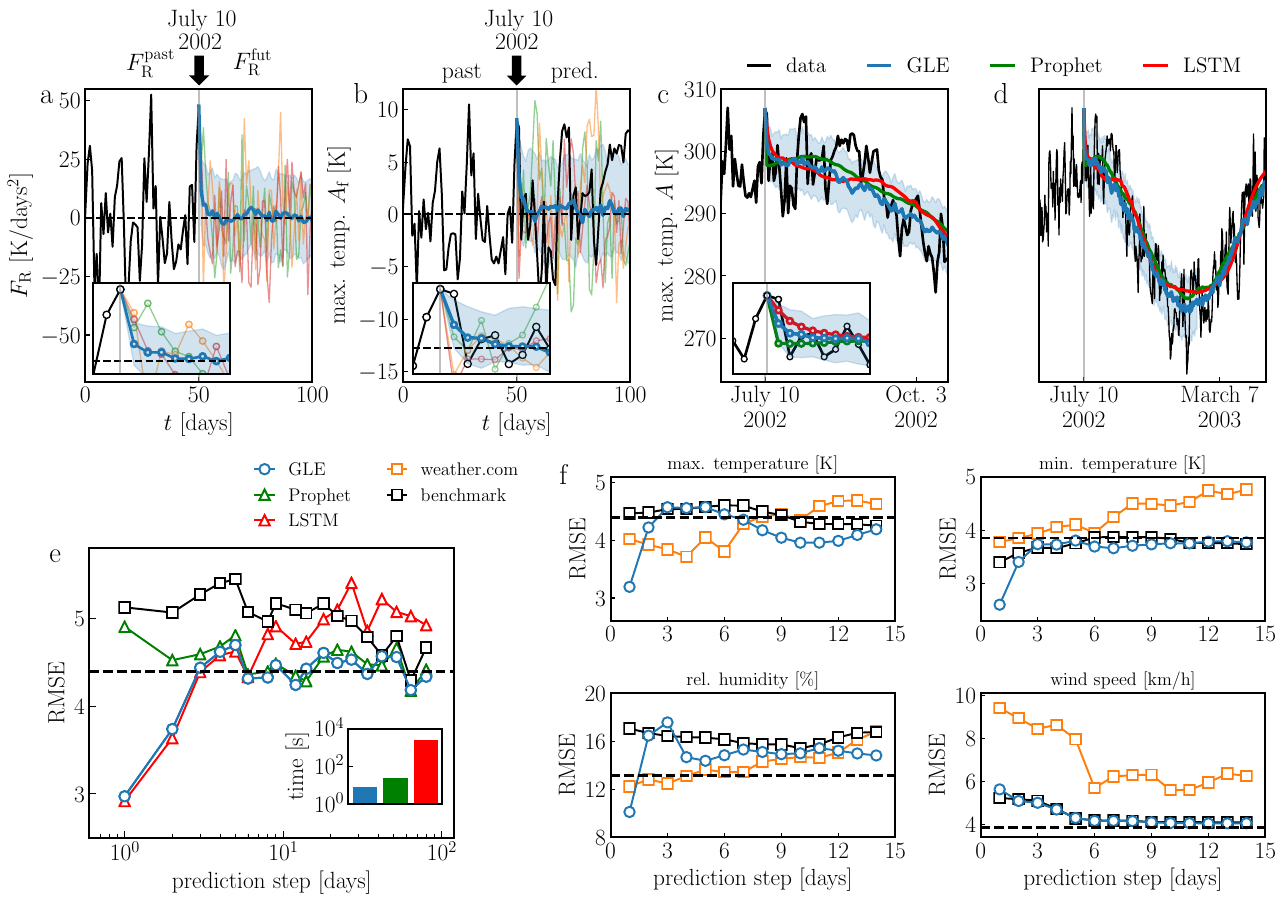} 
		\caption{{\bf\sffamily  Prediction of weather data}.  {\bf\sffamily a}-{\bf\sffamily d}: Illustration of the  GLE prediction scheme (see Methods \ref{sec:GLE}), here for the maximal temperature in Berlin-Tegel shown  in Fig.~\ref{fig:maxTemp1}. 
		{\bf\sffamily a}: The black line shows the past random force $F_\text{R}$, obtained from  equation~(\ref{eq:fr}) in the Supplementary Information \ref{app:SFRF},
		 until July 10, 2002. The colored lines  start on July 10, 2002  and correspond to different realizations of the future random force $F_\text{R}$, 
the blue line shows the mean over  $N_\text{p}$ = 100 realizations  of the future random force.  
The inset visualizes the first 8 prediction steps. The blue-shaded area denotes the standard deviation of the mean prediction.
{\bf\sffamily b}: The black line shows the actual filtered temperature trajectory $A_\text{f}(t)$,
the colored lines  starting on July 10, 2002 show different predictions of the filtered temperature using the random-force trajectories in ({\bf\sffamily a}). 
The blue line represents the average over $N_\text{p}$ = 100 predictions. 
{\bf\sffamily c and d}: Comparison of the predicted  maximal air temperature in Berlin using  an LSTM  neural network (red), 
the Facebook Prophet model  (green),  and the GLE (blue) starting from July 10, 2002 with  the actual data (black). 
 The trajectory until July 10, 2002 was used as a training set for all models. The inset shows the first 20 prediction steps.
			{\bf\sffamily e}: Comparison of the root-mean-squared error (RMSE) of all prediction models as a function of  prediction length. Every RMSE is computed from $N_\text{s}$ = 100 predictions starting from  randomly chosen times between July 19, 1991, and April 3, 2014, using equation~(\ref{eq:RMSE}). Black data show the  benchmark based on  prediction with  a single cosine  fitted to the training data. The inset compares the computation time on a computer with a 16-core CPU  for a prediction over 360 days using different methods. 
{\bf\sffamily f}:  RMSE of 14-day forecasts collected from the   \textit{weather.com}  website (orange) for the maximal and the minimal air temperature, the relative air humidity and the maximal wind speed in Berlin-Tegel, compared with the RMSE of the GLE predictions and the benchmark. The RMSE  are averages over $N_\text{s}$ = 60 predictions  starting from different times between March 9 and May 25, 2020.
The black-dashed horizontal  lines in ({\bf\sffamily e}) and ({\bf\sffamily f})  denote the standard deviation $\sqrt{\langle A_\text{f}^2\rangle}$.
}
		\label{fig:maxTemp2}
	\end{figure*}
	\begin{figure*}[hbt!]
		\centering
		\includegraphics[width=\textwidth]{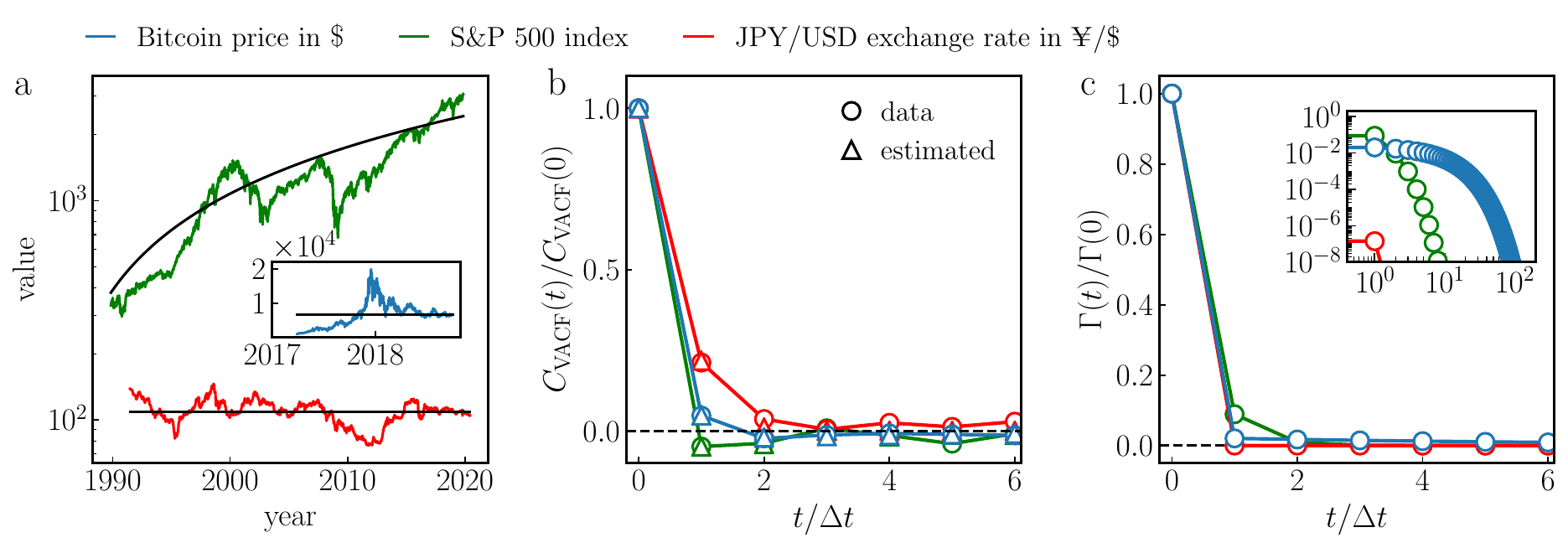}
		\caption{{\bf\sffamily GLE analysis of financial data}. {\bf\sffamily a}: Time-series data  $A(t)$ of the minutely resolved Bitcoin price, the daily resolved Standard \& Poor's 500 (S\&P 500) stock index  and the weekly resolved exchange rate between the Japanese Yen (JPY) and US Dollar (USD).  The black lines are the trends  $A(t)-A_\text{f}(t)$ obtained  from the data using convolution low-pass filtering 
(see Methods \ref{sec:filtering}).  The difference between the data and the trends, the filtered financial data $A_\text{f}(t)$, is used for further GLE analysis. 
			{\bf \sffamily b, c}:  Discrete VACFs $C_{\rm VACF} ( t)$ (circles) of $A_\text{f}(t)$, together with the GLE predictions (triangles) 
using  the discrete parameter estimation procedure described in Supplementary Information \ref{app:mitterwallner}, and the corresponding memory kernels $\Gamma(t)$. 
Note that the time axes are divided by the time resolutions $\Delta t$, for Bitcoin $\Delta t = 1$ minute, for S\&P 500 $\Delta t = 1$ day, and for JPY/USD $\Delta t = 1$ week, respectively. 
The inset in ({\bf\sffamily c}) shows the same data  on a logarithmic vertical scale.}
		\label{fig:finance1}
	\end{figure*}
For the  illustration of our prediction method, we  set July 10, 2002 as the division between past and future. 
For the  prediction, we generate $N_\text{p}$ = 100  realizations of the discrete future  random force $F_\text{R}$ that are conditioned on the past  $F_\text{R}$, obtained by inversion of equation~(\ref{eq:mori_GLE})  \cite{carof_two_2014}
(see Supplementary Information \ref{app:SFRF} for details). 
 Three  future $F_\text{R}$ realizations  are shown in Fig.~\ref{fig:maxTemp2}a (colored lines), 
together with the mean over all $N_\text{p}$ = 100 future  realizations (thick blue line), the  standard deviation (blue-shadowed area) and the past $F_\text{R}$ (black line). 
In the inset, it is seen that the mean over the future $F_\text{R}$ does not vanish and, therefore,  contributes  to the predicted mean future temperature.
In Fig.~\ref{fig:maxTemp2}b,   three predictions for  $A_\text{f}$ using   the random forces  in Fig.~\ref{fig:maxTemp2}a  are presented (light colored lines),
which fluctuate similarly to the actual $A_\text{f}$ trajectory (black); the mean  predicted $A_\text{f}$  (blue line) decays after a few days as seen in the inset. 
In Fig.~\ref{fig:maxTemp2}c,d,  we compare the  mean  temperature $A$ predicted from our GLE (blue) with ML LSTM (red) and Prophet (green) predictions (see Methods \ref{sec:LSTM} and \ref{sec:prophet}). 
 All predictions follow the seasonality of the actual temperature trajectory (black) in (d) but at shorter times deviations are clearly seen in (c), which are due to the stochasticity of the data.
The  prediction accuracy is quantified   in Fig.~\ref{fig:maxTemp2}e, where we compare the root-mean-squared error (RMSE) of  $N_\text{s}$ predictions starting 
 at randomly chosen times $t_i$  between July 19, 1991, and April 3, 2014,  according to 
	\begin{equation}
		\label{eq:RMSE}
		\text{RMSE}( t_j) = \sqrt{\frac{1}{N_\text{s}}\sum_{i=1}^{N_\text{s}} \Bigl\lbrack A(t_i+ t_j) - {A}_\text{pred}( t_i+t_j )\Bigr\rbrack^2},
	\end{equation}
 where we choose $N_\text{s}=100$ and  ${A}(t_i+t_j)$ denotes the actual temperature trajectory while  ${A}_\text{pred}(t_i+t_j)$ denotes the prediction starting at time $t_\text{i}$
 (see Supplementary Information \ref{app:performance_comp},  \ref{app:performance_without_filtering} and \ref{app:model_sys}
  for a discussion  of our  prediction method and   the predictability analysis of a model system). 	
 As a benchmark, we also show  prediction results based on the fit of a single cosine to the data. 
 We see that all prediction tools outperform the benchmark, except LSTM at long time. 
 At short times, Prophet performs poorly compared to LSTM and GLE, which show almost equal performance. 
 For longer  prediction times, all RMSEs  converge to the standard deviation  $ \langle A_\text{f}^2 \rangle^{1/2}$ (broken horizontal line)
 and  all methods  only  predict seasonal effects.  
 The computation time for the  various prediction methods in the inset of Fig.~\ref{fig:maxTemp2}e shows that  the GLE 
 is around $10^3$ times faster than LSTM and $10$ times faster than Prophet. 
Conversely, despite their significantly longer computation time, neither Prophet nor LSTM produce better predictions  than the GLE. \\
\indent In Fig.~\ref{fig:maxTemp2}f, we compare GLE predictions to the two-week  \textit{weather.com}  forecast for
   the maximal temperature,   minimal temperature,  relative humidity and  wind speed 
(all in Berlin, see Supplementary Information \ref{app:kernels_meteo} for the extracted memory kernels). 
For the calculation of  the RMSEs  in equation~(\ref{eq:RMSE}) we use  $N_\text{s}$ = 60 predictions with randomly chosen starting points between March 9 and May 25, 2020. 
It is seen that  the GLE outperforms \textit{weather.com} over the entire prediction range for the minimal temperature and the wind speed;
even for the  maximal temperature and the humidity  the GLE  is more accurate than  \textit{weather.com} for the single-day prediction. 
This is surprising, because the GLE predictions are each based on only a single time-series, namely the  trajectory of the predicted quantity  at the 
location where it is predicted, and it uses no knowledge of the physical principles governing the weather except that it follows Hamiltonian dynamics. 
Importantly, all predictions differ not much from the benchmark RMSE, the latter being a measure of the weather data variance around the seasonal trend,
which reflects that the persistence time $\tau_{\rm per}$,  relaxation time  $\tau_{\rm rel}$ and 
memory time $\tau$ in Table~\ref{tab:fits_kernel_tmax} are rather short compared to $\Delta t$, such that noise effects dominate 
the  weather observables studied here.

\subsection*{ GLE Analysis and Forecasting of Financial Data}   \label{S:Finance}

	\begin{figure*}[hbt!]
		\centering
		\includegraphics[width=\textwidth]{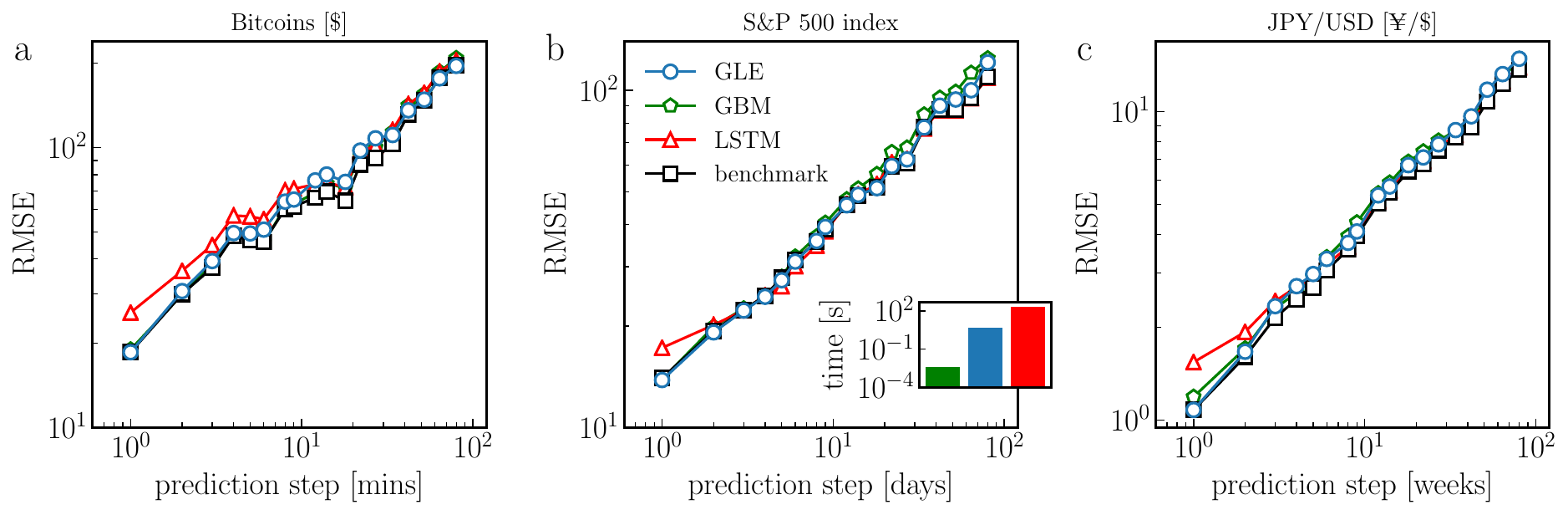}
		\caption{{\bf\sffamily  Prediction of  financial data}. Shown are the root-mean-squared prediction  errors (RMSE)  of the Bitcoin price in $\$$ ({\bf\sffamily a}), 
of the S\&P 500 stock market index in ({\bf\sffamily b}) and of  the exchange rate between Japanese Yen and US Dollar in  ({\bf\sffamily c}) as a function of prediction length
using the GLE (blue), the geometric Brownian motion model (GBM, green) and an LSTM network (red). 
Every  RMSE is the mean over $N_\text{s}$ = 100 different starting times between September 15, 2017 and April 20, 2018, between November 6, 1997 and May 21, 2013, and between April 7, 2000 and July 18, 2014 for Bitcoins, S\&P 500 index and JPY/USD, respectively. 
We compare with the RMSE of a completely  agnostic benchmark model, where the prediction coincides with the  last known value. 
In the inset of ({\bf\sffamily b}), we compare the computation time on a computer with a 16-core CPU for  360  prediction steps using the three different methods.}
		\label{fig:finance2}
	\end{figure*}
In Fig.~\ref{fig:finance1}a, we show trajectories $A(t)$ of three different  financial data,  the minutely resolved Bitcoin price in $\$$ (blue), 
the daily resolved  Standard \& Poor's 500  stock index   (S\&P 500, green) and the  weekly resolved exchange rate between the Japanese Yen and the US Dollar (JPY/USD, red). 
Since the spectral analysis reveals no oscillatory trends (see Methods \ref{sec:filtering}), 
we only apply a low-pass filter to remove the slow data components and to  obtain $A_\text{f}(t)$. 
The memory kernels extracted using  the Volterra method indicate very short exponential decay, see Supplementary Information \ref{app:results_volterra}.
As for the weather data,  by fitting the discrete VACF from the GLE with an exponential memory kernel,  equation~(\ref{eq:meteo_kernel}),  to the data in Fig.~\ref{fig:finance1}b, 
we obtain the memory kernels in (c), which fully account for discretization effects as explained in Supplementary Information \ref{app:mitterwallner}.	
The memory kernels for all systems in Fig.~\ref{fig:finance1}c  consist of a dominant $\delta$-peak at $t=0$ and a very weak exponential contribution, so that the memory  almost entirely decays  in the  first time step,
the  exponential decay time  is  $\tau = 6.11\:\Delta t$, $\tau =  0.43\:\Delta t$ and $\tau =  0.11\:\Delta t$  with $\Delta t=1$ minute, $\Delta t=1$ day, $\Delta t=1$ week 
 for Bitcoin, S\&P 500, and USD/JPY, respectively
(all fit parameters are summarized in Supplementary Information \ref{app:fit_kernels_params}). 
Interestingly, only the  Bitcoin memory kernel  persists  over several time steps, as more clearly seen  in the inset of Fig.~\ref{fig:finance1}c;
the S\&P 500 stock index and the USD/JPY exchange rate are essentially Markovian on their discretization level. 
Thus, our S\&P 500  and  USD/JPY  results are consistent with the efficient market hypothesis,
 a widely accepted  financial theory  that  states that any price pattern could be used to make a profit as soon as the pattern is recognized,
which would  thereby lead to an immediate elimination of the pattern \cite{dimson_brief_1998}. 
Hence, the past does not  contain any information about the future beyond the present price  
and financial data dynamics is Markovian. We conclude that the Bitcoin data does not fully comply with the efficient market hypothesis,
 which might hint at  speculative contributions. \\
 \indent In Fig.~\ref{fig:finance2} we show RMSEs for financial   forecasts (again using $N_\text{p}$ = 100 random force predictions) 
at $N_\text{s}$ = 100 randomly chosen starting points for the Bitcoin price, the S\&P 500 index value and the USD/JPY exchange rate,
comparing GLE predictions (blue) with  LSTM (red)  and geometric Brownian motion (GBM, green) predictions \cite{osborne_brownian_1959}. 
The GBM is a variation of the standard Markovian Brownian motion (BM) model  for a  strictly positive fluctuating variable, as appropriate for  financial data 
(for details on GBM see Methods \ref{sec:GBM}). 
We find no significant differences between the GLE and GBM RMSEs, 
which is expected since the data essentially  is Markovian.
Importantly, both GLE and GBM  perform not better than  the benchmark method (black), 
 which here is taken as the completely agnostic  time-independent forecast corresponding to the present value of $A(t)$. 
This is  anticipated from the GLE analysis, since persistence time $\tau_{\rm per}$, relaxation time $\tau_{\rm rel}$ 
and memory time $\tau$ are smaller than or of the order of
 the observation time step $\Delta t$, see Table \ref{tab:fit_finance} in Supplementary Information   \ref{app:fit_kernels_params}.
Our results also show  that LSTM is not suited for financial predictions of  data, as it looses even against the agnostic benchmark. 
Still, LSTM predictions need about $10^2$ more computational resources than GLE predictions, see inset of Fig.~\ref{fig:finance2}b. 
In summary, the GLE has a clear advantage over LSTM in terms of prediction accuracy and computational demand;
in contrast to GBM, it yields the  persistence, relaxation and memory times of financial data and, thereby, shows  above which time scale the GBM model and the 
efficient market hypothesis start to hold.

\section*{Outlook}
	
With the only assumption  that  the dynamics of the environment is Hamiltonian, we have introduced the GLE as an accurate prediction tool for
arbitrary time-series data, accounting for data discretization, persistence and memory in the data, colored environmental stochasticity   as well as data trends and seasonal effects. 
The extracted GLE parameters are directly interpretable and indicate when simpler models, that neglect memory, relaxation or persistence, as quantified
by the  times $\tau$, $\tau_{\rm rel}$ and $\tau_{\rm per}$, respectively,  are applicable.
The computational demand is substantially  less than ML methods of equal prediction quality, relevant in light of the need for environmental sustainability.

In the future it will be interesting to apply  GLE prediction  to multivariate data and to see whether the simultaneous analysis of correlated data,
as  in recent ML weather prediction    \cite{lam2023learning}, will improve prediction accuracy.
In fact, the GLE  can be interpreted as a time-continuous memory perceptron, which in  multivariate applications 
can  be organized in arrays with hidden layers, very much like standard ML architectures.
  It would, furthermore, be interesting to combine GLE with  state-of-the-art ML  implementations, such as attention-based transformer methods \cite{vaswani2017attention,wu2021autoformer}. 

	\newpage

	\bibliographystyle{unsrt}
	\bibliography{refs}
	\newpage
	\section*{Methods}
%total words: 2807 < 3000 words
\subsection{Filtering  of Phase-Space Observables}
\label{sec:filtering}

We here introduce the basic concepts of the filter formalism for a general phase-space observable $A(\omega,t)$, for a detailed description we  refer to \cite{netz24filtering}.
The starting point is the general time-independent Hamiltonian for a general many-body system of $N$ interacting particles or atoms with masses $m_\text{n}$ in three-dimensional space
\begin{eqnarray}
	\label{eq:Hamiltonian}
	H (\omega) &= \sum_{n=1}^N \frac{\textbf{p}_\text{n}^2}{2 m_\text{n}} + V ( \{ \textbf r_\text{n}\}),
\end{eqnarray}
where $\omega = ( \{ \textbf r_\text{n}\},  \{ \textbf p_\text{n}\}  )$ is a point in the 6$N$-dimensional phase space given by the Cartesian particle positions  $\{ \textbf r_\text{n}\}$ 
and the  conjugate momenta  $ \{ \textbf p_\text{n}\}$. 
$V ( \{ \textbf r_\text{n}\})$ is the potential containing all interactions between the particles and possible external potentials. 
Using  the Liouville operator
\begin{eqnarray}
	\label{eq:Liouville}
	{\cal L}(\omega)&= \sum_{n=1}^N  \sum_{\alpha=x,y,z}  
	\left(   \frac{ \partial H }{\partial p_\text{n}^\alpha }  \frac{ \partial  }{\partial r_\text{n}^\alpha }  -
	\frac{ \partial H }{\partial r_\text{n}^\alpha }  \frac{ \partial  }{\partial p_\text{n}^\alpha } 
	\right),
\end{eqnarray}
the  $6N$-dimensional  Hamilton equation of motion can be expressed as $\dot{\omega}(t) = {\cal L}(\omega)  \omega(t)$, where $\omega(t)$ is the  phase-space location of the system at time $t$ 
with the corresponding phase-space velocity $\dot{\omega}(t) = {\rm d} \omega(t) /   {\rm d} t$.
We denote a general scalar observable as a Schrödinger-type (i.e. time-independent) phase-space function $A_\text{S}(\omega)$, 
which can correspond to  the position of a particle or a molecule, to the reaction coordinate describing  a chemical reaction or the folding of a protein.
For a phase space large enough to contain all relevant atoms, the formalism also works for 
  more complex observables such as the position or state of a living organism,  a meteorological  or a  financial property.
The time-dependent mean of the observable $A_\text{S}(\omega)$ can be written as \cite{zwanzig_nonequilibrium_2001}
\begin{align}
	\label{eq:mean}
	a(t) &  \equiv\int {\rm d} \omega  \, A_\text{S}(\omega) \rho(\omega,t), \nonumber \\
	&=  \int {\rm d} \omega  \, A_\text{S}(\omega) e^{-(t-t_0){\cal L}(\omega)}\rho(\omega,t_0).
\end{align}
The time-dependent phase-space density $\rho(\omega,t)$ obeys the 
Liouville equation $\dot{\rho}(\omega,t) = - {\cal L} (\omega)  \rho(\omega,t)$,
which has the formal solution $ \rho(t) = e^{-(t-t_0){\cal L}(\omega)}\rho(\omega,t_0)$.
Since the Liouville operator is anti-self adjoint, it follows that
\begin{align}
	\label{eq:mean2}
	a(t) &=  \int {\rm d} \omega  \, \rho(\omega,t_0)  e^{(t-t_0){\cal L}(\omega)} A_\text{S}(\omega) ,  \nonumber \\
	& = \int {\rm d} \omega  \, \rho(\omega,t_0)   A(\omega, t),
\end{align}
where $A(\omega, t)$ is the Heisenberg observable
\begin{align}
	\label{eq:Heisenberg}
	A(\omega, t)  &\equiv e^{(t-t_0){\cal L}(\omega)}  A_\text{S}(\omega),
\end{align}
satifying the equation of motion 
\begin{align}
	\label{eq:Heisenberg2}
	\dot{A}(\omega, t) & = {\cal L}(\omega) A(\omega, t).
\end{align}
Since our modest objective is to predict deviations around slow transient and seasonal dynamics, 
we consider the convolution-filtered  observable mean, which using equation~(\ref{eq:mean2}) can be  written as

\begin{figure*}%[hbt!]
	\centering
	\includegraphics[width=0.9\linewidth]{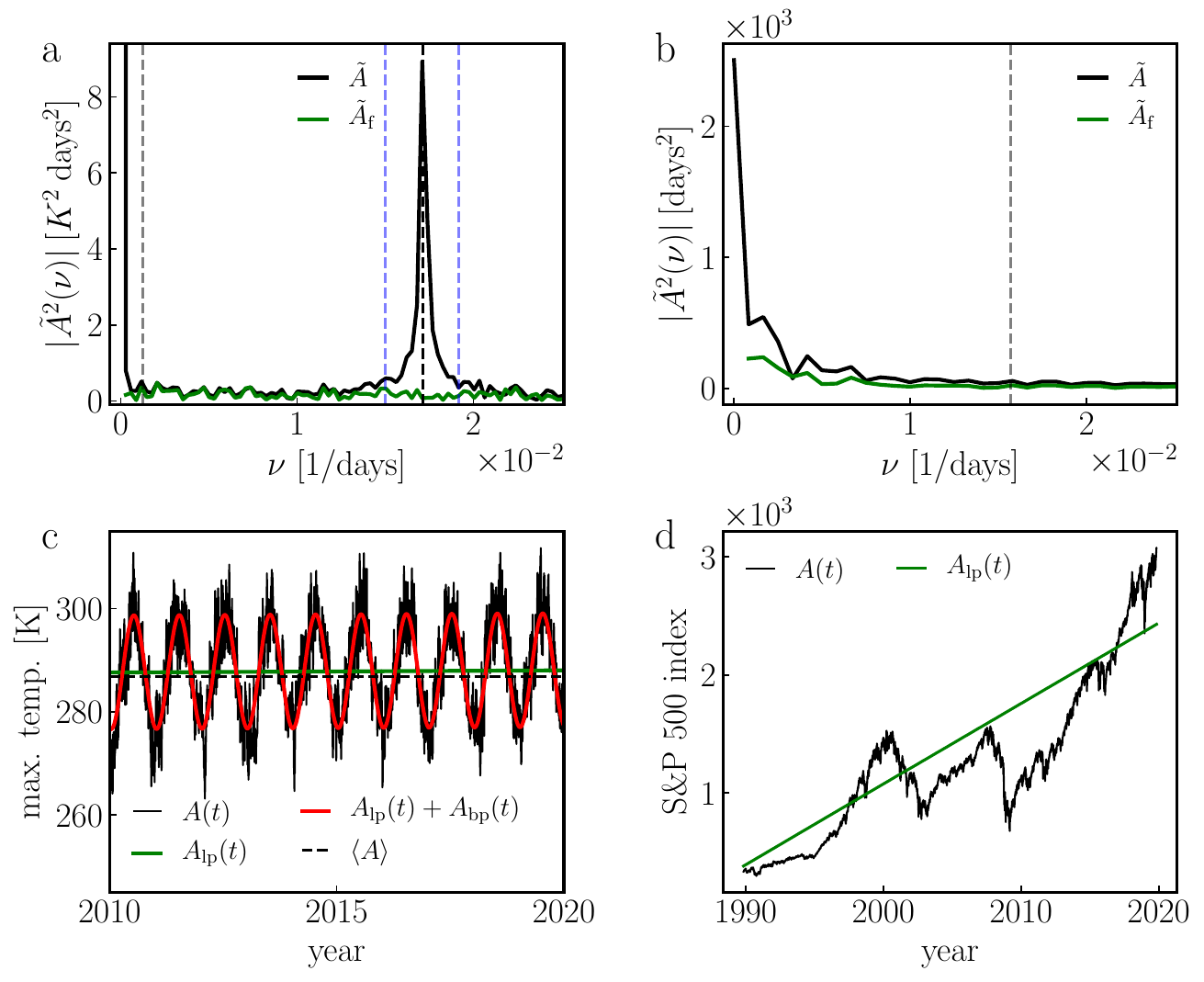}
	\caption{{\bf\sffamily Filtering of weather and financial time-series data}.
		{\bf\sffamily a}, {\bf\sffamily b}: Power spectra of the Fourier-transformed time-series data, $|\Tilde{A}^2(\nu)|$ (black line), 
for the maximal temperature of Berlin ({\bf\sffamily a}) and the Standard $\&$ Poor's 500 index ({\bf\sffamily b}). The green lines denote the filtered data $A_\text{f}$ after removing the low- and band-pass contributions $ \Tilde{A}_\text{lp}$ and $ \Tilde{A}_\text{bp}$ using equations~(\ref{eq:filtered_var}), (\ref{eq:low-pass}) and (\ref{eq:band-pass}). 
For the low-pass filter, we choose a cutoff  $\lambda_\text{lp}$ = 796 days for the temperature data and $\lambda_\text{lp}$ = 64 days for S$\&$P 500 
(vertical gray dashed lines). The vertical black dashed line denotes  the yearly oscillation with a period of  365 days.  For the band-pass filter, 
we choose bandwidths of $\lambda_\text{m}^{-1} = \sqrt{2}10 \pi /(\Delta t N_\text{traj})$ for all weather data, shown  as vertical blue dashed lines. $N_\text{traj}$ is the trajectory length.
For the financial data no periodic effects are seen.
		{\bf\sffamily c}, {\bf\sffamily d}: Time-series data $A(t)$ (black) for the maximal temperature of Berlin ({\bf\sffamily c}) and the Standard $\&$  Poor's 500 index ({\bf\sffamily d}) together with the trend $A_\text{lp}(t)$ (green) and, for the temperature data, the  filter complement $A(t) - A_\text{f}(t) = A_\text{lp}(t) + A_\text{bp}(t)$ (red). 
The horizontal  black dashed line in ({\bf\sffamily c}) denotes  the mean of the  temperature data.}
	\label{fig:filtering_examples_meteo_finance}
\end{figure*}

\begin{align}
	\label{eq:meanfilter4}
	a_\text{f}(t) 
	&\equiv  \int_{-\infty}^{\infty} {\rm d} s \, f(s)  \int {\rm d} \omega  \, \rho(\omega,t_0)   A(\omega, t-s), \nonumber \\
	&=  \int {\rm d} \omega  \,  \rho(\omega,t_0)   \int_{-\infty}^{\infty} {\rm d} s f(s)  A(\omega, t-s), \nonumber \\
	&=  \int {\rm d} \omega  \,  \rho(\omega,t_0)  A_\text{f}(\omega, t ),
\end{align}
where $f(s)$ is a general filter function.
In the last line we defined the filtered Heisenberg observable 
\begin{align}
	\label{eq:meanfilter5}
	A_\text{f}(\omega, t) & \equiv  \int_{-\infty}^{\infty} {\rm d} s \, f(s)  A(\omega, t-s),  
	\\ &=  \int_{-\infty}^{\infty} {\rm d} s \, f(s)  e^{(t-s-t_0) {\cal L}} A_\text{S}(\omega), \label{eq:meanfilter6}
\end{align}
which by construction has the same time-propagation properties as the unfiltered Heisenberg observable in 
equations~(\ref{eq:Heisenberg}) and (\ref{eq:Heisenberg2}), 
allowing us to derive a GLE for the filtered Heisenberg observable $ A_\text{f}(\omega, t)$ in the exact same way as for the ordinary (unfiltered)  
observable $ A(\omega, t)$ (see Methods \ref{sec:Mori}).

Using the definition in equation~(\ref{eq:meanfilter5}), we decompose the unfiltered 
Heisenberg observable $ A(\omega, t)$  into the filtered Heisenberg observable
$A_\text{f}(\omega, t)$ and its complement $A^*_\text{f}(\omega, t) = A(\omega, t) - A_\text{f}(\omega, t)$ according to 
\begin{align}
	\label{eq:decomp}
	A(\omega, t) &= A_\text{f}(\omega, t)+  A^*_\text{f}(\omega, t).
\end{align}
This decomposition is exact and
one  can use several filters to remove all unwanted components from $A(\omega, t)$. 
   We  employ a low-pass filter $\tilde{f}_\text{lp}(\nu)$ to remove   slow trends, which in Fourier space is defined as
\begin{align}
	\label{eq:low-pass}
	\Tilde{f}_\text{lp}(\nu) &= e^{-\lambda_\text{lp}^2\nu^2/2},
\end{align}
where $\lambda_\text{lp}^{-1}$ is the  cutoff frequency  and
  we define Fourier transformations as $\tilde{x}(\nu) = \int_{-\infty}^{\infty} dt\:e^{-i\nu t} x(t)$.
  We remove seasonalities using a sum of band-pass filters
\begin{align}
	\Tilde{f}_\text{bp}(\nu) &= \sum_{m=1}^{M} \Tilde{f}_\text{bp}^m(\nu), \\
	&= \sum_{m=1}^M \left( \frac{e^{-\lambda_\text{m}^2(\nu - \nu_\text{m})^2/2} + e^{-\lambda_\text{m}^2(\nu+\nu_\text{m})^2/2}}{2e^{-\lambda_\text{m}^2\nu_\text{m}^2/2}}\right), 	\label{eq:band-pass}
\end{align}
where $\nu_\text{m}$ are the filter-frequencies and $\lambda_\text{m}^{-1}$ the bandwidths. From equation~(\ref{eq:filtered_var}), which 
in Fourier space  reads
  $\Tilde{A}_\text{f}(\nu) = \tilde{f}(\nu)\Tilde{A}(\nu)$, 
and using the total filter $\Tilde{f}(\nu) = 1 - \Tilde{f}_\text{lp}(\nu) - \Tilde{f}_\text{bp}(\nu)$, 
we obtain  $\Tilde{A}_\text{f}(\nu) $ and from that  $A_\text{f}(t)$  by back-Fourier transformation, 
which we model and forecast with the Mori-GLE in equation~(\ref{eq:mori_GLE}). 

In Fig.~\ref{fig:filtering_examples_meteo_finance} we explain how we choose the low-pass and band-pass filter settings
for the data examined in this work.
In Fig.~\ref{fig:filtering_examples_meteo_finance}a and b, we show the power spectra of the Fourier-transformed time-series data, $|\Tilde{A}^2(\nu)|$ (black line), of the maximal temperature of Berlin (a) and the Standard $\&$ Poor's 500 index (b). A large component around $\nu= 0$ is visible for both data sets, which reflects a combination of a constant
offset and slow trends in the data. We apply the low-pass filter in equation~(\ref{eq:low-pass}) to eliminate this component from the data. We choose a cutoff of $\lambda_\text{lp}$ = 796  days for the temperature data and $\lambda_\text{lp}$ = 64 days  for S$\&$P 500, denoted as vertical gray dashed lines in Fig.~\ref{fig:filtering_examples_meteo_finance}a and b.
For all  financial data we examined,  the spectrum  $\Tilde{A}$ does not exhibit periodic  features at finite frequencies,
as exemplified  in Fig.~\ref{fig:filtering_examples_meteo_finance}b, therefore band-pass filtering is not needed. 
  In contrast, for the weather data we observe pronounced peaks at a finite frequency $\nu_1$,
 as seen   in Fig.~\ref{fig:filtering_examples_meteo_finance}a, which reflects  the expected yearly oscillation 
 of 365 days (vertical black dashed line). 
 We select all peaks with an  intensity    above  10$\%$ of the maximal value of the  power spectrum after low-pass filtering. 
 Except for the wind speed, where we find a second periodic  component with a frequency of 
  $\nu_2 = 2 \pi/54$ days$^{-1}$,  all other weather data only exhibit the yearly periodicity.
  For removing the periodic data contributions  with  the band-pass filter in equation~(\ref{eq:band-pass}),
we choose bandwidths of $\lambda^{-1}_\text{m} = \sqrt{2}10\pi/ (N_\text{traj} \Delta t)$ (where $N_\text{traj}$ is the trajectory length),
  illustrated as vertical blue dashed lines in Fig.~\ref{fig:filtering_examples_meteo_finance}a.
In Fig.~\ref{fig:filtering_examples_meteo_finance}c and d, we compare the original data $A(t)$ (black) 
with  the  low-pass filtered trend $A_\text{lp}(t)$ (green) and, for the temperature data,
with  the filter complement  $A(t) - A_\text{f}(t) = A_\text{lp}(t) + A_\text{bp}(t)$ (red).
 The filtered  data satisfies $\langle{A}_\text{f}(t)\rangle=0$    due to   low-pass  filtering.
 Interestingly, the low-pass filtered trend in Fig.~\ref{fig:filtering_examples_meteo_finance}c 
reveals an increase of the daily maximal temperature by 0.42 K  from 2010 to 2020,
reflecting global warming.

For data forecasting, we fit  the band-pass filtered part $A_\text{bp}(t)$ by a sum of periodic functions given by
\begin{eqnarray}
	\label{eq:seas_part}
	A_\text{bp}^\text{fit}(t) = \sum_{m=1}^M \Bigl\lbrack \alpha_\text{m} \cos{(\frac{2\pi t}{T_\text{m}}+ \phi_\text{m})}  \Bigr\rbrack
\end{eqnarray}
using a least-squares fit,
where initial values of the amplitudes $a_\text{m}$, periods $T_\text{m}$, and phases $\phi_\text{m}$ of the $M$ components are 
determined from   the frequency spectrum of $\tilde{A}(\nu)$.
The low-pass filtered part  $A_\text{lp}(t)$ is fitted by a slowly varying periodic  component 
$A^\text{fit}_\text{lp}(t) = A_\text{lp}^0 + \alpha_\text{lp} \cos{(2\pi t/ T_\text{lp}+\phi_\text{lp})}$. 
The fitted function $A^\text{fit}_\text{lp}(t) + A^\text{fit}_\text{bp}(t)$ is then used to 
forecast $A(t) = A_\text{f}(t) + A_\text{lp}(t) + A_\text{bp}(t)$.
In Supplementary Information \ref{app:filtering_examples}, we  discuss the effect of filtering on weather and financial data in more detail.

\subsection{Derivation of the Mori-GLE}
\label{sec:Mori}
The acceleration of a general Heisenberg observable can be decomposed according to equation~(\ref{eq:decomp}) as 
\begin{align}
	\label{eq:decomp2}
	\ddot A(\omega, t) &= \ddot A_\text{f}(\omega, t)+  \ddot A^*_\text{f}(\omega, t).
\end{align}
This decomposition is  exact  and the  two parts on the right-hand side of equation~(\ref{eq:decomp2}) can be analyzed  separately
and using different approaches. 
In the following, we apply the projection formalism  on 
$\ddot A_\text{f}(\omega, t)$ \cite{netz24filtering}. 
We introduce a projection operator ${\cal P}$ acting on a general phase-space function
and its complementary operator ${\cal Q}$, defined  by the relation $1= {\cal Q} + {\cal P}$.
 From equation~(\ref{eq:meanfilter6}) we obtain  the time propagator relation for the filtered observable 
 \begin{align}
	\label{eq:Heisenfilter}
	e^{(t-t_P)  {\cal L}}A_\text{f}(\omega,t_P) &= e^{(t-t_P){\cal L}}  \int_{-\infty}^{\infty} {\rm d} s \, f(s)  
	e^{(t_P-s-t_0) {\cal L}} A_\text{S}(\omega),\nonumber \\
	&= \int_{-\infty}^{\infty} {\rm d} s \, f(s)  e^{(t-s-t_0) {\cal L}} A_\text{S}(\omega), \nonumber \\
	&=A_\text{f}(\omega,t),
\end{align}
and  the analogue of the Liouville equation for filtered Heisenberg observables
\begin{align}
	\label{eq:Heisenfilter2}
	\dot{A}_\text{f}(\omega, t) & = {\cal L} A_\text{f}(\omega, t).
\end{align}
Inserting $1= {\cal Q} + {\cal P}$  and using equation~(\ref{eq:Heisenfilter2}),  we obtain from equation~(\ref{eq:Heisenfilter})
\begin{align}
	\label{eq:GLE1}
	& \ddot A_\text{f}(\omega, t)= e^{(t-t_\text{P})  {\cal L}} ( {\cal P}+ {\cal Q}) {\cal L}^2 A_\text{f}(\omega,t_\text{P}),  \nonumber \\
	&= e^{(t-t_\text{P})  {\cal L}}  {\cal P} {\cal L}^2 A_\text{f}(\omega,t_\text{P}) + e^{(t-t_\text{P})  {\cal L}}  {\cal Q} {\cal L}^2 A_\text{f}(\omega,t_\text{P}), 
\end{align}
where $t_\text{P}$ defines the time at which the projection is performed, which in principle can 
differ from the time $t_0$ at which the time propagation of the  Heisenberg 
variable in equation~(\ref{eq:Heisenfilter}) starts.
Inserting the Dyson operator 
decomposition 
for the propagator $e^{(t-t_\text{P})  {\cal L}}$,
\begin{align}
	\label{eq:GLE2}
	e^{(t-t_\text{P})  {\cal L}}= e^{(t-t_\text{P})Q{\cal L}} + \int_0^{t-t_\text{P}}\mathrm{d}s\,e^{(t-t_\text{P}-s){\cal L}}P {\cal L} e^{sQ {\cal L}},
\end{align}
into the second term on the right-hand side  in equation~(\ref{eq:GLE1}),
yields the GLE in general form 
\begin{align}
	\label{eq:GLE3}
	\ddot A_\text{f}(\omega, t) & =  e^{(t- t_\text{P})   {\cal L}}  {\cal P}  {\cal L}^2 A_\text{f}(\omega,t_\text{P}) + F_\text{R}(\omega,t) \nonumber \\
	& + \int_0^{t-t_\text{P}} \rm{d}s\, e^{(t-t_\text{P}-s) {\cal L}} {\cal P}  {\cal  {\cal L}}  \textit{F}_\text{R}(\omega,s+t_\text{P}),
\end{align}
with $F_\text{R}(\omega,t)$ given by
\begin{align}
	\label{eq:F_operator}
	F_\text{R}(\omega,t)&  \equiv  e^{(t-t_\text{P}) {\cal Q} {\cal L}} {\cal Q}{\cal L}^2 A_\text{f}(\omega, t_\text{P}).
\end{align}
Equation~(\ref{eq:GLE3}) is an exact decomposition of 
 $\ddot A_\text{f}(\omega, t)$ into three terms: The first term describes the evolution of   $\ddot A_\text{f}(\omega, t)$ in the
 relevant subspace spanned by ${\cal P}$
  and reflects the deterministic mean force due to a potential, while the third term represents non-Markovian friction effects. The second term $F_\text{R}(\omega,t)$ 
includes all effects that are not contained in the other two terms.   
The form of the GLE in equation~(\ref{eq:GLE3}) depends on the specific form of the projection operator 
$\cal P$.  Here we choose the Mori projection, which  applied on a Heisenberg observable $B(\omega,t)$ reads 
 \cite{mori1965transport, vroylandt2022derivation, ayazhybrid2022}
\begin{align}
	\label{eq:mori_projection}
	&{\cal P}  B(\omega,t) = \langle   B(\omega,t)  \rangle
	+  \frac{\langle   B(\omega,t)  {\cal L} A_\text{f} (\omega, t_\text{P})  \rangle}  {\langle ( {\cal L}A_\text{f} (\omega, t_\text{P}) )^2 \rangle} 
	{\cal L}A_\text{f}  (\omega, t_\text{P})  \nonumber \\
	&  +\frac{\left\langle   B(\omega,t)  (A_\text{f} (\omega, t_\text{P}) - \langle A_\text{f}(\omega)\rangle )\right\rangle}  {\langle (A_\text{f}  (\omega, t_\text{P}) - \langle A_\text{f}(\omega) \rangle )^2 \rangle} 
	(A_\text{f}  (\omega, t_\text{P}) - \langle A_\text{f}(\omega) \rangle ).
\end{align}
 The expectation value of an arbitrary phase-space function $X(\omega) $ 
with respect to a time-independent projection distribution 
$ \rho_{\rm P}(\omega)$ is defined as
\begin{align}
	\langle   X(\omega)  \rangle = 
	\int {\rm d}\omega  X (\omega) \rho_{\rm P}(\omega),
\end{align}
where we choose $ \rho_{\rm P}(\omega)$ to be the equilibrium canonical distribution
$\rho_{\rm P}(\omega)= e^{-H(\omega)/(k_BT)}/ Z $ with  $Z$ being the partition function. 
The Mori projection is linear,
i.e. for two arbitrary observables $B(\omega,t)$ and $C(\omega,t')$ it satisfies
${\cal P}\left(c_1 B(\omega,t) + c_2 C(\omega,t')\right)=c_1 {\cal P} B(\omega,t)  + c_2 {\cal P} C(\omega,t')$,
it is idempotent, i.e.  ${\cal P}^2= {\cal P}$, and it is self-adjoint, i.e. it satisfies  the relation
$\langle B(\omega,t) {\cal P} C(\omega,t')  \rangle = \langle C(\omega,t') {\cal P} B(\omega,t)   \rangle$ \cite{netz24filtering}.
From these properties it follows that the complementary  projection
operator ${\cal Q}=1-{\cal P}$ is also linear,  idempotent and self-adjoint.
Thus, ${\cal P}$ and ${\cal Q}$ are orthogonal to each other, i.e. 
${\cal P}{\cal Q}= 0 = {\cal Q}{\cal P}$ \cite{netz24filtering}. 
With all these properties, the GLE in equation~(\ref{eq:GLE3}) becomes,
choosing $B(\omega,t) = A_\text{f}(\omega,t)$,
\cite{mori1965transport}
\begin{align} \label{eq:mori_GLE2}
	\ddot A_\text{f}(\omega, t) & = -  k   (A_\text{f}(\omega,t) - \langle  A_\text{f}(\omega)\rangle)  + F_\text{R}(\omega,t) \nonumber \\
	& - \int_0^{t-t_\text{P}} {\rm d}s\, \Gamma(s) \dot A_\text{f}(\omega, t-s).
\end{align}
The parameter $k$ is the potential stiffness defined by 
\begin{align} \label{GLEK}
	k &= \frac{\langle ( {\cal L}A_\text{f} (\omega,t_\text{P}) )^2 \rangle}
	{\langle (A_\text{f} (\omega,t_\text{P}) - \langle A_\text{f} (\omega) \rangle)^2 \rangle}.
\end{align}
The memory kernel $\Gamma(t)$ is related to the second moment of the  force $F_\text{R}(\omega,t)$ defined in equation~(\ref{eq:F_operator})
\begin{align} 
	\Gamma(t) = 
	\frac{\langle F_\text{R}(\omega,0) F_\text{R}(\omega, t) \rangle}{\langle ( {\cal L}A_\text{f} (\omega,t_\text{P}) )^2 \rangle}.
	\label{eq:mori_memory}
\end{align}
Due to the specific form of the Mori projection in equation~(\ref{eq:mori_projection}),
several expectation values involving $F_\text{R}(\omega,t)$ vanish \cite{netz24filtering}, namely 
$\langle F_R(\omega,t)\rangle = \langle   F_\text{R}(\omega,t)A_\text{f}(\omega,t_\text{P}) \rangle =
\langle   F_\text{R}(\omega,t){\cal L}A_\text{f}(\omega,t_\text{P}) \rangle =0$, which 
are important properties for extracting GLE parameters from time-series data (see Supplementary Information \ref{app:volterra}).
For practical applications, one  models the force $F_R(\omega,t)$ as a stochastic process  with 
zero mean and a second moment given by equation~(\ref{eq:mori_memory}),
 leading to equations~(\ref{eq:FDT2}) and  (\ref{eq:mori_GLE}), where we omitted the phase-space dependence of all variables
 and set  the projection time to $t_\text{P} = 0$, see \cite{netz24filtering} for more details. 
Often, the distribution of $F_\text{R}(\omega,t)$ is assumed to be Gaussian and higher-order moments of $F_\text{R}(\omega,t)$ are neglected.
For non-linear systems this assumption can not hold, 
since $F_\text{R}(\omega,t)$ is the only term in the Mori-GLE that accounts for non-linearities \cite{ayazhybrid2022, vroylandt2022derivation}. 
In our case, the filtering renders the data effectively Gaussian, so non-Gaussian contributions  in $F_\text{R}(\omega,t)$
can indeed be neglected.

\subsection{Predictions Using the GLE}
\label{sec:GLE}
Here we  present the prediction algorithm for a general discretized trajectory $A(t)$. First, the filtered part $A_\text{f}(t)$ is calculated from  $A(t)$ 
using the filtering procedure  explained in Methods \ref{sec:filtering}.
Our forecasting scheme directly solves the GLE in equation~(\ref{eq:mori_GLE}) for future times. 
Having a discrete trajectory $A_\text{f}(t)$ up to a given time $t_l$,  we compute
 the discrete velocity autocorrelation function 
 $C^{\dot{A}_\text{f}\dot{A}_\text{f}}(i \Delta t) = \langle \dot{A}_\text{f}(\Delta t/2)\dot{A}_\text{f}((i+1/2)\Delta t) \rangle$ from the discrete velocities 
 $\dot{A}_\text{f}((i+1/2)\Delta t) = (A_\text{f}((i+1)\Delta t) - A_\text{f}(i\Delta t))/\Delta t$.
 First, the memory kernel is obtained by an extraction  
  using the Volterra method, explained in Supplementary Information \ref{app:volterra},
  which demonstrates that the  kernel functional form in equation~(\ref{eq:meteo_kernel}) is suitable  for all data sets.
Due to discretization effects,  the kernel obtained from the Volterra method does not correspond to the continuum-limit kernel which appears in the GLE.  
The parameters $k$, $B$ and the memory kernel $\Gamma(t)$ valid in the continuum  limit are obtained using the discrete estimation method, 
as  described in Supplementary Information \ref{app:mitterwallner}. 
Next, the past random forces $F_\text{R}^\text{p}$  are  computed
from the GLE and the past trajectory $A_\text{f}(t)$, see Supplementary Information \ref{app:SFRF}) for details. 
We then draw a set of future random forces $F_\text{R}^\text{f}$ using a Gaussian sampling method
and  the relation in equation~(\ref{eq:FDT2}) between the random force and the memory kernel.
In Supplementary Information \ref{app:fdt_disc}, we show how to correct for discretization effects and  how to compute 
 past random forces $F_\text{R}^\text{p}$ that correspond to the continuum  limit, which we then evaluate at
discrete time steps and  use for the computation of  realizations of the  discrete future random force $F_\text{R}^\text{f}$. 
Using the random-force correction explained  in Supplementary Information \ref{app:fdt_disc},
the future random forces become Gaussian,  as discussed in Supplementary Information \ref{app:non_gauss},
which shows that low-pass filtered observables are Gaussian as well.
This is interesting since unfiltered  complex data,
especially  financial data \cite{gopikrishnan1999scaling}, 
do often  exhibit pronounced non-Gaussian behavior. 

We discretize the GLE in equation~(\ref{eq:mori_GLE}) as
\begin{equation}
	\Ddot{A}_\text{f}^\text{i}= - k A_\text{f}^\text{i} - \Delta t \sum_{j=0}^\text{i} \omega_\text{i,j}\Gamma_\text{j}  \Dot{A}_\text{f}^\text{i-j} +F_\text{R}^\text{i},
	\label{eq:ODE_discret}
\end{equation}
and solve this discretized form  for $A_\text{f}$ using a Runge-Kutta 4$^{\text{th}}$-order (RK4) integrator.
The discrete velocities and accelerations are computed via finite differences from the discretized trajectory $A_\text{f}^\text{i} =  A_\text{f}(i\Delta t)$
as $\Dot{A}^\text{i}_\text{f} = \Dot{A}_\text{f}(i\Delta t) = \bigl[A_\text{f}\bigl((i+1)\Delta t\bigr) - A_\text{f}\bigl((i-1)\Delta t\bigr)\bigr]/(2\Delta t)$ and $\Ddot{A}^\text{i}_\text{f} = \Ddot{A}_\text{f}(i\Delta t) = \bigl[A_\text{f}\bigl((i+1)\Delta t\bigr) - 2A_\text{f}(i\Delta t) + A_\text{f}\bigl((i-1)\Delta t\bigr)\bigr]/(\Delta t)^2$.
The expression $\omega_\text{ij} = 1 - \delta_\text{ij} /2 - \delta_\text{0j} /2$ corresponds to the trapezoidal rule for the integral in equation~(\ref{eq:mori_GLE}). 
Starting with the last known position $A_\text{f}^l$ at time $t_l =  l \Delta t$, 
we determine a realization of the future random force $F_\text{R}^\text{l+1}$ and obtain the next trajectory point $A_\text{f}^\text{l+1}$ 
according to  equation~(\ref{eq:ODE_discret}).
For multi-step predictions, we treat the predicted value $A_\text{f}^\text{l+1}$ as past value for the next prediction step $A_\text{f}^\text{l+2}$, and so forth. 
The predictions using  $N_\text{p}$ different realizations of the random force are then averaged to obtain the mean prediction of the future $A_\text{f}$. 
We choose $N_\text{p} = 100$ since the performance does not change significantly for higher numbers of realizations 
(see Supplementary Information \ref{app:performance_avg_size}). 
To accelerate the numerical solution of equation~(\ref{eq:ODE_discret}), we truncate the memory kernel after a length of $\tau_\text{t}$, which is the time after the memory kernel has decayed to zero. For the weather data, as well as for the financial data, we choose a value of $\tau_\text{t} = 10\Delta t$.

 With  our extraction method  we determine  the  kernel $\Gamma(t)$, the stiffness $k$, and the mean-squared velocity $B$ in the continuum limit;
 we also calculate the continuum-limit past   random force at  discrete time steps 
 and use that to calculate  the  continuum-limit future  random force at discrete time steps.
 The numerical solution of the discretized GLE in equation~(\ref{eq:ODE_discret}) using the RK4 method indeed produces  trajectories
 that correspond to  dynamics in the continuum limit, as shown in Supplementary Information \ref{app:test_RK4}. 
Our treatment of discretization effects is thus consistent.
 
 In Supplementary Information \ref{app:POC} we demonstrate the self-consistency of our prediction technique 
by showing that the predicted future  trajectory  $A_\text{f}$ has the same  properties as the past trajectory, 
i.e. that it exhibits the same position and velocity autocorrelation functions and the same extracted memory kernel. 
In Supplementary Information \ref{app:model_sys}, we demonstrate that the prediction  using  a non-Markovian random force 
is superior to using a Markovian random force
  when memory times  are  longer than the discretization  time $\Delta t$.

 Finally, the predicted filtered  trajectory $A_\text{f}(t)$ is added to the fitted and extrapolated  slow trends and periodic contributions 
 $A_\text{lp}(t) + A_\text{bp}(t)$ (see Methods \ref{sec:filtering}) to yield  the prediction of the unfiltered observable $A(t)$.

\subsection{Long Short-Term Memory}
\label{sec:LSTM}
Long short-term memory (LSTM) networks are special types of
recurrent artificial neural networks mainly used for
processing and modeling time-dependencies in time-series data \cite{hochreiter1997long}.
 Our LSTM model consists of three layers: An input layer with the dimension of previous steps $n_{\text{prev}}$, followed by an LSTM hidden layer \cite{schuster1997bidirectional} with an adjustable number of neural units $n_{\text{nu}}$, and an output layer with output dimension $n_{\text{out}}=1$
  with a rectified linear unit (ReLu) activation function.

In the training process, we use the full available historical data until the starting point of the prediction as the training set. 
The output value $t_{\text{n}_{\text{prev}}+1}$ of the neural network, which is generated from a sample of previous time steps $t_0,..., t_{\text{n}_{\text{prev}}}$, 
is compared with the actual value. 
As a loss function, the mean-squared error between the target and the obtained output values is minimized using the
backpropagation algorithm RMSprop \cite{wilson_marginal_2018}, which adjusts the weights and biases of the neural network. In the training process, we set the epoch number to 2000. The training process is aborted earlier if the loss does not change significantly after 200 epochs (early stopping callback). In addition, we set the batch size to 128. To prevent overfitting, we include a dropout layer with an adjustable dropout rate between the hidden and the
output layer \cite{srivastava_dropout_2014}. 

 We normalize the data in the training set to their minimum and maximum values, such that the value range lies between 0 and 1. For generating forecasts of multiple time steps after the last known step, we use the optimized LSTM network to predict a step into the future and feed the prediction back at the end of the time window while cutting out the first observation at the
beginning of the window.  
As discussed in Supplementary Information \ref{app:perfornance_lstm},
to achieve a performance comparable to the GLE, for the financial data, we choose an input sample length of $n_{\text{prev}}$ = 10 previous values, $n_{\text{nu}}$ = 32 neural units in the hidden layer and a dropout rate of 0.15, which gives a total number of 8,769 parameters to be optimized during  training. 
For the  weather data in Berlin-Tegel in Fig.~\ref{fig:maxTemp2}, we choose  $n_{\text{prev}}$ = 200, $n_{\text{nu}}$ = 32  and a dropout rate of 0.15.

\subsection{Facebook Prophet}
\label{sec:prophet}
In the Prophet forecast algorithm, the time-series data is modeled as
\begin{equation}
	\label{eq:Prophet}
	A(t) =A_\text{T}(t) + A_\text{S}(t) + A_\text{f}(t),
\end{equation}
where $A_\text{S}(t)$ is the seasonal variation, $A_\text{T}(t)$ is the trend, and $A_\text{f}$ is the model residual
(here we neglect holiday effects).
We model the trend by a linear function, as suggested by our low-pass filtering results in Methods \ref{sec:filtering}.
The seasonal variations are described by a  Fourier series according to 
\begin{equation}
	A_\text{S}(t) = \sum_{m=1}^M \Bigl\lbrack a_\text{m} \cos{(\frac{2\pi mt}{P})} + b_\text{m} \sin{(\frac{2\pi mt}{P})}\Bigr\rbrack,
\end{equation}
where we have $2M$ Fourier coefficients $\beta = [a_1,b_1,...,a_\text{M},b_\text{M}]$, and $P$ is here the yearly seasonality. 
The parameters of $A_\text{T}(t)$  and  $A_\text{S}(t)$   in equation~(\ref{eq:Prophet}) are estimated
 using their likelihood and improved iteratively by the maximum aposteriori estimation (MAP) optimization scheme. 
Prophet uses the quasi-Newton optimization method L-BFGS. 
For a detailed optimization description, we refer to the original publication \cite{taylor_forecasting_2018}.
Using the fit results for  $A_\text{T}(t)$  and  $A_\text{S}(t)$, we can extrapolate the function $A(t)$ into the future.

\subsection{Data Sets} \label{sec:data}
All time-series data shown in this work was  obtained  free of charge from the Internet at the time this paper was written. 
The JPY/USD exchange rate trajectory from June 21, 1991 to June 12, 2020 was downloaded from the website \textit{quandl.com}. 
We obtained the S\&P 500 index data from November 6, 1989 to November 5, 2019 on the website \textit{stooq.com}. The minutely resolved Bitcoin price from March 23, 2017 to October 2, 2018 was provided by the Bitfinex application programming interface  (\textit{bitfinex.com}). The weather data from January 1, 1963 to May 4, 2021 in Berlin-Tegel was downloaded from the historical data platform of Deutscher Wetterdienst (\textit{opendata.dwd.de}). We downloaded daily forecasts of \textit{weather.com} using a web crawler.

\subsection{Geometric Brownian Motion\label{sec:GBM}}
The geometric Brownian motion (GBM) model  is a variant of the Brownian motion (BM) model \cite{mandelbrot_variation_1963}.
 Unlike BM, the fluctuating variable in GBM is strictly  positive, an important property when modeling financial assets. 
 GBM describes a Markov process, i.e. it fulfills the efficient market hypothesis.  Consider the random variable $x(t)$ that describes Brownian motion with a drift ${\mu-\sigma^2/2}$ and diffusion coefficient $\sigma$, which  obeys the stochastic differential equation $\mathrm{d}x(t) = (\mu-\sigma^2/2)\,\mathrm{d}t + \sigma\, \mathrm{d}W(t)$ with $W(t)$ being a Wiener process.
 The random variable $A(t)=A(0)e^{x(t)}$ then describes  geometric Brownian motion. 
Using Ito's Lemma, it can be shown that $A(t)$  obeys the stochastic differential equation $\mathrm{d}A(t) = \mu\mathrm{d}t+\sigma\mathrm{d}W(t)$,
 which has the solution $A(t) = A(0) \exp\left[(\mu - \sigma^2/2)t + \sigma W(t)\right]$. For $t,s>0$ and $t>s$ the solution
can be written  as $A(t)=A(s)\exp\left[(\mu-\sigma^2/2)(t-s) + \sigma(W(t)-W(s))\right]$, which in discretized form reads
\begin{eqnarray}
	A_\text{n+1} = A_\text{n} e^{(\mu-\sigma^2/2)\Delta t + \sigma\,\sqrt{\Delta t}\, Z_\text{n}},
\end{eqnarray}
where all $Z_\text{n}$ are normally distributed random numbers. Different ways to estimate the drift and diffusion coefficients exist. We make use of the first two moments of $A$,
which are given by
\begin{align}
	\ln\left(\frac{\langle A\rangle}{A_0}\right) &= \mu + \sigma^2/2, & \frac{\langle A^2\rangle - \langle A\rangle^2}{\langle A^2\rangle} &= 1- e^{-\sigma^2}.
\end{align}
As for the GLE forecast, a GBM forecast corresponds to  the mean over  $N_\text{p}=100$ different realizations.

\section*{Code and Data Availability}
The custom computer codes and the datasets generated and analyzed during the current study are available from the corresponding author on request. A Python package for the GLE extraction and prediction introduced in the present work is available at GitHub: https://github.com/hkiefer/mempred.
\section*{Acknowledgements}
We gratefully acknowledge support by the Deutsche Forschungsgemeinschaft (DFG) via Grant No. SFB 1114 and No. SFB 1449, by the European Research Council (ERC) under the European Union’s Horizon 2020 Research and Innovation Program under Grant Agreement No. 835117 and by the Infosys Foundation. We gratefully acknowledge computing time on the HPC clusters at the Physics Department and ZEDAT, FU Berlin.
\vspace{5mm}
\section*{Author Contributions}
H.K., D.F., and R.R.N. designed the simulations and analysis. H.K., A.K., C.A., J.O.D., and R.R.N. conceived the theory. H.K. and D.F. performed the simulations and analysis. H.K. designed the figures. All authors discussed the results, analyses, and interpretations. H.K. and R.R.N. wrote the paper with input from all authors.

\section*{Competing Interests}
The authors declare no competing interests.
\section*{Additional information}

\noindent \textbf{Supplementary Information} is attached as seperated file.\\
\textbf{Correspondence} and requests for materials should be addressed to Roland R. Netz.

\newpage
%\begin{widetext}
%\appendix
%%%%%%%%%% Merge with supplemental materials %%%%%%%%%%
\pagebreak
\newpage
\widetext
\begin{center}
	\textbf{\large Supplementary Information for: Predictability Analysis and Prediction of Discrete Weather and Financial Time-Series Data with a Hamiltonian-Based Filter-Projection Approach}\\
	\vspace{5mm}
	\textbf{Henrik Kiefer$^1$, Denis Furtel$^1$, Cihan Ayaz$^1$, Anton Klimek$^1$, Jan O. Daldrop$^1$, Roland R. Netz$^{1,2}$}\\
	\vspace{1mm}
	\textit{$^1$ Freie  Universität  Berlin,  Department  of  Physics,  Arnimallee  14,  14195  Berlin,  Germany}\\
	\textit{$^1$ Centre for Condensed Matter Theory, Department of Physics, Indian Institute of Science, Bangalore 560012, India}\\
	
\end{center}

%%%%%%%%%% Merge with supplemental materials %%%%%%%%%%
%%%%%%%%%% Prefix a "S" to all equations, figures, tables and reset the counter %%%%%%%%%%
\setcounter{equation}{0}
\setcounter{figure}{0}
\setcounter{table}{0}
\setcounter{page}{1}
\makeatletter
\makeatletter
\renewcommand{\thepage}{S\arabic{page}}
\renewcommand{\thesection}{S\arabic{section}}
\renewcommand{\theequation}{S\arabic{equation}}
\renewcommand{\thefigure}{S\arabic{figure}}

	\section{Memory Kernel Extraction with the Volterra Method}
\label{app:volterra}
Numerous methods for extracting the memory kernel $\Gamma(t)$ from a given one-dimensional trajectory $A(t)$ exist \cite{schmitt_analyzing_2006, gottwald_parametrizing_2015, lei2016data,jung2017iterative}, one of which is a numerically robust approach making use of time-correlation functions \cite{daldrop_butane_2018, lee_multi-dimensional_2019, berne1970calculation, straub1987calculation}. Multiplying equation~(\ref{eq:GLE}) in Supplementary Information \ref{app:approx_gle} by the initial velocity $\dot{A}(0)$, ensemble averaging the result and leveraging the condition $\langle \dot{A}(0)F_\text{R}(t)\rangle=0$ \cite{mori1965transport} yields a Volterra equation of first kind
\begin{eqnarray}
	\label{eq:Volterra}
	C^{\Dot{A}\Ddot{A}}(t) = - \int_0^\text{t} dt'\: \Gamma(t')C^{\Dot{A}\Dot{A}}(t-t') - C^{\Dot{A}\nabla U[A(t)]}(t),
\end{eqnarray}
where time-correlation functions are denoted as $C^{AB}(t)=\langle A(0)B(t)\rangle$ and $C^{\dot{A}\dot{A}}(t) = C_\text{VACF}(t)$ is the VACF used in the main text. Discretizing equation~(\ref{eq:Volterra}) and using the trapezoidal rule for the memory integral, we obtain
\begin{eqnarray}
	C_\text{i}^{\Dot{A}\Ddot{A}} = - \Delta t\sum_\text{j=0}^{i-1}\omega_\text{i,j} \Gamma_\text{j} C^{\Dot{A}\Dot{A}}_\text{i-j} - C^{\Dot{A}\nabla U}_\text{i},
\end{eqnarray}
where $C^{AB}_\text{i} = \langle A_0B_\text{i}\rangle$ is the discretized two-point correlation function, and $\omega_\text{i,j} = 1 - \frac{\delta_\text{0,j}}{2} -  \frac{\delta_\text{i,j}}{2}$ follows from the trapezoidal rule. The subscript $i$ refers to the discrete time-series step $i\Delta t$. Finally, after rearranging, we obtain the iterative formula
\begin{eqnarray}
	\label{eq:extraction}
	\Gamma_\text{i} = - \frac{2}{\Delta t C^{\Dot{A}\Dot{A}}_0}\Bigl(C^{\Dot{A}\nabla U}_\text{i}+C_\text{i}^{\Dot{A}\Ddot{A}} + \frac{\Delta t}{2} \Gamma_0 C^{\Dot{A}\Dot{A}}_\text{i} +  \Delta t\sum_{j=1}^{i-1}\Gamma_\text{j} C^{\Dot{A}\Dot{A}}_\text{i-j}\Bigr),
\end{eqnarray}
for the discrete memory kernel, using the correlation functions from the discrete velocities 
$\Dot{A}_\text{i} = \Dot{A}(i\Delta t) = \bigl[A\bigl((i+1)\Delta t\bigr) - A\bigl((i-1)\Delta t\bigr)\bigr]/(2\Delta t)$ and accelerations $\Ddot{A}_\text{i} = \Ddot{A}(i\Delta t) = \bigl[A\bigl((i+1)\Delta t\bigr) - 2A(i\Delta t) + A\bigl((i-1)\Delta t\bigr)\bigr]/(\Delta t)^2$.
From the time derivative of equation~(\ref{eq:Volterra}) at $t=0$, we obtain the expression for the initial value of the memory kernel $\Gamma_0$
\begin{eqnarray}
	\Gamma_0 = \frac{C^{\Ddot{A}\Ddot{A}}_{0} + C^{\Ddot{A}\nabla U}_{0}}{C^{\Dot{A}\Dot{A}}_{0}}.
\end{eqnarray}
The correlation $C^{\dot{A}\nabla U}_\text{i}$ is calculated by
interpolation of the effective potential \cite{terrell1992variable}, which is derived from the probability distribution $\rho(A)$, according to $U(A) = -\langle \dot{A}^2\rangle\:$log$\:\rho(A)$.\\ 
\indent From the Mori-GLE for the filtered observable $A_\text{f}$ in equation~(\ref{eq:mori_GLE}) in the main text, where $\nabla U(A_\text{f}) = k(A_\text{f}-\langle A_\text{f} \rangle)$, the iterative equation for the discrete memory kernel is derived as similarly for equation~(\ref{eq:extraction}), and is given by

\begin{eqnarray}
	\label{eq:extraction2}
	\Gamma_\text{i} = - \frac{2}{\Delta t C^{\Dot{A}_\text{f}\Dot{A}_\text{f}}_0}\Bigl(k C^{\Dot{A}_\text{f}A_\text{f}}_\text{i}+C_\text{i}^{\Dot{A}_\text{f}\Ddot{A}_\text{f}} + \frac{\Delta t}{2} \Gamma_0 C^{\Dot{A}_\text{f}\Dot{A}_\text{f}}_\text{i} +  \Delta t\sum_{j=1}^{i-1}\Gamma_\text{i-j} C^{\Dot{A}_\text{f}\Dot{A}_\text{f}}_\text{j}\Bigr),
\end{eqnarray}
with the initial value given by 
\begin{eqnarray}
	\Gamma_0 = \frac{C^{\Ddot{A}_\text{f}\Ddot{A}_\text{f}}_{0} + k C^{\Ddot{A}_\text{f}A_\text{f}}_{0}}{C^{\Dot{A}_\text{f}\Dot{A}_\text{f}}_{0}}. 
\end{eqnarray}
Note that we set $\langle A_\text{f} \rangle = 0$, which holds for high-pass filtered observables.
The harmonic strength $k$ is defined by the ratio between the variances of velocity and position, i.e. $k = \langle \Dot{A}_\text{f}^2 \rangle/\langle A_\text{f}^2 \rangle $.  Note that in previous work, an extraction method for the running integral of the memory kernel $G(t)$, i.e. $G(t) = \int_0^t ds\:\Gamma(s)$, was used \cite{kowalik2019memory, ayaz2021non}, which contains fewer correlation functions to compute, but is inconvenient in the presence of delta-components in the memory kernel \cite{klimek2023data,tepper2024accurate}.
\section{Memory Kernel Extraction with the Discrete Estimation Method}
\label{app:mitterwallner}
\begin{figure*}[hbt!]
	\centering
	\includegraphics[width=1\linewidth]{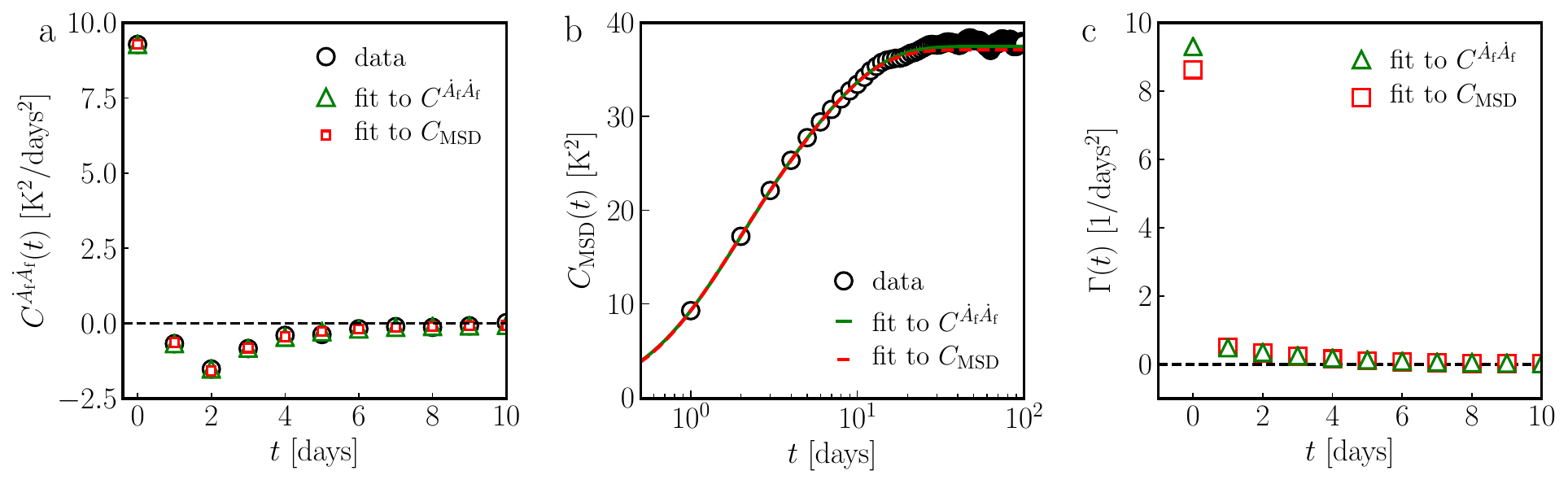}
	\caption{Test of the discrete estimation method for the maximal temperature in Berlin. We compare data for the autocorrelation function  $C^{\Dot{A}_\text{f}\Dot{A}_\text{f}}(t)$ ({\bf\sffamily a}) and the mean-squared displacement $C_{\text{MSD}}(t)$ ({\bf\sffamily b}) with GLE results parameterized using the discrete estimation method. In ({\bf\sffamily c})  we show the memory kernel $\Gamma(t)$ extracted from data using two different approaches, where we either fit the discrete MSD $C_{\text{MSD}}(t)$ using equation~(\ref{eq:vacf2}) (red) or the VACF $C^{\Dot{A}_\text{f}\Dot{A}_\text{f}}(t)$ using equation~(\ref{eq:vacf3}) (green).}
	\label{fig:berlin_max_temp_time_kernel_mitterwallner_compare}
\end{figure*}
To account for discretization effects and to estimate the memory kernel in the continuum limit, we utilize the procedure proposed by Mitterwallner \textit{et al.} \cite{mitterwallner2020non}. The VACF, $C^{\Dot{A}_\text{f}\Dot{A}_\text{f}}(t) = \langle \dot{A}_\text{f}(0)\dot{A}_\text{f}(t) \rangle$,  follows from the Mori-GLE with a memory kernel having the generic form in equation~(\ref{eq:meteo_kernel}) in the main text, which we justify by a fit of the memory kernel extracted from the Volterra method in Supplementary Information \ref{app:volterra} (shown in Fig.~\ref{fig:maxTemp1}d in the main text and in Supplementary Information \ref{app:results_volterra}). 
The analytic expression of the VACF follows from the mean-squared displacement $C_{\text{MSD}}(t) = \langle [A_\text{f}(t) - A_\text{f}(0)]^2 \rangle$ by $C^{\Dot{A}_\text{f}\Dot{A}_\text{f}}(t) = \partial^2 C_{\text{MSD}}(t) /\Bigl(2 \partial t^2\Bigr)$, where $C_{\text{MSD}}(t)$ is given by
\begin{eqnarray}
	\label{eq:vacf2}
	C_{\text{MSD}}(t) = \frac{B}{\tau^2} \left( \sum_{i=1}^{3}\frac{e^{-\sqrt{-\nu_\text{i}^2} t} - 1}{\sqrt{-\nu_\text{i}^2} \prod_{j\neq i} (\nu_\text{i}^2-\nu_\text{j}^2) } \left[ k_1 + k_2 \nu_\text{i}^2 \right] \right) 
\end{eqnarray}
and the constants $\nu_\text{i}$, $k_1$ and $k_2$ are determined by the parameters $a$, $b$, $\tau$ and $k$.
For a derivation of $C_{\text{MSD}}(t)$ from the memory kernel in equation~(\ref{eq:meteo_kernel}) for the Mori-GLE, we refer to Supplementary Information \ref{app:msd_harm}. 

Assuming that the discrete data of the observable $A_\text{f}(i\Delta t)$ is not affected by experimental localization noise \cite{mitterwallner2020non}, the discrete VACF of the data, computed from the discrete velocities $\dot{A}_\text{f}((i+1/2)\Delta t) = (A_\text{f}((i+1)\Delta t) - A_\text{f}(i\Delta t))/\Delta t$, follows from the discrete second derivative of the MSD $C_{\text{MSD}}(i \Delta t)$
\begin{eqnarray}
	\label{eq:vacf3}
	C^{\dot{A}_\text{f}\dot{A}_\text{f}}(i \Delta t) = \frac{C_{\text{MSD}}[(i+1) \Delta t] - 2 C_{\text{MSD}}(i \Delta t) + C_{\text{MSD}}[(i-1) \Delta t]}{2 (\Delta t)^2},
\end{eqnarray}
where $C_{\text{MSD}}(i \Delta t)$ is given by equation~(\ref{eq:vacf2}) at discrete time steps. The initial value is obtained via $C^{\dot{A}_\text{f}\dot{A}_\text{f}}(0) =C_{\text{MSD}}(\Delta t)/(\Delta t)^2$.
We estimate the parameters $a$, $b$, and $\tau$ of the memory kernel and the parameters $B$ and $k$ from the GLE via a fit of equation~(\ref{eq:vacf3}) to the VACF of the data. Equation~(\ref{eq:vacf3}) is fitted using the Levenberg-Marquardt algorithm implemented in scipy v.~1.6 \cite{2020SciPy-NMeth}. Here, we use only the data points from the VACF before the values decay to zero. Since we need to determine five parameters from a limited number of data points, the initial values for all parameters are chosen suitably from a fit of the memory kernel extracted using the Volterra method. We constrain the parameter space to boundary values given by (in units summarized in Supplementary Information \ref{app:fit_kernels_params})
\begin{align*}
	0 < a < 10, \\
	-10 < b < 10, \\
	0 < \tau < 100, \\
	0 < k < 10, \\
	0 < B < 1000.
\end{align*}
Using this procedure, the fitting uncertainties are found to be small, for the maximal temperature in Berlin we obtain $a = (4.31 \pm 0.31 )$ days$^{-1}$, $b = (2.07 \pm 1.26 )$ days$^{-1}$, $\tau = (3.04 \pm 1.78 )$ days, $k = (1.57 \pm 0.19)$ days$^{-2}$ and $B = (29.46 \pm 1.42 )$ K$^{2}$/days$^{2}$. 

For the data shown in the main text, we extracted the GLE parameters from fits of the VACF in equation~(\ref{eq:vacf3}) to the data as described so far. Alternatively, a fit of the data to the MSD in equation~(\ref{eq:vacf2}) is also possible. In Fig.~\ref{fig:berlin_max_temp_time_kernel_mitterwallner_compare}, we show that for the maximal temperature in Berlin, this alternative method leads to almost identical results for the memory kernel. 
\section{Mean-Squared Displacement for Delta + Single-Exponential Memory Kernel in Harmonic Confinement}
\label{app:msd_harm}
Here, the mean-squared displacement (MSD) is analytically calculated for the GLE in equation~(\ref{eq:mori_GLE}) using equation~(\ref{eq:FDT2}) in the main text. Applying a Fourier transformation, $\tilde{x}(\nu) = \int_{-\infty}^{\infty} dt\:e^{-i\nu t} x(t)$, the GLE can be written as
\begin{eqnarray}
	\label{eq:ft_gle}
	\tilde{x}(\nu) = \tilde{\chi}(\nu)\tilde{F}_R(\nu) \,,
\end{eqnarray}
where $\tilde{\chi}(\nu)$ is the Fourier transformation of the position response function, which takes the form
\begin{eqnarray}
	\label{eq:def_response_fun}
	\tilde{\chi}(\nu) = \left(k -\nu^2 + i\nu\tilde{\Gamma}_+(\nu) \right)^{-1},
\end{eqnarray}
with $\tilde{\Gamma}_+(\nu)$ being the half-sided Fourier transformation of the memory kernel defined via
\begin{eqnarray}
	\label{eq:def_halfside_kernel_transform}
	\tilde{\Gamma}_+(\nu) = \int_{0}^{\infty} \Gamma(t) e^{-i\nu t} dt \,.
\end{eqnarray}
The Fourier transformation of the position correlation function $C^{A_\text{f}A_\text{f}}(t) = \langle A_\text{f}(0) A_\text{f}(t)\rangle $ can be written as
\begin{eqnarray}
	\label{eq:ft_pos_corr}
	\tilde{C}^{A_\text{f}A_\text{f}}(\nu) = B\tilde{\chi}(\nu) \tilde{\Gamma}(\nu) \tilde{\chi}(-\nu) \,,
\end{eqnarray}
where we made use of equation~(\ref{eq:FDT2}) in the main text. Equation~(\ref{eq:ft_pos_corr}) can be rewritten as
\begin{eqnarray}
	\tilde{C}^{A_\text{f}A_\text{f}}(\nu) = -\frac{B}{i\nu} \left( \tilde{\chi}(\nu) - \tilde{\chi}(-\nu) \right) \,,
\end{eqnarray}
which leads to the MSD
\begin{eqnarray}
	C_{\text{MSD}}(t) &=& 2 (C^{A_\text{f}A_\text{f}}(0) - C^{A_\text{f}A_\text{f}}(t)) \,, \nonumber\\
	&=& B \int_{-\infty}^{\infty} \frac{d\nu}{\pi} \frac{e^{i\nu t} - 1}{i\nu}\left( \tilde{\chi}(\nu) - \tilde{\chi}(-\nu) \right) \label{eq:msd_integral} \,.
\end{eqnarray}
Inserting the Fourier transformation of the memory kernel in equation~(\ref{eq:meteo_kernel}) in the main text into equation~(\ref{eq:def_response_fun}), we find the response function to be
\begin{eqnarray}
	\label{eq:response_model}
	\tilde{\chi}(\nu) = \left(k -\nu^2 + i\nu a + \frac{i \nu b \tau^{-1}}{i\nu +\frac{1}{\tau}} \right)^{-1} \,,
\end{eqnarray} 
which leads, by inserting into equation~(\ref{eq:msd_integral}), to
\begin{eqnarray}
	\label{eq:msd_int_inserted}
	C_{\text{MSD}}(t) = B \int_{-\infty}^{\infty} \frac{d\nu}{\pi} \frac{(e^{i\nu t} - 1) (k_1 + k_2\nu^2)}{c_0 + c_1 \nu^2 + c_2 \nu^4 + c_3\nu^6} \,.
\end{eqnarray}
The constants $c_\text{i}$ and $k_\text{i}$ are defined by
\begin{eqnarray}
	c_0 =&& k^2 \,, \\
	c_1 =&& (a^2 - 2 k + 2 a b + ( b \tau^{-1} + k)^2 \tau^2) \,,\\
	c_2 =&& (1 + (a^2 - 2 ( b \tau^{-1} + k)) \tau^2) \,,\\
	c_3 =&& \tau^2 \,,\\
	k_1 =&& -2 (a + b)\,, \\
	k_2 =&& -2 a \tau^2 \label{eq:constants_msd_int_last}\,.
\end{eqnarray}
We rewrite equation~(\ref{eq:msd_int_inserted}) as
\begin{eqnarray}
	\label{eq:msd_int_inserted_denom}
	C_{\text{MSD}}(t) = B \int_{-\infty}^{\infty} \frac{d\nu}{\pi} \frac{(e^{i\nu t} - 1) (k_1 + k_2\nu^2)}{\tau^2 (\nu^2-\nu_1^2)(\nu^2-\nu_2^2)(\nu^2-\nu_3^2)} \,,
\end{eqnarray}
where $\nu_\text{i}^2$ are the solutions of $\nu^2$ to the polynomial equation
\begin{eqnarray}
	\label{eq:polynomial_eq}
	c_0 + c_1 \nu^2 + c_2 \nu^4 +c_3\nu^6 = c_3 (\nu^2-\nu_1^2)(\nu^2-\nu_2^2)(\nu^2-\nu_3^2)=0\,.
\end{eqnarray}
Next, we use the partial fraction decomposition
\begin{eqnarray}
	\frac{1}{\prod_{i=1}^{3}(\nu^2-\nu_\text{i}^2)} &=&  \sum_{i=1}^{3} \frac{1}{(\nu^2-\nu_\text{i}^2)\prod_{j\neq i}(\nu_\text{i}^2-\nu_\text{j}^2)},
	\label{eq:partial_fraction_decomp}
\end{eqnarray}
to rewrite equation~(\ref{eq:msd_int_inserted_denom}) as a sum of terms proportional to $(\nu^2-\nu_\text{i}^2)^{-1}$. Using the solution of the integral
\begin{eqnarray}
	\int_{-\infty}^{\infty}\frac{e^{i\nu t} -1}{\nu^2-\nu_\text{i}^2}d\nu  &=& \frac{\pi (e^{-\sqrt{-\nu_\text{i}^2}t} - 1)}{\sqrt{-\nu_\text{i}^2}},
	\label{eq:integrals_Roots}
\end{eqnarray}
for $t>0$ with the condition $\text{Re}(\nu_\text{i}^2)<0 \lor \text{Im}(\nu_\text{i}^2)\neq0 \land \text{Im}(\sqrt{\nu_\text{i}^2})\neq0$, we can rewrite the integral of the MSD equation~(\ref{eq:msd_int_inserted}) as
\begin{eqnarray}
	C_{\text{MSD}}(t) = \frac{B}{\tau^2} \left( \sum_{i=1}^{3}\frac{e^{-\sqrt{-\nu_\text{i}^2} t} - 1}{\sqrt{-\nu_\text{i}^2} \prod_{j\neq i} (\nu_\text{i}^2-\nu_\text{j}^2) } \left[ k_1 + k_2 \nu_\text{i}^2 \right] \right) \,.
	\label{eq:msd_final_Res}
\end{eqnarray}
Using input values for $a$, $b$, $\tau$ and $k$, we obtain the values for $\nu_\text{i}$ by solving equation~(\ref{eq:polynomial_eq}) numerically.

\section{Generalized Langevin Equation for an Observable with Non-Harmonic Potential of Mean force}
\label{app:approx_gle}

\begin{figure*}[hbt!]
	\centering
	\includegraphics[width=0.5\textwidth]{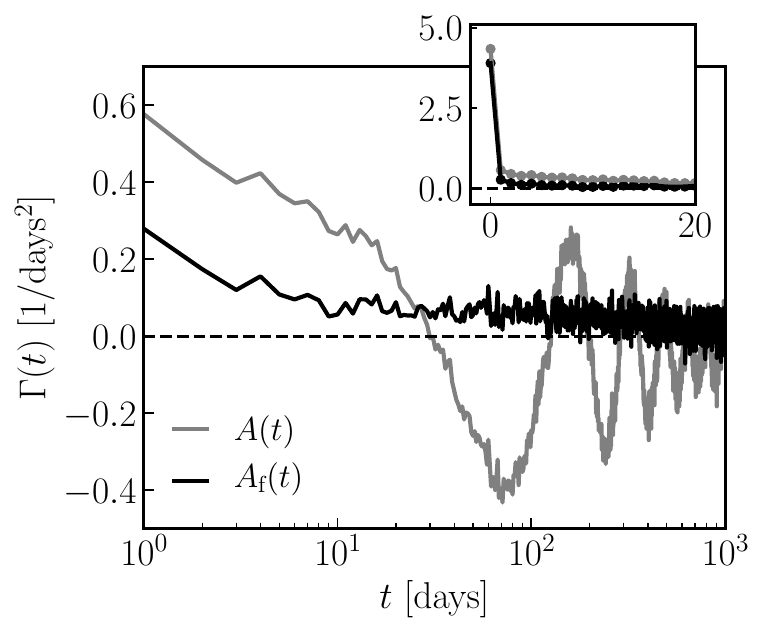}
	\caption{Memory kernel $\Gamma(t)$ extracted using the Volterra method based on equation~(\ref{eq:GLE}) and equation~(\ref{eq:mori_GLE}) (see Supplementary Information \ref{app:volterra} for details) from the full ($A(t)$, gray) and filtered ($A_\text{f}(t)$, black) trajectory shown in Fig.~\ref{fig:maxTemp1} in the main text. The inset shows the results on a linear time scale.}
	\label{fig:maxTemp_compare}
\end{figure*}

The generalized Langevin equation (GLE) for a general observable $A(t)$, including a non-harmonic potential of mean force $U(A)$ and a friction force linear in the velocity $\dot{A}(t)$, is obtained by non-linear hybrid projection \cite{darve_computing_2009,ayazhybrid2022, vroylandt2022derivation} and is given by
\begin{align}
	\label{eq:GLE}
	\ddot{A}(t) = -\nabla U\left[A(t)\right] - \int_0^\text{t}\mathrm{d}s\,\Gamma(t-s)\dot{A}(s) + F_\text{R}(t),
\end{align}
where the second moment of the random force $F_\text{R}(t)$ is approximately connected to the memory kernel $\Gamma(t)$ via 
\begin{align}
	\label{eq:FDT}
	\langle F_\text{R}(t) F_\text{R}(t^\prime)\rangle \backsimeq	 \langle \dot{A}^2\rangle \Gamma(|t-t^\prime|).
\end{align}
The GLE in equation~(\ref{eq:GLE}) differs in one crucial point from the GLE obtained via Mori's formalism in equation~(\ref{eq:mori_GLE}) in the main text: The potential of mean force (PMF) $U\left[A(t)\right]$ \cite{darve_calculating_2001} is defined by 
\begin{align}
	\label{eq:PMF}
	U\left[A(t)\right] = -\langle\dot{A}^2\rangle \log \rho\left[ A(t)\right],
\end{align}
whereas in the Mori-GLE it is given by $U\left[A(t)\right] = - \frac{k}{2}\left( A(t) - \langle A \rangle \right)^2$ with $k=\langle \dot{A}^2 \rangle /\langle A \rangle$.
From a given time-series $A(t)$, all parameters of equation~(\ref{eq:GLE}) can be computed via a Volterra extraction method \cite{ayaz2021non} as explained in Supplementary Information \ref{app:volterra}. 
Comparing equation~(\ref{eq:GLE}) for a general observable $A(t)$ with equation~(\ref{eq:mori_GLE}) in the main text for a filtered variable $A_\text{f}(t)$, it becomes clear that the memory kernels will differ significantly.

In Fig.~\ref{fig:maxTemp_compare}, we compare the memory kernels for the maximal temperature data in Berlin extracted from the unfiltered data $A(t)$ using equation~(\ref{eq:GLE}) and from the filtered data $A_\text{f}(t)$ using equation~(\ref{eq:mori_GLE}) in the main text. It becomes evident that both memory kernels have a dominant $\delta(t)$-contribution of comparable size (see inset). The memory kernel of $A(t)$ using equation~(\ref{eq:GLE}) has a significant periodic contribution with a periodicity of around half a year, originating from the seasonality of the data of 365 days, which is absent in the memory kernel of the filtered observable $A_\text{f}$ using equation~(\ref{eq:mori_GLE}). 

\section{Impact of Filtering Details on Memory of Weather and Financial Data}
\label{app:filtering_examples}

\begin{figure*}%[hbt!]
	\centering
	\includegraphics[width=0.9\linewidth]{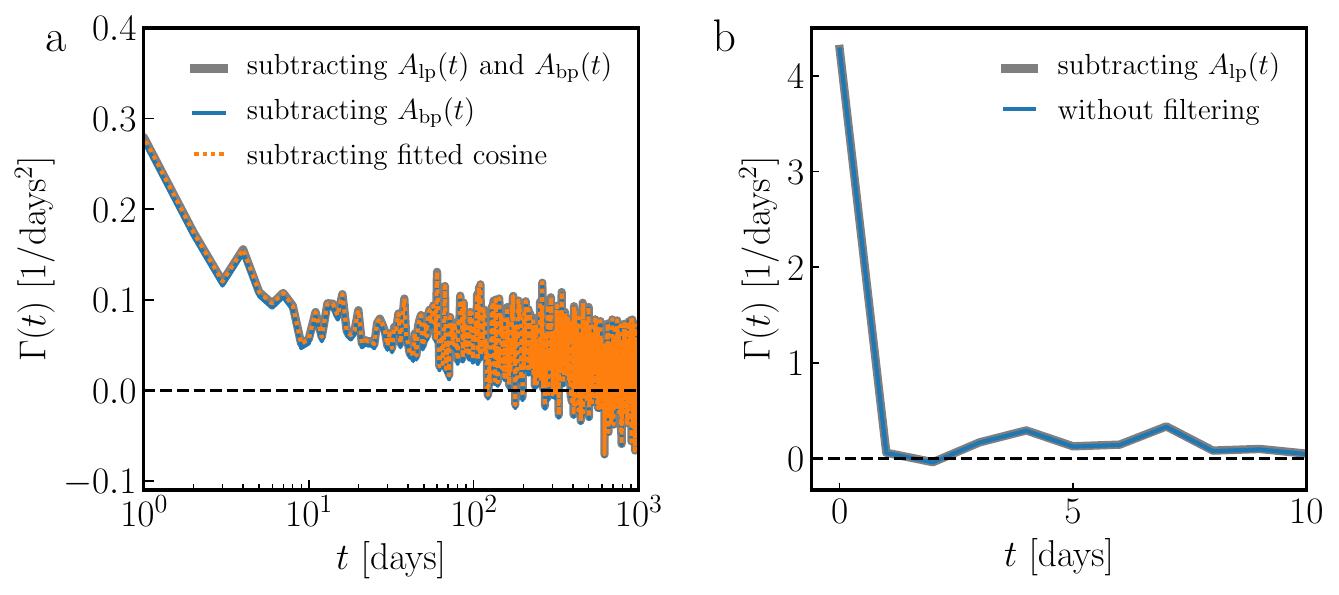}
	\caption{Influence of details of the applied filters on the memory extracted using the Volterra extraction method. {\bf\sffamily a}: Memory kernel $\Gamma(t)$ of $A_\text{f}(t)$ for the maximal temperature in Berlin (gray), obtained by subtracting the low- and band-pass contributions, together with $\Gamma(t)$ of $A_\text{f}(t)$  which we obtain by only subtracting the band-pass (equation~(\ref{eq:band-pass}) in the main text) contribution (blue) and $\Gamma(t)$ obtained by subtracting a fitted single cosine from $A(t)$ (orange).  {\bf\sffamily b}: $\Gamma(t)$ for the S\&P 500 index with and without subtracting low-pass (equation~(\ref{eq:low-pass})) contributions from $A(t)$.  }
	\label{fig:filtering_examples_meteo_finance_effect_filtering}
\end{figure*}

\begin{figure}%[hbt!]
	\centering
	\includegraphics[width=\linewidth]{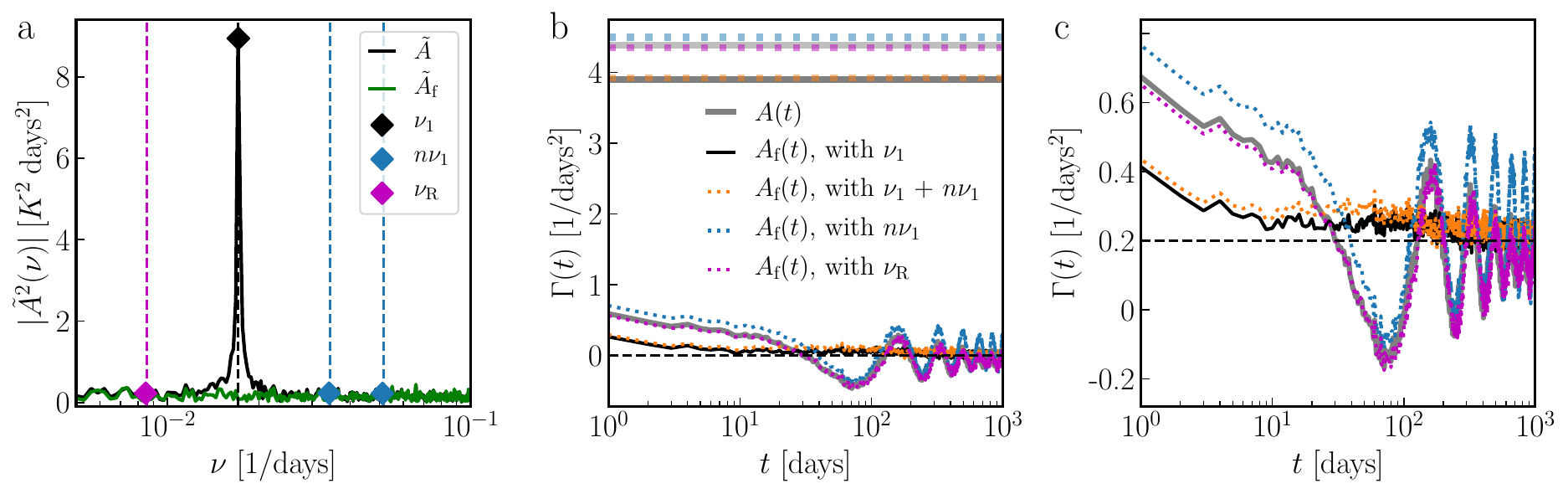}
	\caption{Effect of filter frequencies on the memory extracted from $A_\text{f}(t)$. {\bf\sffamily a}: Spectrum of the Fourier transformed daily maximal temperature trajectory in Berlin before and after filtering (given in Fig.~\ref{fig:filtering_examples_meteo_finance}a in the main text). The blue symbols denote the position of higher harmonics of the dominant peak at $\nu_1$ = 2$\pi$/365 days (2$\nu_1$ and 3$\nu_1$). The purple symbol denotes the position of a randomly chosen frequency $\nu_\text{R}$. {\bf\sffamily b}: Memory kernels extracted using the Volterra method (Supplementary Information \ref{app:volterra}) from $A_\text{f}(t)$ obtained by equation~(\ref{eq:band-pass}) in the main text only at $\nu_1$ (black), at additional higher harmonics $2\nu_\text{1}$ and $3\nu_\text{1}$(orange), only at higher harmonics (blue) and at the randomly chosen frequency $\nu_\text{R}$ (purple). The gray line is the result without filtering. The horizontal lines denote $\Gamma(0)$ of the respective memory kernels. {\bf\sffamily c}: Zoomed-in version of the results in ({\bf\sffamily b}).}
	\label{fig:filtering_examples_meteo_effect_frequency}
\end{figure}

In Fig.~\ref{fig:filtering_examples_meteo_finance_effect_filtering}a, we show that details of the high-pass filtering process (equation~(\ref{eq:low-pass}) in the main text) does not influence the memory extracted from $A_\text{f}(t)$; the extracted memory kernel does not change when omitting the high-pass filter and only using the band-stop filter (equation~(\ref{eq:band-pass}) in the main text) to obtain  $A_\text{f}(t)$ from  $A(t)$. The high-pass filter also does not affect the memory of financial data (Fig.~\ref{fig:filtering_examples_meteo_finance_effect_filtering}b). Furthermore, we show that the memory time is not perturbed by details of the band-stop filtering, since the memory kernel of $A_\text{f}(t)$ obtained by band-stop filtering is similar to the one from a trajectory created by subtracting a fitted single cosine function from $A(t)$ (red line in Fig.~\ref{fig:filtering_examples_meteo_finance_effect_filtering}a). This underpins the robustness of the extrapolation of the seasonalities using equation~(\ref{eq:seas_part}) in the main text.
\\ \indent In Fig.~\ref{fig:filtering_examples_meteo_effect_frequency}, we examine the effect of the choice of filter frequencies on memory effects in the $A_\text{f}(t)$ trajectory. When using higher harmonics $n\nu_\text{1}$ in addition to the dominant frequency $\nu_1$ for band-stop filtering the maximal temperature in Berlin, the memory kernel does not change substantially. The exclusive use of higher harmonics, or, generally, frequencies away from peaks, does not remove the seasonal contribution, as the memory kernel is similar to that of the unfiltered trajectory $A(t)$.

\section{Predictability of Time-Series Data}
\label{app:predictability}
Using the GLE, the predictability of a time-series can be quantified by analyzing the different characteristic time scales of the system; they determine which terms in the equation of motion are relevant and determine the future values of the trajectory.  If we set the memory time $\tau$ in equation~(\ref{eq:meteo_kernel}) for the GLE in equation~(\ref{eq:mori_GLE}) in the main text to zero, we obtain the Langevin equation (LE), $\Ddot{A}_\text{f}(t) = - (a+b) \dot{A}_\text{f}(t) - k A_\text{f}(t) + F_\text{R}(t)$ and $\langle  F_\text{R}(t) F_\text{R}(t')\rangle = 2B(a+b)\delta(t-t')$. The mean dynamics is described by the harmonic oscillator, $A_\text{f}(t) = A_\text{f}(0)e^{-t w}$, where $w_{1,2} = \frac{(a+b)}{2}\pm\sqrt{\frac{(a+b)^2}{4} - k}$. In the overdamped regime $(a+b)^2/4 \gg k$,  $w_{1,2}$ are purely real and $A_\text{f}$ exponentially decays with the relaxation time $\tau_\text{rel} = w_2^{-1} = (a+b)/k$. In the underdamped regime $(a+b)^2/4 \ll k$, $w_{1,2}$ are complex, $w_{1,2} = \frac{(a+b)}{2} \pm i k^{1/2} \sqrt{1-\frac{(a+b)^2}{4k}}$, and $A_\text{f}^2$ decays with the persistence time $\tau_\text{per} = (a+b)^{-1}$. The imaginary part $i k^{1/2} \sqrt{1-\frac{(a+b)^2}{4k}}$ describes the oscillations in $A_\text{f}$ with oscillation time $T_0$, i.e. $\frac{T_0}{2\pi} = 1/\left(k^{1/2} \sqrt{1-\frac{(a+b)^2}{4k}}\right) \backsimeq k^{-1/2} $.\\
\indent The boundary between the over- and underdamped regime can be expressed in terms of $\tau_\text{per}$, $\tau_\text{rel}$ and $T_0$. The dynamics of $A_\text{f}$ is overdamped if $2\tau_\text{per} < \frac{T_0}{2\pi}$ which is analogous to $\tau_\text{rel}/4 > \tau_\text{per}$ or $\tau_\text{rel} > T_0/\pi$.
In the overdamped regime, the friction term in the LE dominates over the potential term, as $(a+b)^2 (\dot{A}_\text{f})^2 \sim (a+b)^2 B$ and $k^2 (A_\text{f})^2 \sim k B $, where we used $\langle \dot{A}^2_\text{f}\rangle = B$ and $k = \langle \dot{A}^2_\text{f}\rangle/ \langle A^2_\text{f}\rangle$. In the underdamped regime, the potential term dominates. The relevance of the random force $F_\text{R}(t)$ depends on the time resolution $\Delta t$, since $\langle F_\text{R}^\text{i} F_\text{R}^\text{i+j} \rangle = \frac{2B(a+b)}{\Delta t}\delta_{0j}$ and, therefore, $(F_\text{R})^2 \sim B(a+b)/\Delta t$. 
In the overdamped regime, we choose $\Delta t = \tau_\text{per}$ and, therefore, have $(F_\text{R})^2 \sim (a+b)^2B$. Hence, friction and random force are equally important. In the underdamped regime, we choose $\Delta t = \tau_\text{rel}$ and have $(F_\text{R})^2 \sim Bk$. We, therefore, find that the potential and random force are equally important. \\
\indent The data can be predicted up to $\tau_\text{per}$ just by extending the current velocity $\dot{A}_\text{f}(t)$ into the future, regardless of whether the system is in the under- or overdamped regime; this defines the persistence-prediction scenario. In the overdamped regime, persistent prediction is possible up to times $t \backsimeq \tau_\text{per} \ll \tau_\text{rel}$. For larger times $\tau_\text{per} < t < \tau_\text{rel}$, prediction is possible for rare states, i.e. when $A_\text{f} \gg \sqrt{\langle A_\text{f}^2 \rangle} = \sqrt{B/k}$, by realizing that relaxation from a rare state proceeds according to $A_\text{f}(t) \backsimeq A_\text{f}^0 e^{-t/\tau_\text{rel}}$; this defines the relaxation-prediction scenario. For $A_\text{f} \leq  \sqrt{\langle A_\text{f}^2 \rangle}$ and $t > \tau_\text{per}$, Markovian friction or a Markovian random force does not help to predict, and prediction is only possible if the random force is non-Markovian. \\
\indent To discuss the effect of memory on the predictability of data, we assume that the memory kernel in the GLE has the generic form in equation~(\ref{eq:meteo_kernel}) in the main text. Here, prediction is possible up to the memory time $\tau$ in both the short- ($\tau < \tau_\text{rel}$) and long-memory ($\tau > \tau_\text{rel}$) time regimes; defining the memory-prediction scenario. In the underdamped regime ($\tau_\text{per} > \tau_\text{rel}/4$), memory-based prediction should be possible and is useful only for $\tau > \tau_\text{per}$ because for $\tau < \tau_\text{per}$, persistence-based prediction can be employed. For longer memory times ($\tau > \tau_\text{per}$), the predictability is expected to increase with increasing memory time. The relevance of the memory time for prediction vastly depends on the ratio between the delta and exponential component in equation~(\ref{eq:meteo_kernel}). If we assume $\tau > \tau_\text{per}$ and integrate the memory kernel over the persistence time, we obtain
\begin{equation}
	\int_{0}^{ \tau_\text{per}} dt\:\Gamma(t) = 2a + \frac{b  \tau_\text{per}}{\tau},
\end{equation}
from which we define the non-Markovian fraction
\begin{equation}
	\label{eq:memory_fraction}
	\xi = \frac{b \tau_\text{per}/\tau}{2a + b  \tau_\text{per}/{\tau}}, 
\end{equation} 
and which quantifies the contribution of exponential memory on a prediction. From fitting parameters for the maximal temperature given in Table~\ref{tab:fits_kernel_tmax} in the main text, we calculate a non-Markovian fraction of $\xi \approx 0.012 \ll 1$. Thus, although the memory time $\tau = 3.04$ days is higher than the time resolution, the single-exponential memory component does not contribute significantly to data predictability, which is validated by the comparison between GLE and Langevin predictions in Supplementary Information \ref{app:performance_comp}. Moreover, for weather data, the random force is irrelevant for the prediction since only the non-Markovian fraction of $\xi \approx 0.012$ can be predicted, which agrees with our results in Fig.~\ref{fig:pred_comps} in Supplementary Information \ref{app:performance_comp}. Although we find a negligible effect of memory on the predictability, the characteristic time scales $\tau$, $\tau_\text{per}$ and $\tau_\text{rel}$, listed in Table~\ref{tab:fits_kernel_tmax} in the main text, are rather similar, which necessitates analysis with the full GLE.

\section{Conditional Sampling of Random Forces}
\label{app:SFRF}
The technique of Markovian embedding \cite{ceriotti2010colored, li2017computing, kappler2018memory} is useful for generating trajectories that are described by the GLE \cite{ayaz2021non, brunig2022time, brunig2022pair}. Conditional probability sampling is needed when striving for trajectory predictions starting from an already-known part of a trajectory. This procedure ensures the validity of equation~(\ref{eq:FDT2}) in the main text for all time steps by considering past random forces before sampling future ones. In the following paragraphs, we introduce general notations for Gaussian processes and a method to generate future random forces from a conditional Gaussian probability distribution governed by a memory kernel.

\subsection{Sampling from Gaussian Distributions}
\label{sec:sampling_gaussian}

We consider the multivariate Gaussian set of random variables $\mathbf{F}_\text{R} = \left(F_\text{R}^1, F_\text{R}^2, \dots, F_\text{R}^N\right)^\text{T}$ at times $t_1<\dots < t_N$ with corresponding means $\boldsymbol\mu = \left(\mu_1, \mu_2, \dots, \mu_N\right)^\text{T}$  as well as a covariance matrix $\hat{C} \in \mathbb{R}^{N\times N}$ with entries $C_\text{ij} = C(t_i, t_j)$
\begin{eqnarray}
	\label{eq:cov_fun}
	C_\text{ij} = \int_{-\infty}^\infty \int_{-\infty}^\infty d F_\text{R}^\text{i}\:d F_\text{R}^\text{j}\:	P_{F_\text{R}}(F_\text{R}^\text{i},F_\text{R}^\text{j} ;t_\text{i},t_\text{j} ) (F_\text{R}^\text{i}-\mu_\text{i})(F_\text{R}^\text{j}-\mu_\text{j}),
\end{eqnarray}
governed by the joint distribution
\begin{align}
	\label{eq:Normal}
	P_{F_\text{R}}\left(F_\text{R}^1,F_\text{R}^2,\dots,F_\text{R}^N;t_1,t_2,\dots,t_N\right) &= \frac{1}{\sqrt{2\pi|\hat C|}}\,e^{-\frac{1}{2}(\mathbf{F}_\text{R}-\boldsymbol\mu)^\text{T}\cdot\: \hat{C}^{-1}(\{t_\text{N}\})\cdot \:(\mathbf{F}_\text{R}-\boldsymbol\mu)},\\
	\nonumber &= \mathcal{N}\left(\boldsymbol \mu, \hat{C}(\{t_\text{N}\})\right), 
\end{align}
where $|..|$ denotes the determinant of a matrix. From equation~(\ref{eq:cov_fun}), it follows that the covariance matrix is symmetric. Furthermore, we assume that the covariance matrix is positive definite, i.e. for all non-zero eigenvectors $\mathbf{x}$ we have $\mathbf{x}^\text{T}  \hat{C} \mathbf{x} > 0$ and thus the square root in equation~(\ref{eq:Normal}) is real.
Any random process $\mathbf{F}_\text{R}$ with correlated entries can be decomposed into a random uncorrelated process $\mathbf{G}$ governed by a stationary Gaussian process with zero mean and unit variance using the relation
\begin{eqnarray}
	\label{eq:trafo}
	\mathbf{F}_\text{R} = \hat{L} \mathbf{G} + \boldsymbol \mu,
\end{eqnarray}
with $\hat L \hat L^\text{T}=\hat{C}$ being the Cholesky decomposition of the lower triangular matrix $\hat{L} \in \mathbb{R}^{N\times N}$. We prove the validity of the transformation in equation~(\ref{eq:trafo}) in Supplementary Information \ref{app:proofs_sampling}. In the context of equation~(\ref{eq:FDT2}) in the main text, the covariance matrix $\hat{C} $ is determined by the memory kernel $\Gamma(t)$ after discretization, as described in the following section.

\subsection{Conditional Gaussian Process}
\label{sec:cond_gaussian}
We start with a given past random force sequence $\mathbf{F}_\text{R}^\text{p}$ of length $N_\text{past}$ determined by $\mathcal{N}\left(\boldsymbol \mu, \hat{C}(\{t_{\text{N}_\text{past}}\})\right)$ and compute the conditional probability of obtaining the future $N_\text{fut}$ forces $\mathbf{F}_\text{R}^\text{f}$ given the past ones \cite{dennis_s_bernstein_author_matrix_2009, williams2006gaussian}. The values of $\mathbf{F}_\text{R}^\text{p}$ are computed from the observable trajectory $A_\text{f}(t)$ by discretizing the GLE in equation~(\ref{eq:mori_GLE}) in the main text  with a discretization step $\Delta t$ and solving the memory integral with the trapezoidal rule 
\begin{eqnarray}
	\label{eq:fr}
	F_\text{R}^\text{p} \sqrt{\Delta t/\delta t}= \ddot{A}_\text{f}^\text{i,c} + \frac{\Delta t}{2} \Gamma_0 \dot{A}_\text{f}^\text{i,c}+ \frac{\Delta t}{2} \Gamma_\text{i} \dot{A}_\text{f}^\text{0,c} + \Delta t \sum_{j=1}^{i-1} \Gamma_\text{j} \dot{A}_\text{f}^\text{i-j,c} + k A_\text{f}^\text{i},
\end{eqnarray}

where we compute the discrete velocities according to
$\dot{A}_\text{f}^\text{i} = \dot{A}_\text{f}(i\Delta t) = \bigl[A_\text{f}\bigl((i+1)\Delta t\bigr) - A_\text{f}\bigl((i-1)\Delta t\bigr)\bigr]/(2\Delta t)$ and accelerations according to $\ddot{A}_\text{f}^\text{i} = \ddot{A}_\text{f}(i\Delta t) = \bigl[A_\text{f}\bigl((i+1)\Delta t\bigr) - 2A_\text{f}(i\Delta t) + A_\text{f}\bigl((i-1)\Delta t\bigr)\bigr]/(\Delta t)^2$ from the trajectory $A_\text{f}^\text{i} =  A_\text{f}(i\Delta t)$. Since the random force determined by inverting equation~(\ref{eq:mori_GLE}) is significantly affected by discretization effects, the discrete velocities and accelerations are corrected to their respective continuous counterparts ($\dot{A}_\text{f}^\text{i,c}$ and $\ddot{A}_\text{f}^\text{i,c}$), while $A_\text{f}^\text{i}$ remain the same,  as explained in Supplementary Information \ref{app:fdt_disc}. $\delta t \leq \Delta t$ is the time step of the quasi-continuous representation of  $F_\text{R}^\text{p}$, which enters in equation~(\ref{eq:fr}) to ensure validity of equation~(\ref{eq:FDT2}) in the continuous limit. For the proper choice of $\delta t$, we refer to Supplementary Information \ref{app:fdt_disc}. \\
\indent The future values $\mathbf{F}_\text{R}^\text{f}$ of length $N_\text{fut}$ are determined by the conditional distribution following from equation~(\ref{eq:Normal}) as
\begin{equation}
	\label{eq:conditional}
	P_{F_\text{R}}\left(\mathbf{F}_\text{R}^\text{f} | \mathbf{F}_\text{R}^\text{p}\right) = \frac{1}{\sqrt{2\pi |\hat{C}_\text{ff}-\hat{C}_\text{pf}^\text{T} \hat{C}_\text{pp}^{-1}\hat{C}_\text{pf} |}} \cdot\:  e^{-\frac{1}{2}\left(\mathbf{F}_\text{R}^\text{f} - \hat{C}_\text{pf}^\text{T}\hat{C}_\text{pp}^{-1}\mathbf{F}_\text{R}^\text{p} \right)^\text{T} \left( \hat{C}_\text{ff}-\hat{C}_\text{pf}^\text{T} \hat{C}_\text{pp}^{-1}\hat{C}_\text{pf} \right)^{-1}\left( \mathbf{F}_\text{R}^\text{f} - \hat{C}_\text{pf}^\text{T}\hat{C}_\text{pp}^{-1}\mathbf{F}_\text{R}^\text{p} \right)},
\end{equation}
where the covariance matrix is partitioned into the block matrices $\hat{C}_\text{pp}  \in \mathbb{R}^{N_\text{past}\times N_\text{past}}, 	\hat{C}_\text{pf}  \in \mathbb{R}^{N_\text{past}\times N_\text{fut}}$ and $\hat{C}_\text{ff}  \in \mathbb{R}^{N_\text{fut}\times N_\text{fut}}$ using the discrete memory kernel of length $N =N_\text{past} + N_\text{fut}$ corresponding to \cite{dennis_s_bernstein_author_matrix_2009}

\begin{eqnarray}
	\label{eq:part_cov}
	\hat{C} = \left(\begin{array}{c c}
		\hat{C}_\text{pp} & \hat{C}_\text{pf}\\
		\hat{C}_\text{pf}^\text{T} & \hat{C}_\text{ff}
	\end{array}\right) =B \left(\begin{array}{c c c c c}
		\Gamma_0 & \Gamma_1 & \Gamma_2 & \cdots & \Gamma_\text{N-1}\\
		\Gamma_1 & \Gamma_0 & \Gamma_1 & \cdots & \Gamma_\text{N-2}\\
		\Gamma_2 & \Gamma_1 & \Gamma_0 & \cdots & \Gamma_\text{N-3}\\
		\vdots   &          &          &        &       \vdots\\
		\Gamma_\text{N-1} & \Gamma_\text{N-2} & \Gamma_\text{N-3} & \cdots & \Gamma_0    
	\end{array}\right).
\end{eqnarray}
We derive equation~(\ref{eq:conditional}) in Supplementary Information \ref{app:proofs_sampling2}. The conditional distribution in equation~(\ref{eq:conditional}) is related to the joint distribution in equation~(\ref{eq:Normal}) via the Bayesian theorem, i.e. $P_{F_\text{R}}\left(\mathbf{F}_\text{R}^\text{f} , \mathbf{F}_\text{R}^\text{p}\right) = P_{F_\text{R}}\left(\mathbf{F}_\text{R}^\text{p}\right)	P_{F_\text{R}}\left(\mathbf{F}_\text{R}^\text{f} | \mathbf{F}_\text{R}^\text{p}\right)$. The conditional probability distribution in equation~(\ref{eq:conditional}) can be rewritten as a non-conditional Gaussian distribution according to equation~(\ref{eq:Normal}) with shifted mean $\boldsymbol\mu$ and covariance $\hat{C}$
\begin{equation}
	\label{eq:non-cond_gaussian}
	\mathcal{N}(\hat{C}_\text{pf}^\text{T}\hat{C}_\text{pp}^{-1}\mathbf{F}_\text{R}^\text{p}, \hat{C}_\text{ff}-\hat{C}_\text{pf}^\text{T} \hat{C}_\text{pp}^{-1}\hat{C}_\text{pf}),
\end{equation} 
which shows that the sampling procedure of conditional future random forces is equivalent to the sampling from an unconditional Gaussian process using equation~(\ref{eq:trafo}) \cite{williams2006gaussian, pavliotis2014stochastic}. Using a random vector $\mathbf{G}$ drawn from an uncorrelated Gaussian distribution with zero mean and unit variance and with the desired prediction length, performing a matrix multiplication with $\hat L$ from the Cholesky decomposition $\hat L \hat L^\text{T}=\hat{C}_\text{ff}-\hat{C}_\text{pf}^\text{T} \hat{C}_\text{pp}^{-1}\hat{C}_\text{pf}$ and adding the mean $\boldsymbol \mu=\hat{C}_\text{pf}^\text{T}\hat{C}_\text{pp}^{-1}\mathbf{F}_\text{R}^\text{p}$ yields the sought set of future random forces $\mathbf{F}_\text{R}^\text{f}$.

Note that this method works only if the covariance matrix in equation~(\ref{eq:part_cov}) is positive definite, as is always the case for the data we consider, since the Cholesky decomposition is otherwise not possible. If an extracted memory kernel neither fulfills this requirement nor is properly represented by the fitting function in equation~(\ref{eq:meteo_kernel}) in the main text, it can be parameterized by a sum of 
monotonically decreasing exponential functions to enforce positive definiteness  \cite{ceriotti2010colored, li2017computing, ayaz2021non}.

%\newpage
\section{Fitting parameters of the Memory Kernels from Weather and Financial Data}
\label{app:fit_kernels_params}

\begin{table}[!htbp]
	\centering
	\caption{Fitting parameters of the memory kernel in equation~(\ref{eq:meteo_kernel}) and characteristic time scales for the weather data shown in Fig.~\ref{fig:maxTemp1} and Fig.~\ref{fig:meteo_kernels} using the discrete estimation method explained in Supplementary Information \ref{app:mitterwallner}.}
	\begin{ruledtabular}
		\begin{tabular}{c c c c c} 
			
			value & max. air  & min. air  & rel. air & max. wind \\ 
			& temp. [K] & temp. [K] & humidity [$\%$]  & speed [km/h] \\
			\hline
			
			$a$ [days$^{-1}$] & 4.31 & 5.09 & 6.83 & 4.11 \\
			$b$ [days$^{-1}$] & 2.07 & 5.33 & 9.94 & 10.39 \\
			$\tau$ [days] & 3.04 & 5.92 & 9.56 & 12.67 \\
			$k$ [days$^{-2}$] & 1.57 & 1.84 & 4.68 & 2.57 \\
			$B$ [$A_\text{f}$/days$^{2}$] & 29.46 & 28.80 & 555.51 & 96.58 \\
			$\tau_{\rm per}=1/(a+b)$  [days] & 0.16 & 0.1 &  0.06 & 0.07  \\
			$\tau_{\rm rel}=(a+b)/k$  [days] & 4.06 & 5.66 & 3.58 & 5.64 \\
			$ \langle A_\text{f}^2 \rangle^{1/2}= (B/k)^{1/2} $ [$A_\text{f}$]  & 4.33 & 3.96 & 10.89 & 6.13
			
		\end{tabular}
	\end{ruledtabular}
	\label{tab:fits_kernel}
\end{table}

\begin{table}[!htbp]
	\centering
	\caption{Fitting parameters of the memory kernel in equation~(\ref{eq:meteo_kernel}) and characteristic time scales for the weather data shown in Fig.~\ref{fig:maxTemp1} and Fig.~\ref{fig:meteo_kernels_volterra} using the Volterra method explained in Supplementary Information \ref{app:volterra}.}
	\begin{ruledtabular}
		\begin{tabular}{c c c c c} 
			
			value & max. air  & min. air  & rel. air & max. wind \\ 
			& temp. [K] & temp. [K] & humidity [$\%$]  & speed [km/h] \\
			\hline
			
			$a$ [days$^{-1}$] & 1.85 & 1.90 & 2.28 & 2 \\
			$b$ [days$^{-1}$] & 2.53 & 1 & 1 & 2.26 \\
			$\tau$ [days] & 12.07 & 3.46 & 4.21 & 6.65 \\
			$k$ [days$^{-2}$] & 0.22 & 0.55 & 0.57 & 0.72 \\
			$B$ [$A_\text{f}$/days$^{2}$] & 9.28 & 7.56 & 104.56 & 26 \\
			$\tau_{\rm per}=1/(a+b)$  [days] & 0.23 & 0.34 &  0.3 & 0.23  \\
			$\tau_{\rm rel}=(a+b)/k$  [days] & 19.91 & 5.27 & 5.75 & 5.92 \\
			$ \langle A_\text{f}^2 \rangle^{1/2}= (B/k)^{1/2} $ [$A_\text{f}$]  & 6.49 & 3.71 & 13.54 & 6.01
			
		\end{tabular}
	\end{ruledtabular}
	\label{tab:fits_kernel2}
\end{table}

\begin{table}[!htbp]
	\centering
	\caption{Fitting parameters of the memory kernel in equation~(\ref{eq:meteo_kernel}) and characteristic time scales  for the financial data in Fig.~\ref{fig:finance1} using the discrete estimation method explained in Supplementary Information \ref{app:mitterwallner}.}
	\begin{ruledtabular}
		\begin{tabular}{c c c c} 
			
			value & Bitcoin ($\Delta t = $ 1 min) & S\&P 500 ($\Delta t = $ 1 day) & JPY/USD ($\Delta t = $ 1 week)  \\ 
			\hline
			
			$a$  & 8.09 mins$^{-1}$ & 0.66 days$^{-1}$ & 3.04 weeks$^{-1}$   \\
			$b$ & 2.42 mins$^{-1}$  & 3.28 days$^{-1}$ & 1.3  $\cdot\: 10^{-3}$ weeks$^{-1}$  \\
			$\tau$ & 6.11 mins & 0.43 days & 0.11 weeks   \\
			$k$  & 0.02 mins$^{-2}$ & 0.14 days$^{-2}$ & 1.73 $\cdot\: 10^{-4}$ weeks$^{-2}$   \\
			$B$ & 940.7 $\$$/mins$^{2}$ & 364.98 1/days$^{2}$  & 3.94 ($\yen/\$$)/weeks$^{2}$    \\
			$\tau_{\rm per}=1/(a+b)$   & 0.1 mins & 0.25 days &  0.33  weeks \\
			$\tau_{\rm rel}=(a+b)/k$   & 525.5  mins & 28.14 days & 1.75 $\cdot\: 10^{4}$  weeks \\
			$ \langle A_\text{f}^2 \rangle^{1/2}= (B/k)^{1/2} $   &  216.88 $\$$ & 51.06 & 150.91 $\yen/\$$
			
		\end{tabular}
	\end{ruledtabular}
	\label{tab:fit_finance}
\end{table}

\begin{table}[!htbp]
	\centering
	\caption{Fitting parameters of the memory kernel in equation~(\ref{eq:meteo_kernel}) and characteristic time scales for the financial data in Fig.~\ref{fig:finance_kernels_volterra} using the Volterra method explained in Supplementary Information \ref{app:volterra}.}
	\begin{ruledtabular}
		\begin{tabular}{c c c c} 
			
			value & Bitcoin ($\Delta t = $ 1 min) & S\&P 500 ($\Delta t = $ 1 day) & JPY/USD ($\Delta t = $ 1 week)  \\ 
			\hline
			
			$a$  & 0.17 mins$^{-1}$ & 0.08 days$^{-1}$ & 0.01 weeks$^{-1}$   \\
			$b$ & 0.1 mins$^{-1}$  & 0.1 days$^{-1}$ & 0.2 weeks$^{-1}$  \\
			$\tau$ & 0.73 mins & 0.28 days & 0.55 weeks   \\
			$k$  & 3.49 $\cdot\: 10^{-6}$ mins$^{-2}$ & 2.2 $\cdot\: 10^{-3}$ days$^{-2}$ & 9.6 $\cdot \:10^{-3}$ weeks$^{-2}$   \\
			$B$ & 200.54  $\$$/mins$^{2}$ & 196.29 1/days$^{2}$  & 1.78 ($\yen/\$$)/weeks$^{2}$    \\
			$\tau_{\rm per}=1/(a+b)$   & 3.7 mins & 5.56 days &  4.76  weeks \\
			$\tau_{\rm rel}=(a+b)/k$   & 7.73 $\cdot\: 10^{4}$ mins & 81.82 days & 21.88 weeks \\
			$ \langle A_\text{f}^2 \rangle^{1/2}= (B/k)^{1/2} $   &  7.58 $\cdot\: 10^{3}$ $\$$ & 298.7  & 13.62 $\yen/\$$
			
		\end{tabular}
	\end{ruledtabular}
	\label{tab:fit_finance2}
\end{table}

\newpage
\section{Prediction Performance Using Different Components of the Mori-GLE}
\label{app:performance_comp}
\begin{figure}
	\centering
	\includegraphics[width=0.5\linewidth]{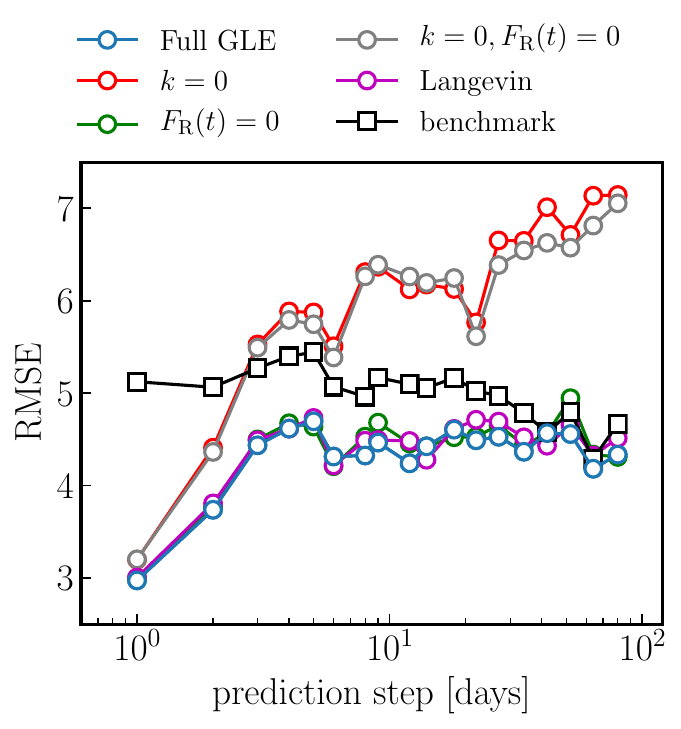}
	\caption{Importance of the different terms in the GLE for prediction. RMSE of the prediction of the daily maximal temperature in Berlin (Fig.~\ref{fig:maxTemp2}d in the main text) using the GLE in equation~(\ref{eq:mori_GLE}) in the main text with different components: The blue data denote the full Mori-GLE in equation~(\ref{eq:mori_GLE}), the red data the unconfined case ($k=0$), the green data the deterministic case ($F_\text{R}(t) =0$) the gray data the unconfined deterministic case ($k=0$ and $F_\text{R}(t) =0$), and the black data the benchmark (compare Fig.~\ref{fig:maxTemp2}e in the main text). Purple symbols denote the prediction performance of the Markovian Langevin equation~(\ref{eq:ODE_discret2}).}
	\label{fig:pred_comps}
\end{figure}

In Fig.~\ref{fig:pred_comps}, we investigate the performance of predicting the daily maximal temperature in Berlin, using the GLE in equation~(\ref{eq:mori_GLE}) with different components. Using the methods discussed in the main text, we employ the complete Mori-GLE for the parameter extraction. The blue data denote the prediction performance using the full GLE, the red data the unconfined case ($k=0$), the green data the deterministic case ($F_\text{R}(t) =0$) and the gray data the unconfined deterministic case ($k=0$ and $F_\text{R}(t) =0$). We see that the benchmark performance using a single cosine fit (black data) is for long prediction times surpassed only by including the linear force $kA_\text{f}(t)$ in the GLE, in agreement with the work of Chorin \cite{chorin_optimal_2000, chorin_non-markovian_2001, chorin_optimal_2002}. 
To investigate the effect of the memory kernel on the prediction quality, we additionally show results for predictions utilizing Markovian friction, i.e. for $\Gamma(t) = 2 (a+b) \delta(t)$ (purple).
In this case, equation~(\ref{eq:ODE_discret}) in the main text takes the form of the discrete Langevin equation
\begin{equation}
	\Ddot{A}_\text{f}^\text{i}= - (a+b) \Dot{A}_\text{f}^\text{i} - k A_\text{f}^\text{i} + F_\text{R}^\text{i},
	\label{eq:ODE_discret2}
\end{equation}
where $\Dot{A}_\text{f}^\text{i} = \Dot{A}_\text{f}(i\Delta t) = \bigl[A_\text{f}\bigl((i+1)\Delta t\bigr) - A_\text{f}\bigl((i-1)\Delta t\bigr)\bigr]/(2\Delta t)$ and $\Ddot{A}_\text{f}^\text{i} = \Ddot{A}_\text{f}(i\Delta t) = \bigl[A_\text{f}\bigl((i+1)\Delta t\bigr) - 2A_\text{f}(i\Delta t) + A_\text{f}\bigl((i-1)\Delta t\bigr)\bigr]/(\Delta t)^2$.
For the prediction based on the Langevin equation, we solve equation~(\ref{eq:ODE_discret2}) numerically using the last known values of $A_\text{f}$ and $\dot{A}_\text{f}$ as initial values and drawing uncorrelated samples of $F_\text{R}^\text{i}$ with variance 
$\langle F_\text{R}^{i}F_\text{R}^{i}\rangle = 2 B (a+b) /\Delta t$. We use the values obtained from the discrete estimation method for the parameters $a$, $b$, $B$, and $k$, listed in Table~\ref{tab:fits_kernel_tmax} in the main text. 

The GLE performs only slightly better than the Markovian Langevin prediction due to the dominating $\delta$-contribution at $t=0$ in the memory kernel.
Thus, the effect of the random force on the prediction quality is not very relevant, as discussed in Supplementary Information \ref{app:predictability}. 
To conclude, we find that the best prediction performance is achieved using all components of the GLE. Although the GLE does not offer a significant prediction performance advantage over the Langevin equation, it is suitable as a general model for the analysis and predictability investigation of time-series data, as we explain in Supplementary Information \ref{app:predictability}.
\section{Performance of the GLE Prediction with and without Filtering}
\label{app:performance_without_filtering}
\begin{figure*}
	\centering
	\includegraphics[width=0.9\linewidth]{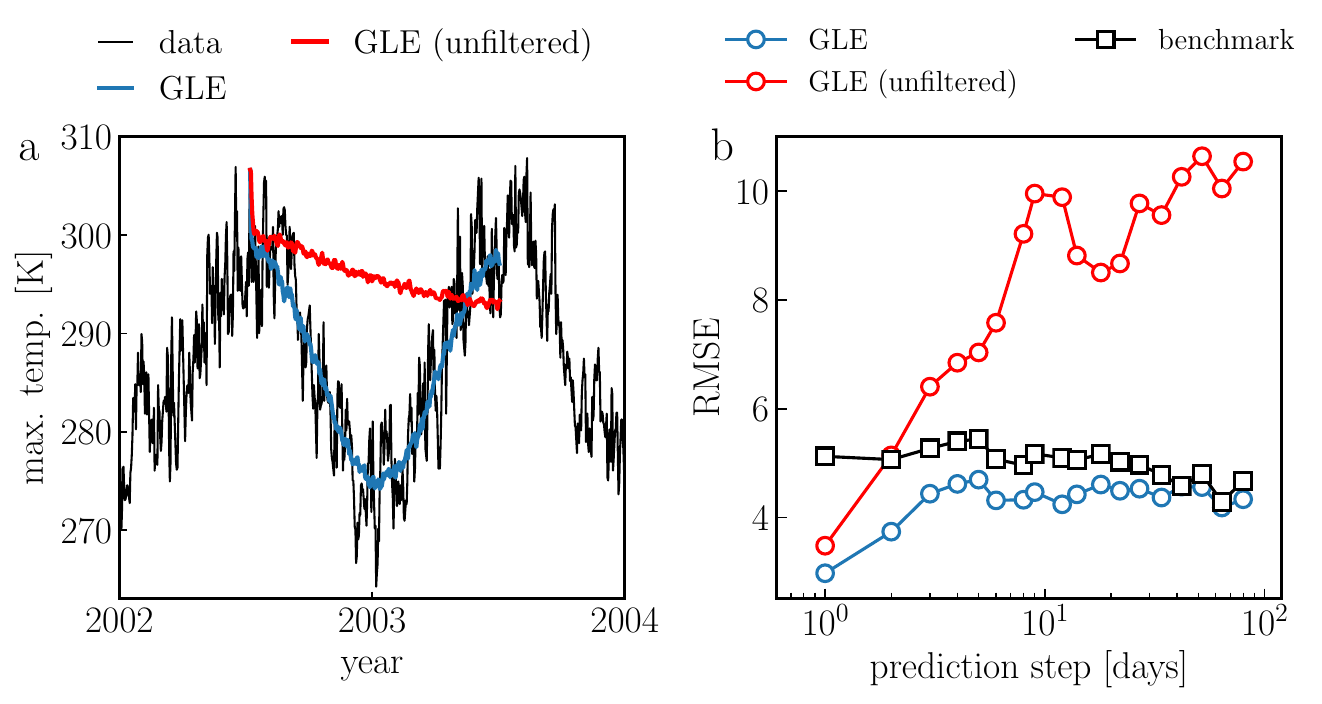}
	\caption{Prediction performance of the GLE with and without filtering. {\bf\sffamily a}: Exemplary prediction of the daily maximal temperature in Berlin, where we employ the unfiltered GLE prediction algorithm (red) based on equation~(\ref{eq:GLE}), using the PMF $U(A)$ and memory kernel $\Gamma(t)$ from the unfiltered trajectory $A$ (given in Fig.~\ref{fig:maxTemp1}c in the main text and Fig.~\ref{fig:maxTemp_compare}). We compare with the prediction using the Mori-GLE for the filtered trajectory $A_\text{f}$ (blue, given in Fig.~\ref{fig:maxTemp2}d in the main text) and the actual data (black). {\bf\sffamily b}: Comparison between the RMSEs for $N_\text{s}$ = 100 starting points (defined in equation~(\ref{eq:RMSE}) in the main text) with (blue) and without (red) filtering the trajectory, together with the benchmark performance (black), compare Fig.~\ref{fig:maxTemp2}e in the main text. }
	\label{fig:predUnfiltered}
\end{figure*}
In Fig.~\ref{fig:predUnfiltered}, we compare the prediction of the daily maximal air temperature in Berlin with and without data filtering prior to extracting the GLE parameters. The prediction in red in Fig.~\ref{fig:predUnfiltered}a is obtained using the approach introduced in Methods \ref{sec:GLE} for the unfiltered trajectory $A(t)$, i.e. we use the discrete version of equation~(\ref{eq:GLE}) and the non-harmonic potential $U(A)$ in Fig.~\ref{fig:maxTemp1}c, the memory kernel obtained from $A(t)$ (given in Fig.~\ref{fig:maxTemp_compare}) and the relation between memory kernel and random force in equation~(\ref{eq:FDT}) to obtain a prediction. For a prediction of $A$, we truncate the memory kernel data after $\tau_t$ = 365 days. The comparison with the prediction based on the filtered data $A_\text{f}(t)$ (blue, same as in Fig.~\ref{fig:maxTemp2}d) and the actual data (black) clearly shows that the memory kernel calculated from the unfiltered data $A$ is not capable of reproducing the seasonality in the actual data (black).  In Fig.~\ref{fig:predUnfiltered}b, we find that the overall prediction performance without filtering is significantly worse than with filtering (compare Fig.~\ref{fig:maxTemp2}e). After the second prediction step, the RMSE is even higher than the benchmark and keeps increasing.

\section{Prediction of Data from a Model System}
\label{app:model_sys}
\begin{figure*}
	\centering
	\includegraphics[width=1\linewidth]{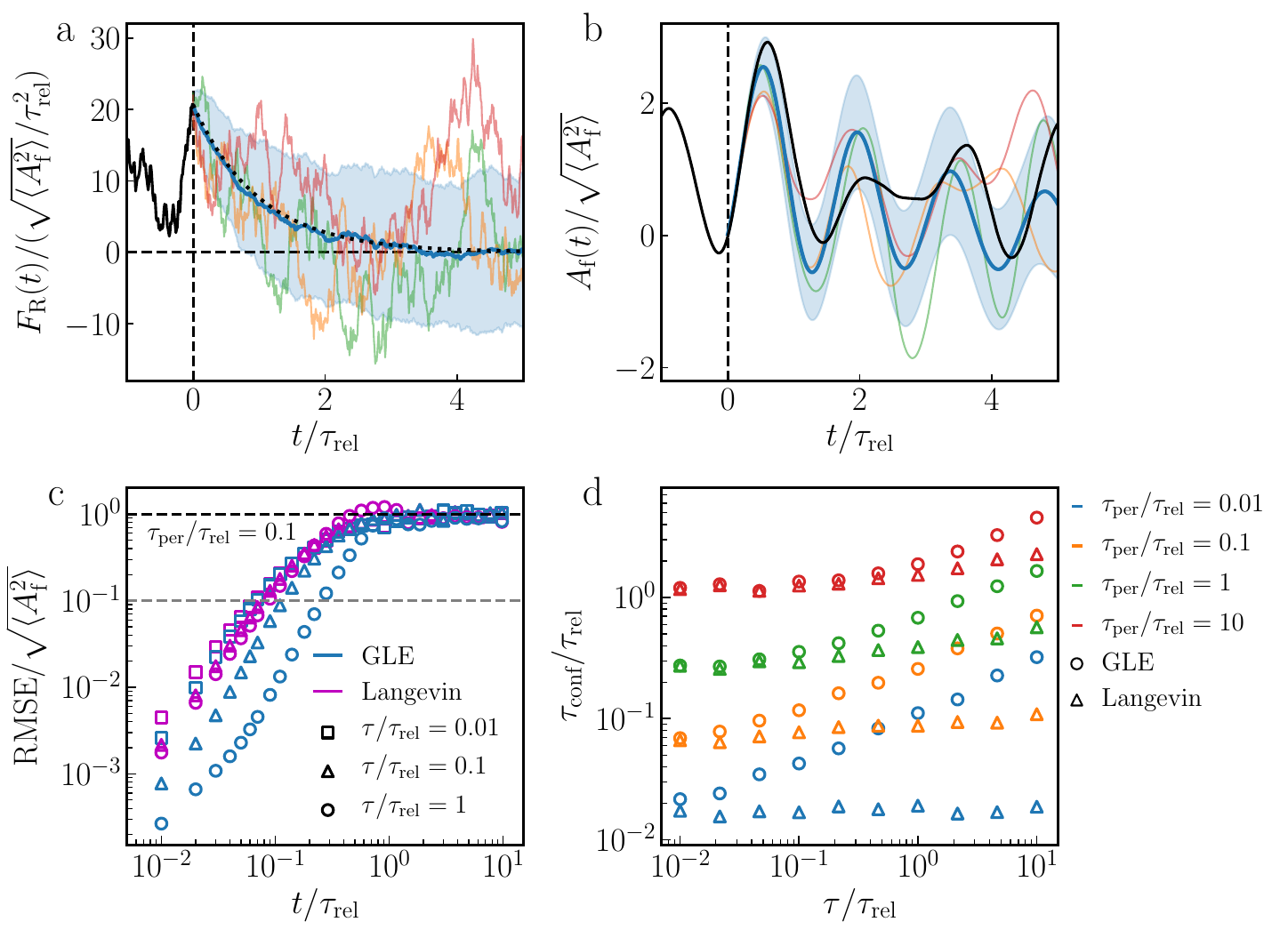}
	\caption{GLE prediction for a model system with exponential memory (defined in equation~(\ref{eq:ODE_system})). Here, we set $\Tilde{b}=1$. {\bf\sffamily a}: Three samples of the predicted random force $F_\text{R}(t)$ using the method introduced in Supplementary Information \ref{app:SFRF}, together with the mean of $N_\text{p}$ = 100 realizations (blue). Here, we choose $\tau = \tau_\text{rel}$ and $\tau_\text{per}/\tau_\text{rel} = 0.1 $. 	Note that the time $t$ is shifted to an arbitrarily chosen initial point (vertical dashed line), the starting point of all predictions. The black-dotted line denotes the prediction of the relaxation of the mean random force, i.e. $\langle F_\text{R}(t) \rangle = F_\text{R}^0e^{-t/\tau}$. The blue-shadowed area illustrates the standard deviation of the mean prediction. {\bf\sffamily b}: Three samples of the resulting predictions for $A_\text{f}(t)$ according to equation~(\ref{eq:ODE_discret}) along with the mean of $N_\text{p}$ = 100 realizations (blue) and the actual trajectory (black), which was generated from the model system according to equation~(\ref{eq:ODE_system}).
		{\bf\sffamily c}: Comparison of the RMSE (equation~(\ref{eq:RMSE}) in the main text) for predictions based on the GLE and Langevin equation (see Supplementary Information \ref{app:performance_comp}), for predictions at $N_\text{s} = 100$ different starting points and for different memory times $\tau$.  The black-dashed line denotes $\sqrt{\langle A_\text{f}^2 \rangle}$, the gray-dashed line denotes $\sqrt{\langle A_\text{f}^2 \rangle}/10$. {\bf\sffamily d:} Confidence time $\tau_\text{conf}$ of the GLE (circles) and Langevin (triangles) predictions defined by the time at which the RMSE in ({\bf\sffamily c}) reaches the threshold of $\sqrt{\langle A_\text{f}^2 \rangle}/10$, for different memory and persistence times.}
	\label{fig:modelsysPOC}
\end{figure*}
As discussed in Supplementary Information \ref{app:predictability}, prediction with the non-Markovian GLE is advantageous over Markovian approaches for long memory times. In fact, the predictability of a trajectory is expected to depend on the persistence time $\tau_\text{per} = (a+b)^{-1}$, the relaxation time $\tau_\text{rel} = (a+b)/k$ and the functional form of the memory kernel. In the following, we test our GLE-based prediction method for a synthetic time-series $A_\text{f}(t)$ that we generate using the GLE in equation~(\ref{eq:mori_GLE}) in the main text.
For simplicity, we assume a memory kernel consisting of the sum of a delta peak  and an exponential contribution, as defined in equation~(\ref{eq:meteo_kernel}) in the main text. Using this form, the GLE in equation~(\ref{eq:mori_GLE}) can be converted into a coupled set of Markovian Langevin equations given by \cite{kappler2018memory, ayaz2021non}
\begin{align}
	\nonumber \Ddot{A}_\text{f}(t) & = -   \frac{b}{\tau}  [A_\text{f}(t) - y(t)] - a\Dot{A}_\text{f}(t)  -k A_\text{f}(t) + \eta_\text{A}(t),\\
	\Dot{y}(t) & = - \frac{1}{\tau} [y(t) - A_\text{f}(t)] + \frac{1}{b} \eta_\text{y}(t).
	\label{eq:ODE_system0}
\end{align}
The random forces $\eta_\text{A}(t)$ and $\eta_\text{y}(t)$ follow Gaussian probability distributions with vanishing means and delta-correlated second moments
\begin{eqnarray}
	\label{eq:Markovian3}
	\langle \eta_\text{A}(t) \eta_\text{A}(0) \rangle =  2Ba \delta(t),
\end{eqnarray}	
and
\begin{eqnarray}
	\label{eq:Markovian3y}
	\langle \eta_\text{y}(t) \eta_\text{y}(0) \rangle =  2Bb \delta(t),
\end{eqnarray}	
respectively.
We simulate equation~(\ref{eq:ODE_system0}) by rescaling times by the relaxation time $\tau_\text{rel} = \frac{(a+b)\langle A_\text{f}^2 \rangle}{B}$ and positions by $\sqrt{\langle A_\text{f}^2 \rangle}$. We define the rescaled time $\tilde{t} = t/\tau_\text{rel}$ and the dimensionless coordinates $\tilde{A}_\text{f}(\tilde{t}) = A_\text{f}(\tau_\text{rel}\tilde{t})/\sqrt{\langle A_\text{f}^2 \rangle}$ and $\tilde{y}(\tilde{t}) = y(\tau_\text{rel}\tilde{t})/\sqrt{\langle A_\text{f}^2 \rangle}$.  The equation~(\ref{eq:ODE_system0}) for the rescaled variables then read \cite{kappler2019non}, using $ k\langle A_\text{f}^2 \rangle/B = 1$,
\begin{align}
	\nonumber \frac{\tau_\text{per}}{\tau_\text{rel}}\Ddot{\Tilde{A}}_\text{f}(\tilde{t}) &= - \tilde{b}\frac{\tau_\text{rel}}{\tau}  [\tilde{A}_\text{f}(\tilde{t}) - \tilde{y}(\tilde{t})] - \tilde{a} \dot{\Tilde{A}}_\text{f}(\tilde{t})-\tilde{A}_\text{f}(\tilde{t}) + \sqrt{\tilde{a}}\tilde{\eta}_\text{A}(\tilde{t}),\\
	\Dot{\tilde{y}}(\tilde{t}) &= - \frac{\tau_\text{rel}}{\tau} [\tilde{y}(\tilde{t}) - \tilde{A}_\text{f}(\tilde{t})] + \sqrt{\tilde{b}^{-1}}\tilde{\eta}_\text{y}(\tilde{t}),
	\label{eq:ODE_system}
\end{align}
where $\tilde{a} = \frac{a}{a+b}$ and $\tilde{b}=\frac{b}{a+b}$. The dimensionless noise terms $\tilde{\eta}_\text{A}(\tilde{t}) =\sqrt{\tilde{a}^{-1}}\sqrt{\langle A_\text{f}^2 \rangle}  \eta(\tau_\text{rel} \tilde{t})/B$ and $\tilde{\eta}_\text{y}(\tilde{t}) =\sqrt{\tilde{b}^{-1}}\sqrt{\langle A_\text{f}^2 \rangle}  \eta_\text{y}(\tau_\text{rel} \tilde{t})/B$ have zero means and variances $\langle \tilde{\eta}_\text{A}(\tilde{t}) \tilde{\eta}_\text{A}(0)\rangle = \tilde{\eta}_\text{y}(\tilde{t}) \tilde{\eta}_\text{y}(0)\rangle = 2 \delta(\tilde{t})$ .
We solve equation~(\ref{eq:ODE_system}) numerically with the RK4 method to obtain a trajectory $A_\text{f}(t)$ with $10^6$ steps and a time step $\delta t=\tau_\text{rel}/100$.  After generating a trajectory $A_\text{f}$, we predefine a randomly selected time $t_l$, separating the trajectory into past and future parts. This future part will be predicted with the method described in Methods \ref{sec:GLE} for the GLE in equation~(\ref{eq:mori_GLE}) and compared with the system's actual trajectory (black), which is the trajectory after the last known step $t_l$. 
We perform simulations for the model system using different memory times $\tau$ and persistence times $\tau_\text{per}$, and varying $\Tilde{a}$ and $\Tilde{b}$, which are related via $\Tilde{a} + \Tilde{b} = 1$. 

\begin{figure*}
	\centering
	\includegraphics[width=1\linewidth]{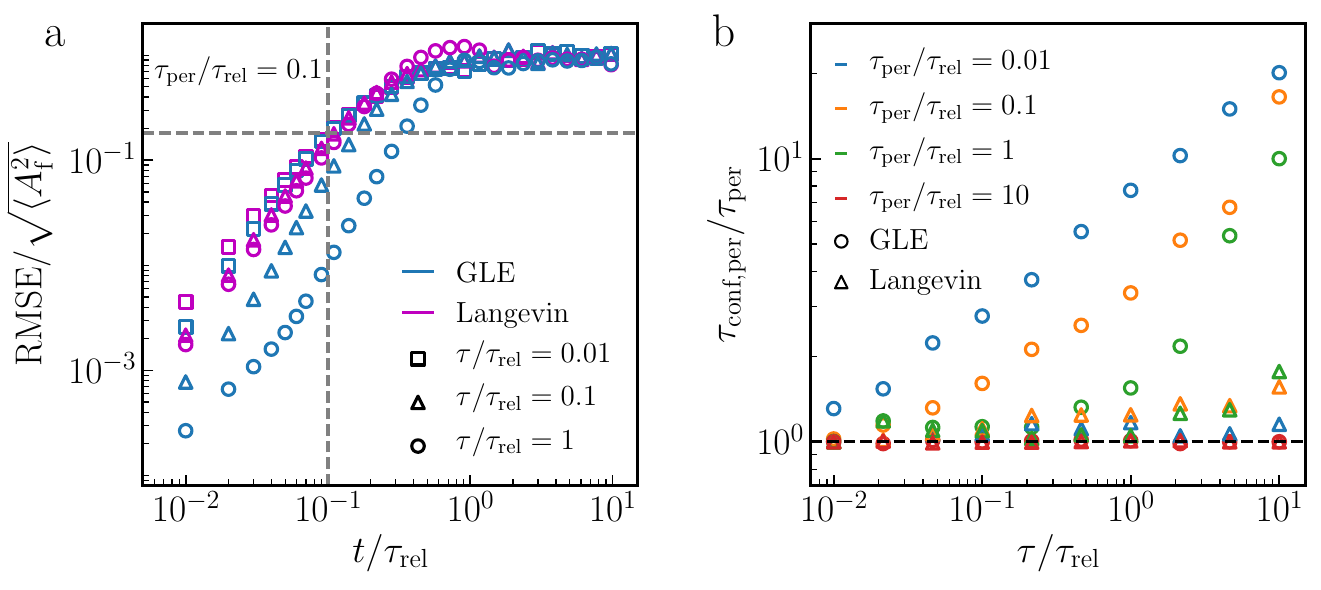}
	\caption{
		{\bf\sffamily a}: Comparison of the RMSE (equation~(\ref{eq:RMSE}) in the main text) for predictions based on the GLE and Langevin equation for $\tau_\text{per}/\tau_\text{rel} = 0.1$ and for different memory times $\tau$ (compare Fig.~\ref{fig:modelsysPOC}c).  The gray-dashed horizontal line denotes the threshold that the RMSE of the prediction using the Langevin equation for a trajectory with short memory time (purple squares) reaches after a prediction time of $t = \tau_{\text{per}}$ (gray-dashed vertical line). {\bf\sffamily b:} Confidence time $\tau_\text{conf,per}$, computed based on the threshold definition in ({\bf\sffamily a}), of the GLE (circles) and Langevin (triangles) predictions for different memory and persistence times.}
	\label{fig:modelsysPOC_collapsed}
\end{figure*}

In Fig.~\ref{fig:modelsysPOC}a and b, we show an exemplary random-force and observable trajectory of the model system with $\tau/\tau_\text{rel} = 1$, $\tau_\text{per}/\tau_\text{rel} = 0.1$ and $\Tilde{b}=1$, together with predictions starting from a randomly chosen step (vertical dashed line). 
The non-vanishing mean of $N_\text{p}$ = 100 random force realizations in (a), which decays as $\langle F_\text{R}(t) \rangle = F_\text{R}^0e^{-t/\tau}$, where $F_\text{R}^0$ is the last time step from $F_\text{R}^\text{p}$, as a consequence from the conditional Gaussian sampling (equation~(\ref{eq:non-cond_gaussian}), demonstrates the importance of random force prediction. 
The mean over $N_\text{p}$ = 100 predictions of the observable $A_\text{f}(t)$ in Fig.~\ref{fig:modelsysPOC}b also does not vanish, and the standard deviation increases with increasing time steps, visualizing the importance of memory effects in the prediction.
\begin{figure*}
	\centering
	\includegraphics[width=1\linewidth]{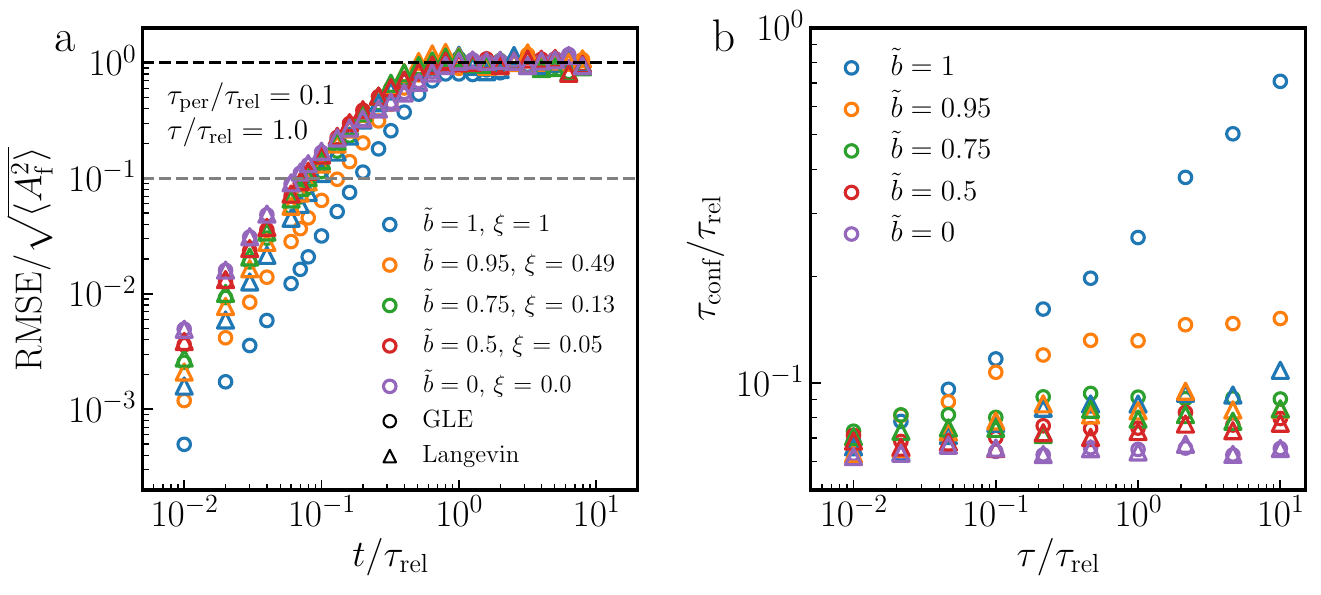}
	\caption{Relevance of the exponential memory component in the memory kernel (given by equation~(\ref{eq:meteo_kernel})) for the model system predictability.  {\bf\sffamily a}: RMSE for predictions obtained from the GLE according to equation~(\ref{eq:RMSE}), for predictions at $N_\text{s} = 100$ different starting points and for different values of $\Tilde{b} = b/(a+b)$ of the model system (see Fig.~\ref{fig:modelsysPOC} for details), using $\tau_\text{per}/\tau_\text{rel} = 0.1$ and $\tau/\tau_\text{rel} = 1$. The corresponding non-Markovian fraction factors $\xi$, defined in equation~(\ref{eq:memory_fraction}), are also given. {\bf\sffamily d:} Confidence time $\tau_\text{conf}$ of GLE and Langevin predictions as a function of the memory time $\tau$ and for different $\Tilde{b}$.}
	\label{fig:modelsysPOC_delta}
\end{figure*}

In Fig.~\ref{fig:modelsysPOC}c, we show the computed RMSE (blue, defined in equation~(\ref{eq:RMSE}) in the main text) for predictions averaged over $N_\text{p}$ = 100 realizations at $N_\text{s}$ = 100 different starting points, in dependence of the prediction depth, for $\tau_\text{per}/\tau_\text{rel} = 0.1$ and for memory times $\tau/\tau_\text{rel} = 0.01, \tau/\tau_\text{rel} = 0.1$, and $\tau/\tau_\text{rel} = 1$. We compare predictions using the GLE with the Langevin prediction scheme, where $\Gamma(t) = 2 (a+b) \delta(t)$,  described in Supplementary Information \ref{app:performance_comp}. The GLE model outperforms the Langevin model for $\tau/\tau_\text{rel} \geq 0.1$ which demonstrates our prediction technique's usability. For times up to $\tau_\text{per}$, persistence in the data is the main mechanism that leads to predictability, as discussed in Supplementary Information \ref{app:predictability}. Hence, the GLE prediction is not superior to a Langevin prediction for $\tau/\tau_\text{rel}\leq 0.1 = \tau_\text{per}/\tau_\text{rel}$. For $\tau > \tau_\text{per}$, non-Markovian effects make the data predictable up to the memory time. This is illustrated by the fact that the GLE prediction remains below $\sqrt{\langle A_\text{f}^2 \rangle}$ up to the memory time. We define the threshold RMSE/$\sqrt{\langle A_\text{f}^2 \rangle} = 0.1$ as the predictability limit below which a prediction is reliable (gray-dashed line). We determine the confidence time $\tau_\text{conf}$, which is the time at which the RMSE reaches the threshold of $\sqrt{\langle A_\text{f}^2 \rangle}/10$, by linear interpolation.
\\ \indent In Fig.~\ref{fig:modelsysPOC}d, it becomes evident that the predictability of the system increases with the memory time and the persistence time, provided that the GLE is used as a predictor. For small memory times ($\tau<\tau_\text{per}$), the persistence time is the time scale that limits the predictability, as discussed in Supplementary Information \ref{app:predictability}. Interestingly, the confidence time for large persistence slightly increases while increasing the memory time when using the Langevin equation for prediction (green and red triangles). We argue that this is due to memory-induced data persistence, which increases the predictability of the actual trajectory.\\ 
\indent For the results shown in Fig.~\ref{fig:modelsysPOC}d, we chose the threshold value of the RMSE such that the advantage of the GLE over the Langevin equation becomes evident if the memory time is long. Alternatively, one can choose a threshold which enables a better discussion of the interplay between persistence and memory times, as demonstrated in Fig.~\ref{fig:modelsysPOC_collapsed}. Since a trajectory with vanishing memory time can, generally, only be predicted up to the persistence time (Supplementary Information \ref{app:predictability}), we choose the RMSE threshold in Fig.~\ref{fig:modelsysPOC_collapsed}a as the value that the RMSE using the Langevin equation for short memory, here $\tau/\tau_\text{rel} = 0.01$ (purple squares), reaches after the persistence time $\tau_\text{per}$ (gray lines in Fig.~\ref{fig:modelsysPOC_collapsed}a). Computing the confidence time $\tau_\text{conf,per}$ from this threshold for the RMSE curves for different persistence times reveals a collapse of the data in Fig.~\ref{fig:modelsysPOC_collapsed}b, after dividing by the persistence time. This presentation of the performance of GLE and Langevin predictions shows that memory effects are only beneficial if they are longer than the persistence time, in perfect correspondance to the discussion in Supplementary Information \ref{app:predictability}.
\\
\indent In Fig.~\ref{fig:modelsysPOC_delta}, we examine the influence of the exponential memory component in the memory kernel on data predictability and show results for different $\Tilde{b} = b/(a+b)$. For a factor of $\Tilde{b} \leq 0.75$, which corresponds to non-Markovian fractions (defined in Supplementary Information \ref{app:predictability}) lower than $\xi= 0.13$, computed using equation~(\ref{eq:memory_fraction}), the RMSE and the confidence time of the GLE prediction nearly coincide with the Langevin prediction. Therefore, memory effects are relevant for a fraction $\Tilde{b}$ above 0.75. 

\section{Memory Kernels of Other Weather Data}
\label{app:kernels_meteo}
\begin{figure*}[hbt!]
	\centering
	\includegraphics[width=\textwidth]{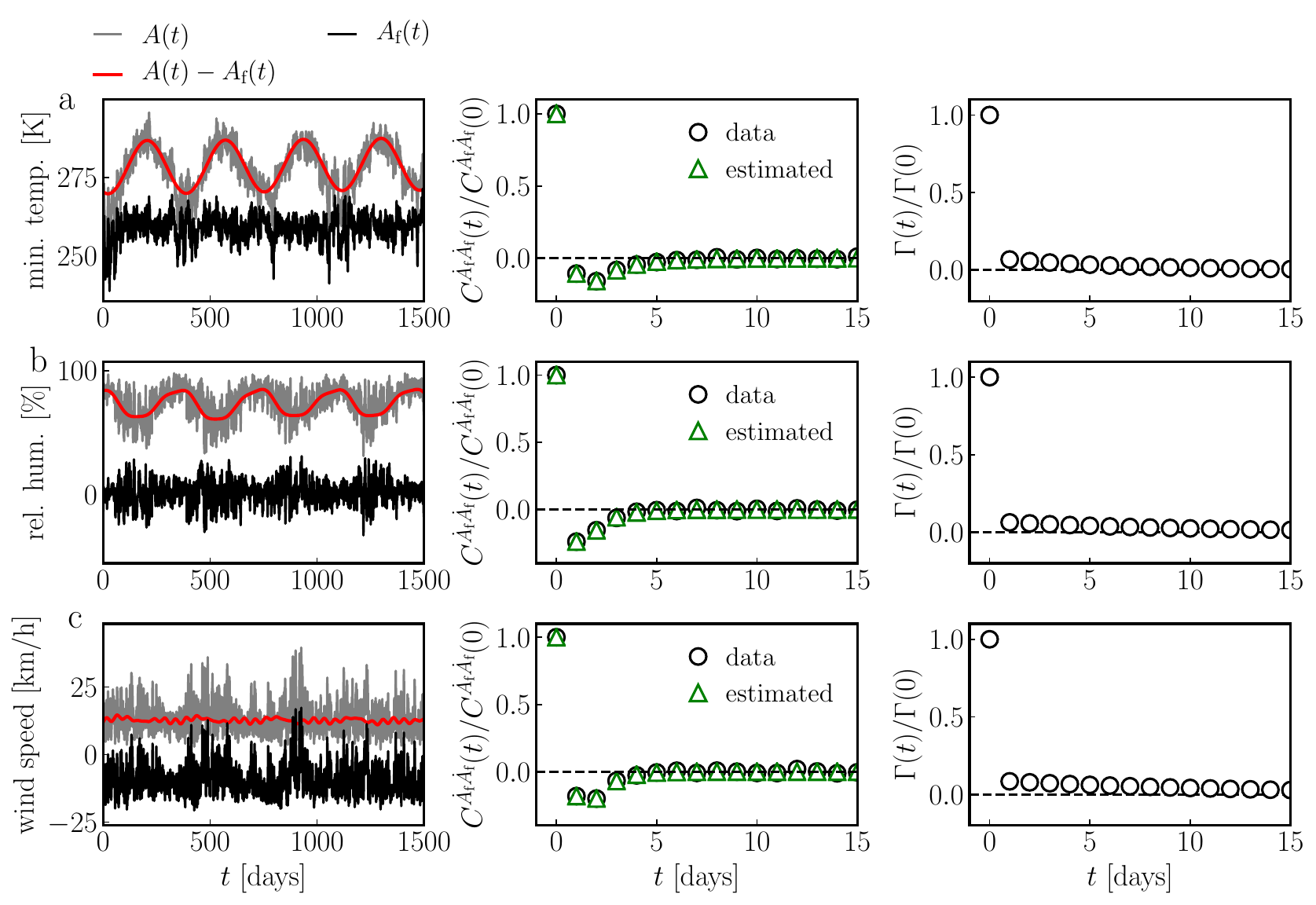}
	\caption{Filtering and memory kernel extraction, using the discrete estimation method, of the weather data investigated in Fig.~\ref{fig:maxTemp2}f in the main text. We show the discrete velocity autocorrelation functions $C^{\Dot{A}_\text{f}\Dot{A}_\text{f}}(t)$ and memory kernels $\Gamma(t)$ (shown in Fig.~\ref{fig:maxTemp1} in the main text for information) for the minimal air temperature ({\bf\sffamily a}), the relative air humidity ({\bf\sffamily b}), and the maximal wind speed ({\bf\sffamily c}). The fitting parameters are summarized in Table~\ref{tab:fits_kernel} in Supplementary Information \ref{app:fit_kernels_params}. Note that $A_\text{f}(t)$ is shifted in y-direction for better visibility.}
	\label{fig:meteo_kernels}
\end{figure*}
In Fig.~\ref{fig:meteo_kernels}, we show the extraction results using the discrete estimation method explained in Supplementary Information \ref{app:mitterwallner} for the weather data investigated in Fig.~\ref{fig:maxTemp2}f in the main text. We proceed similarly as for the maximal temperature results (given in Fig.~\ref{fig:maxTemp1} for more information). All daily weather data sets are dominated by a $\delta$-component at $t=0$, followed by a small exponential decay. The fitting parameters, together with the results for the maximal temperature, are summarized in Table~\ref{tab:fits_kernel} in Supplementary Information \ref{app:fit_kernels_params}.

\section{Extracted Memory Kernels Using the Volterra Method}
\label{app:results_volterra}
\begin{figure}[hbt!]
	\centering
	\includegraphics[width=1\linewidth]{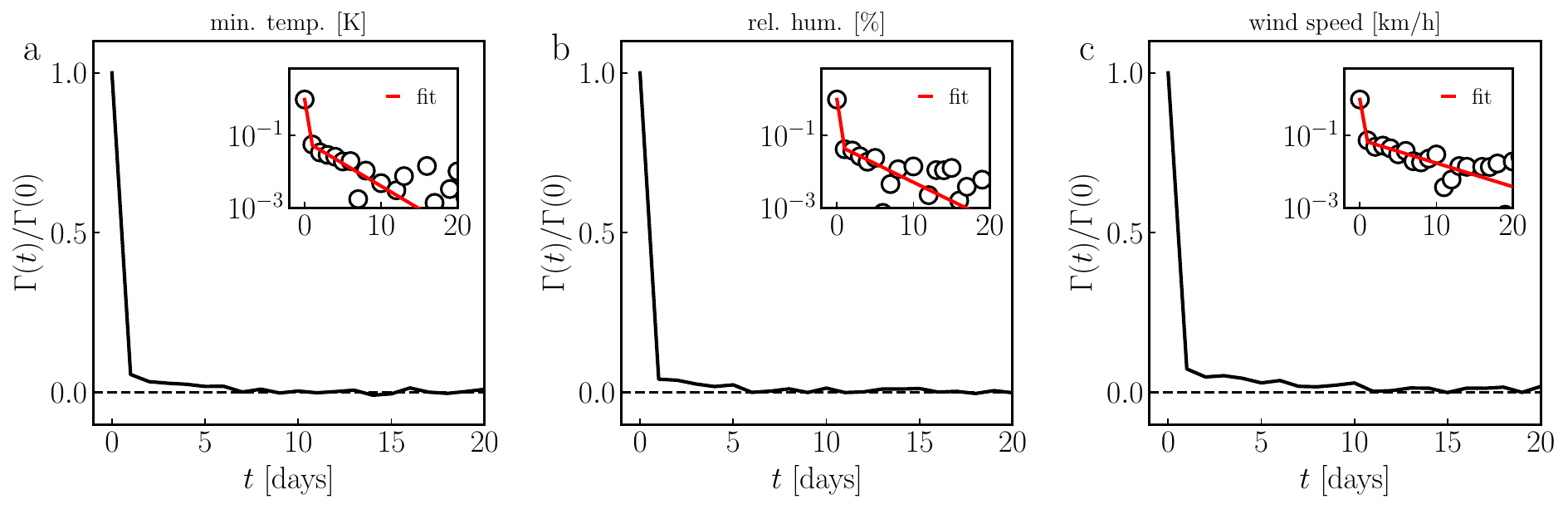}
	\caption{Extracted memory kernel $\Gamma(t)$ of $A_\text{f}(t)$ (see Fig.~\ref{fig:meteo_kernels} for details) for the minimal air temperature ({\bf\sffamily a}), the relative air humidity ({\bf\sffamily b}), and the maximal wind speed ({\bf\sffamily c}), using the Volterra method (Supplementary Information \ref{app:volterra}). The red-solid lines in the insets are fits of the memory kernel according to equation~(\ref{eq:meteo_kernel}) in the main text. The fitting parameters, together with $k$ and $B$, are listed in Supplementary Information \ref{app:fit_kernels_params}.}
	\label{fig:meteo_kernels_volterra}
\end{figure}

\begin{figure}[hbt!]
	\centering
	\includegraphics[width=1\linewidth]{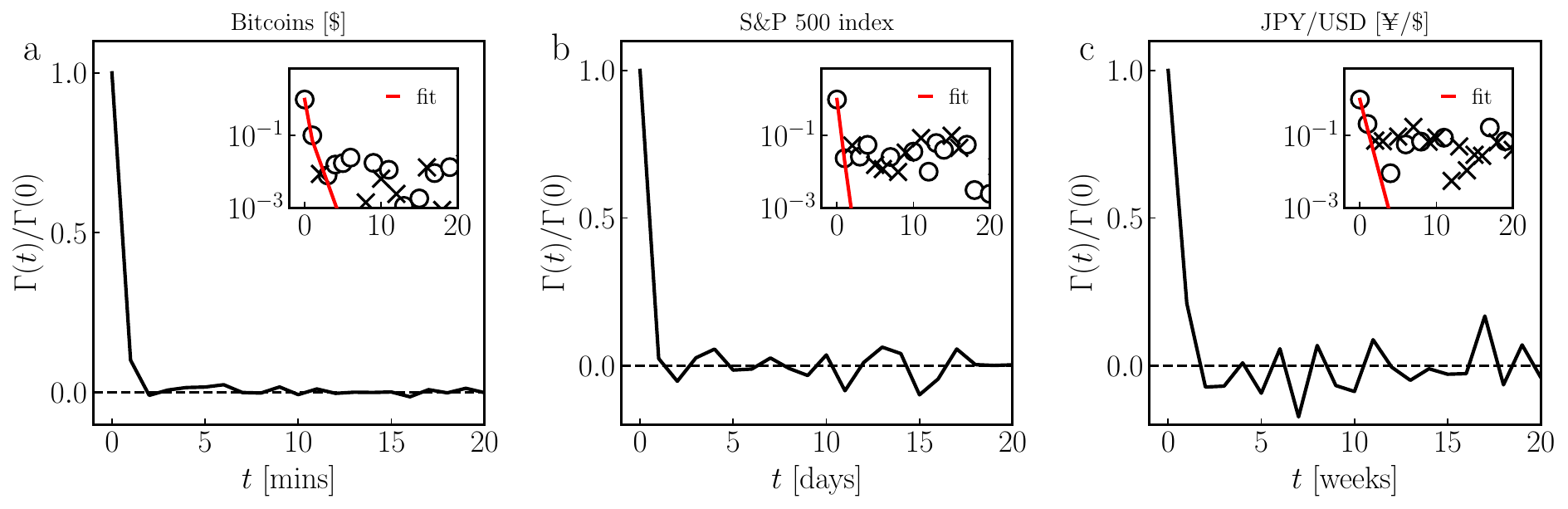}
	\caption{Extracted memory kernel $\Gamma(t)$ of $A_\text{f}(t)$ (see Fig.~\ref{fig:finance1} in the main text for details) for the minutely resolved Bitcoin price ({\bf\sffamily a}), the Standard \& Poor's 500 (S\&P 500) stock index over trading days ({\bf\sffamily b}), and the weekly resolved exchange rate between the Japanese Yen (JPY) and US Dollar (USD) ({\bf\sffamily c}), using the Volterra method (Supplementary Information \ref{app:volterra}). The red-solid lines in the insets are fits of the memory kernel according to equation~(\ref{eq:meteo_kernel}) in the main text. Circles denote positive and crosses negative values. The fitting parameters, $k$ and $B$, are listed in Supplementary Information \ref{app:fit_kernels_params}.}
	\label{fig:finance_kernels_volterra}
\end{figure}

In Fig.~\ref{fig:meteo_kernels_volterra}, we summarize the memory kernels of $A_\text{f}(t)$  for the minimal air temperature, the relative air humidity, and the maximal wind speed (see Fig.~\ref{fig:meteo_kernels} for details) extracted  using the Volterra extraction method explained in Supplementary Information \ref{app:volterra}. 
All memory kernels can be represented by the sum of a dominating $\delta$-component at $t=0$ and an exponential decay. We show data fits according to equation~(\ref{eq:meteo_kernel}) in the insets, denoted as red-solid lines. These fits are the starting point for extracting the memory kernel using the discrete estimation method, whose results we show in Fig.~\ref{fig:meteo_kernels} and which generally differ from the memory kernels extracted using the Volterra method. The fitting parameters, together with $k$ and $B$, are given in Supplementary Information \ref{app:fit_kernels_params}. \\
\indent The memory kernels extracted from the financial data, together with fits according to equation~(\ref{eq:meteo_kernel}), are shown in Fig.~\ref{fig:finance_kernels_volterra}. No distinct exponential decay is visible here,  only a $\delta$-contribution followed by noise. Nevertheless, we choose the model in equation~(\ref{eq:meteo_kernel}) for the discrete estimation method when applying to financial data since this model includes the possibility of a vanishing exponential decay component.

\section{Relation between Random Force and Memory Kernel for Discrete Data}
\label{app:fdt_disc}
\begin{figure*}%[hbt!]
	\centering
	\includegraphics[width=0.9\linewidth]{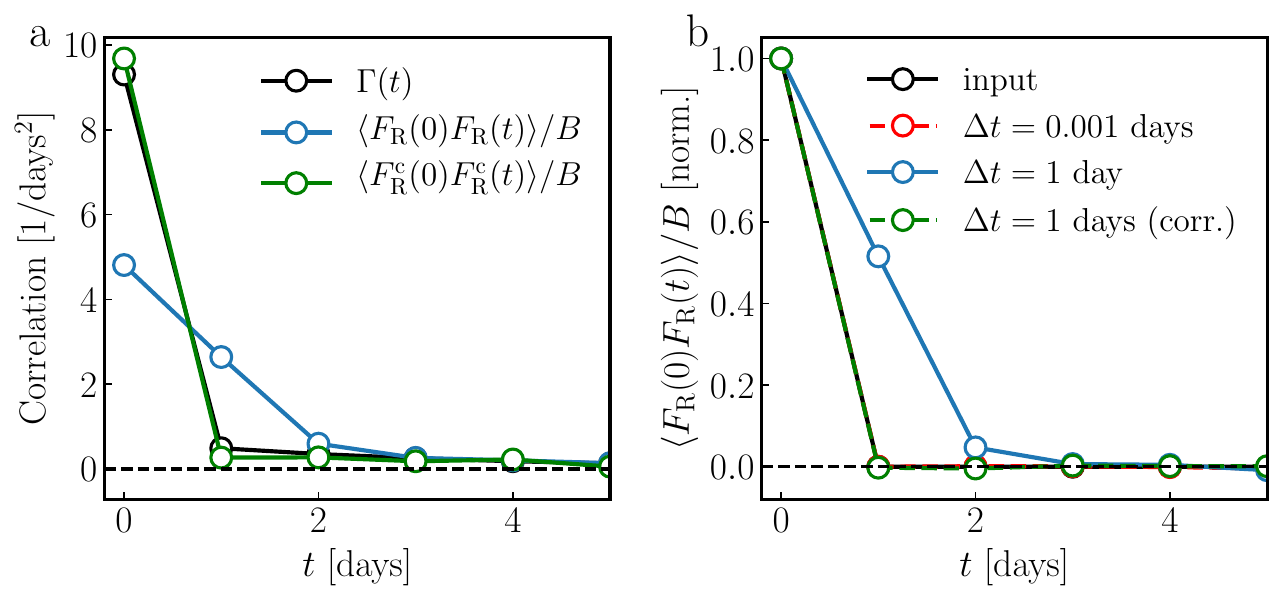}
	\caption{{\bf\sffamily a}: Autocorrelation of the random-force trajectory $F_\text{R}(t)$ from the daily maximal temperature data in Berlin shown in Fig.~\ref{fig:maxTemp1} in the main text computed by equation~(\ref{eq:fr0}) (blue data), together with the memory kernel $\Gamma(t)$ (black), extracted by the discrete estimation method. The green data is the autocorrelation of the corrected data $F_\text{R}^\text{i,c}$ (given in equation~(\ref{eq:fr_corrected})), which is used for the random-force generation technique described in Supplementary Information \ref{app:SFRF}.
		{\bf\sffamily b}: Autocorrelation of the random-force trajectory $F_\text{R}(t)$ (red, for details see text) obtained by equation~(\ref{eq:fr0}) with $\Delta t = 0.001$ days from synthetic data simulated with equation~(\ref{eq:ODE_discret2}) with a time step of $\delta t = 0.001$ days. The fact that the random force obtained from a trajectory with a discretization time step $\Delta t$ = 1 day that is higher than the simulation time step $\delta t = 0.001$ days (blue) does not reproduce the input autocorrelation (black) clearly shows that the violation of equation~(\ref{eq:FDT2}) in ({\bf\sffamily a}) is due to discretization effects. The green data points denote the autocorrelation of the corrected data $F_\text{R}^\text{i,c}$, given in equation~(\ref{eq:fr_corrected}). The black, red, and green data superimpose, and all data sets are normalized by their initial values due to the $\delta$-component at $t=0$.}
	\label{fig:check_fdt_discrete}
\end{figure*}
For the random-force generation in the GLE prediction method, introduced in Supplementary Information \ref{app:SFRF}, we use the memory kernel of the data as a covariance function and thus assume a valid equation~(\ref{eq:FDT2}) in the main text, even for the discrete data. As an example, in Fig.~\ref{fig:check_fdt_discrete}a, we show the autocorrelation of the random-force trajectory from  the daily maximal temperature data in Berlin in Fig.~\ref{fig:maxTemp1} in the main text, obtained in the following way: We compute the random force of the given data by inverting the Mori-GLE in equation~(\ref{eq:mori_GLE}) for the random force 
\begin{eqnarray}
	\label{eq:fr_gle}
	F_\text{R}(t) = \Ddot{A}_\text{f}(t) + \int_0^\text{t}\mathrm{d}s\,\Gamma(t-s)\Dot{A}_\text{f}(s)  + k A_\text{f}(t) ,
\end{eqnarray}
which is discretized using the trapezoidal integral rule, leading to
\begin{eqnarray}
	\label{eq:fr0}
	F_\text{R}^\text{i} = \ddot{A}_\text{f}^\text{i} + \frac{\Delta t}{2} \Gamma_0 \dot{A}_\text{f}^\text{i} + \frac{\Delta t}{2} \Gamma_\text{i} \dot{A}_\text{f}^0 + \Delta t \sum_{j=1}^{i-1} \Gamma_\text{j} \dot{A}_\text{f}^\text{i-j} + k A_\text{f}^\text{i}.
\end{eqnarray}
The velocities $\dot{A}_\text{f}^\text{i}$ and accelerations $\ddot{A}_\text{f}^\text{i} $ in equation~(\ref{eq:fr0}) are obtained by central differences, i.e. $\dot{A}_\text{f}^\text{i} = \dot{A}_\text{f}(i\Delta t) = \bigl[A_\text{f}\bigl((i+1)\Delta t\bigr) - A_\text{f}\bigl((i-1)\Delta t\bigr)\bigr]/(2\Delta t)$ and  $\ddot{A}_\text{f}^\text{i} = \ddot{A}_\text{f}(i\Delta t) = \bigl[A_\text{f}\bigl((i+1)\Delta t\bigr) - 2A_\text{f}(i\Delta t) + A_\text{f}\bigl((i-1)\Delta t\bigr)\bigr]/(\Delta t)^2$.
For the first three time steps, in Fig.~\ref{fig:check_fdt_discrete}a, we see distinct deviations between the autocorrelation function of $F_\text{R}^\text{i}$, calculated according to equation~(\ref{eq:fr0}) (blue data), and the memory kernel (black data), extracted from the data with the discrete estimation method, as explained in Supplementary Information \ref{app:mitterwallner}.\\
\indent To investigate the origin of this deviation, we generate synthetic data from a model system described by a Langevin equation, i.e. for the memory kernel $\Gamma(t) = 2(a+b)\delta(t)$, in a harmonic potential. We generate a quasi-continuous trajectory $A_\text{f}$ with a time resolution of $\delta t = 0.001$ days and a total number of $2\cdot\:10^{7}$ steps by solving equation~(\ref{eq:ODE_discret2}) in Supplementary Information \ref{app:performance_comp} numerically. 
For the parameters, we use the obtained values for the maximal temperature, listed in Table~\ref{tab:fits_kernel} in Supplementary Information \ref{app:fit_kernels_params}. 
Using equation~(\ref{eq:fr0}), we calculate the random force $F_\text{R}^\text{i}$ at discrete time steps $i\Delta t$ from the trajectory $A_\text{f}^\text{i}$. 
In Fig.~\ref{fig:check_fdt_discrete}b, we find that the autocorrelation of the computed random force only agrees with the input memory kernel $\Gamma(t)$ when the time step $\Delta t$ is equal to the simulation time step, $\Delta t = \delta t$ (red data), while it deviates from the input memory kernel $\Gamma(t)$ for a higher time step of $\Delta t = 1$ day (blue), similarly to the results in Fig.~\ref{fig:check_fdt_discrete}a. This indicates that the violation of equation~(\ref{eq:FDT2}) is a discretization effect due to the calculation of velocities and accelerations by finite-time differences in equation~(\ref{eq:fr0}). Note that the red data in (b) is only shown for the discrete time steps of $\Delta t$ = 1 day.\\
\indent  Since the discrete velocities differ from their continuous counterparts, i.e. $\Dot{A}_\text{f}(i\Delta t) \neq \Dot{A}_\text{f}(i1000\delta t)$ for $\Delta t = 1000 \delta t$, we introduce a method to correct discrete velocities and accelerations for this discretization effect based on a given trajectory $A_\text{f}^\text{i}$ and the memory kernel $\Gamma_\text{i}$.
Using the discrete estimation method introduced in Supplementary Information \ref{app:mitterwallner}, we can accurately estimate the continuous velocity autocorrelation function (VACF) at the discrete time steps $C^{\Dot{A}_\text{f}\Dot{A}_\text{f}}(i \Delta t)$ from the MSD in equation~(\ref{eq:vacf2}). From $C^{\Dot{A}_\text{f}\Dot{A}_\text{f}}(t) = \partial^2 C_{\text{MSD}}(t) / \left(2 \partial t\right)^2 $ we obtain
\begin{eqnarray}
	\label{eq:vacf4}
	C^{\Dot{A}_\text{f}\Dot{A}_\text{f}}(t) = \frac{B}{2\tau^2} \left( \sum_{i=1}^{3}\frac{e^{-\sqrt{-\nu_\text{i}^2} t} \sqrt{-\nu_\text{i}^2}}{\prod_{j\neq i} (\nu_\text{i}^2-\nu_\text{j}^2) } \left[ k_1 + k_2 \nu_\text{i}^2 \right] \right) \,,
\end{eqnarray}
from which we compute the effect of discretization on the data VACF as
\begin{eqnarray}
	\label{eq:vacf_diff}
	C^{\Dot{A}_\text{f}\Dot{A}_\text{f}}_\Delta(i\Delta t) = C^{\Dot{A}_\text{f}\Dot{A}_\text{f}}(i\Delta t) - C^{\Dot{A}_\text{f}\Dot{A}_\text{f}}_\text{data}(i\Delta t).
\end{eqnarray}
Using equation~(\ref{eq:vacf_diff}), we correct the discrete velocities $\dot{A}_\text{f}^\text{i}$, based on the autocorrelation $C^{\Dot{A}_\text{f}\Dot{A}_\text{f}}_\text{data}(i\Delta t)$, to values $\Dot{A}_\text{f}^\text{i,c}$ that yield the continuum-limit velocity autocorrelation $C^{\Dot{A}_\text{f}\Dot{A}_\text{f}}(i\Delta t)$. We define the corrected values by adding a correction term to $\Dot{A}_\text{f}^\text{i}$, i.e. $\Dot{A}_\text{f}^\text{i,c} = \Dot{A}_\text{f}^\text{i} + \Dot{A}_\text{f}^{\text{i},\Delta}$, with $\Dot{A}_\text{f}^{\text{i},\Delta}$ being sampled from a conditional Gaussian distribution (explained in Supplementary Information \ref{app:SFRF}) with zero mean and covariance function $C^{\Dot{A}_\text{f}\Dot{A}_\text{f}}_\Delta(i\Delta t)$, determined by equation~(\ref{eq:vacf_diff}).
This leads to the discrete velocity $\Dot{A}_\text{f}^\text{i,c}$ with the correct autocorrelation
\begin{align}
	\label{eq:vacf5}
	\langle \Dot{A}_\text{f}^\text{0,c} \Dot{A}_\text{f}^\text{i,c} \rangle &= \langle \Dot{A}_\text{f}^\text{0} \Dot{A}_\text{f}^\text{i} \rangle + \langle \Dot{A}_\text{f}^{\text{0},\Delta} \Dot{A}_\text{f}^{\text{i},\Delta}  \rangle + \langle \Dot{A}_\text{f}^{\text{0}}  \Dot{A}_\text{f}^{\text{i},\Delta}  \rangle + \langle \Dot{A}_\text{f}^{\text{0},\Delta}  \Dot{A}_\text{f}^{\text{i}}  \rangle \\ 
	&= \langle \Dot{A}_\text{f}^{\text{0}} \Dot{A}_\text{f}^{\text{i}} \rangle + \langle \Dot{A}_\text{f}^{\text{0},\Delta} \Dot{A}_\text{f}^{\text{i},\Delta} \rangle \\ 
	&=C^{\Dot{A}_\text{f}\Dot{A}_\text{f}}(i\Delta t).
\end{align}
Note that this approach only works if the difference VACF $C^{\Dot{A}_\text{f}\Dot{A}_\text{f}}_\Delta(i\Delta t)$ is positive definite. 
\\ \indent The same correction can be applied to the discrete acceleration $\Ddot{A}_\text{f}^{\text{i}}$, i.e. $\Ddot{A}_\text{f}^\text{i,c} = \Ddot{A}_\text{f}^\text{i} + \Ddot{A}_\text{f}^{\text{i},\Delta}$, where the continuous autocorrelation (AACF) at the discrete steps $C^{\Ddot{A}_\text{f}\Ddot{A}_\text{f}}(i \Delta t)$ is given by the curvature of the continuum VACF according to $C^{\Ddot{A}_\text{f}\Ddot{A}_\text{f}}(t) = - \partial^2 C^{\Dot{A}_\text{f}\Dot{A}_\text{f}}(|t|) / \partial t^2$
\begin{eqnarray}
	\label{eq:vacf6}
	C^{\Ddot{A}_\text{f}\Ddot{A}_\text{f}}(t) = -  \frac{B}{2\tau^2} \left( \sum_{i=1}^{3}\frac{e^{-\sqrt{-\nu_\text{i}^2} t} (\sqrt{-\nu_\text{i}^2})^3 - 2 (\sqrt{-\nu_\text{i}^2})^2 \delta(t)}{\prod_{j\neq i} (\nu_\text{i}^2-\nu_\text{j}^2) } \left[ k_1 + k_2 \nu_\text{i}^2 \right] \right) \,.
\end{eqnarray}
$\Ddot{A}_\text{f}^{\text{i},\Delta}$ is sampled from a conditional Gaussian distribution, as explained in Supplementary Information \ref{app:SFRF}, with zero mean and covariance function $C^{\Ddot{A}_\text{f}\Ddot{A}_\text{f}}_\Delta(i\Delta t)$.
Note that the correction term
\begin{equation}
	C^{\Ddot{A}_\text{f}\Ddot{A}_\text{f}}_\Delta(i\Delta t) = C^{\Ddot{A}_\text{f}\Ddot{A}_\text{f}}(i\Delta t) - C^{\Ddot{A}_\text{f}\Ddot{A}_\text{f}}_\text{data}(i\Delta t),
\end{equation}
depends on the time step we use to express the $\delta$-contribution in equation~(\ref{eq:vacf6}) in discretized form. Here, we choose the simulation time step of the quasi-continuous trajectory ($\delta t = 0.001$ days). 

Using the corrected velocities and accelerations, we compute the corrected random-force trajectory according to
\begin{eqnarray}
	\label{eq:fr_corrected}
	F_\text{R}^\text{i,c}  \sqrt{\Delta t/\delta t}= \Ddot{A}_\text{f}^{\text{i,c}} + \frac{\Delta t}{2} \Gamma_0 \Dot{A}_\text{f}^{\text{i,c}} + \frac{\Delta t}{2} \Gamma_\text{i} \Dot{A}_\text{f}^{\text{0,c}} + \Delta t \sum_{j=1}^{i-1} \Gamma_\text{j} \dot{A}_\text{f}^{\text{i-j,c}} + k A_\text{f}^\text{i},
\end{eqnarray}
and find in Fig.~\ref{fig:check_fdt_discrete}b that the corresponding autocorrelation (green) is in agreement with the discrete memory kernel $\Gamma_\text{i}$ (black). Note that $A_\text{f}^\text{i}$ remains uncorrected as the position is not influenced by discretization. Multiplying by  $\sqrt{\delta t/\Delta t}$ to obtain $F_\text{R}^\text{i,c}$ from equation~(\ref{eq:fr_corrected}) is necessary for mapping the $F_\text{R}^\text{i}$ on the continuous random-force trajectory at the discrete time steps $i\Delta t$, since the autocorrelation of the random force includes a $\delta$-peak at $t=0$. 
\\ \indent In Fig.~\ref{fig:check_fdt_discrete}a, we demonstrate that the correction works properly for the maximal temperature data in Berlin. The corrected random force $F_\text{R}^\text{i,c}$ in equation~(\ref{eq:fr_corrected}) is, therefore, used as the starting point $F_\text{R}^\text{p}$ for generating future time steps, as described in Supplementary Information \ref{app:SFRF}. We used $\delta t = \Delta t/5$, which is sufficient to map the data on the continuum limit.
We note in passing that there are alternative methods to correct the autocorrelation of the discrete data, e.g. a whitening approach using the Cholesky decomposition \cite{kessy2018optimal}.
\section{Random-Force Distribution in Weather and Financial Data}
\label{app:non_gauss}

\begin{figure*}
	\centering
	\includegraphics[width=\linewidth]{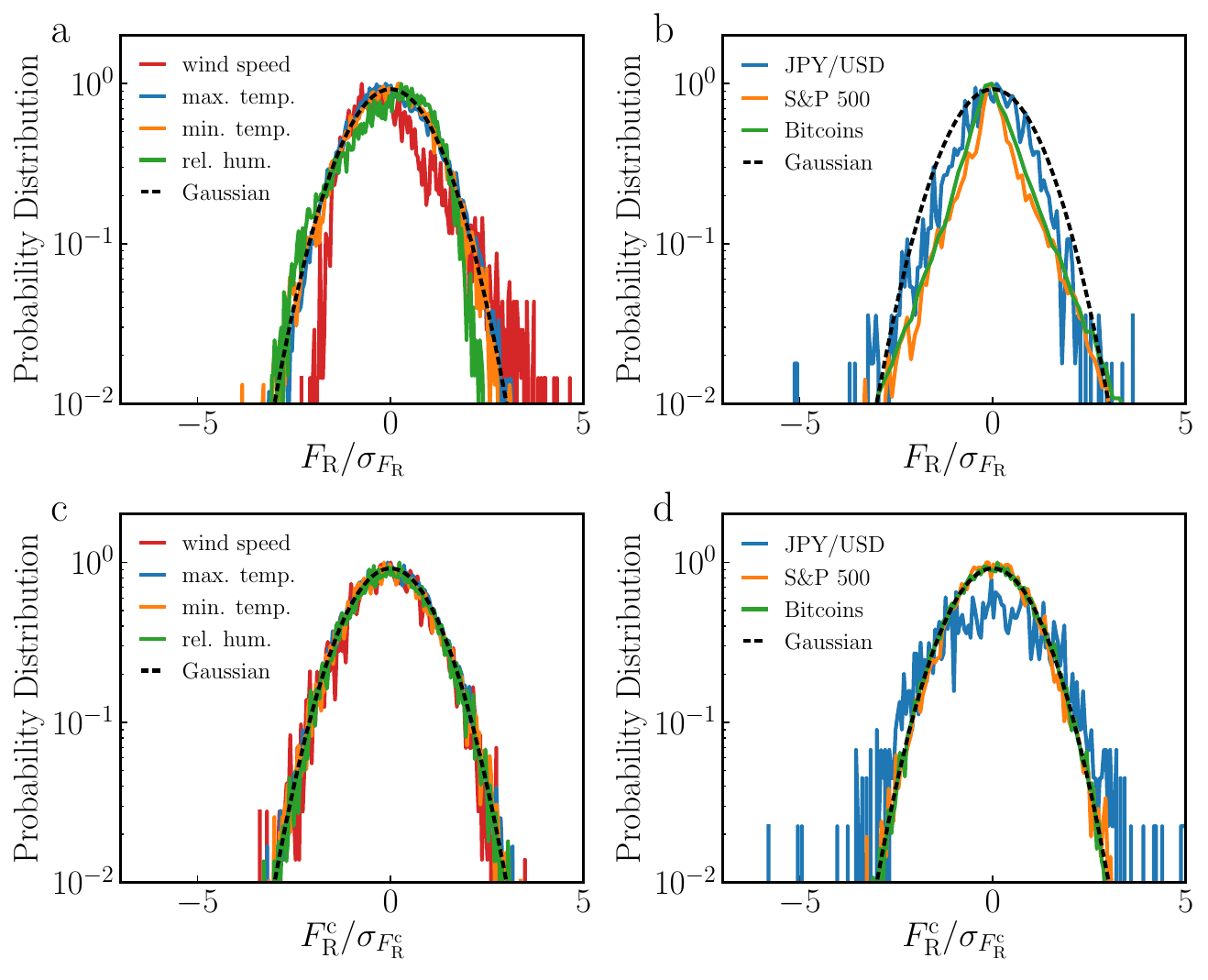}
	\caption{Random-force distributions from weather and financial data. {\bf\sffamily a}: Probability distributions of the random force from the weather data shown in the main text and Supplementary Information \ref{app:kernels_meteo}. We compute the random-force trajectories using equation~(\ref{eq:fr0}) in Supplementary Information \ref{app:fdt_disc}. The black-dashed line denotes a Gaussian distribution with zero mean and unit variance. Note that the bins are normalized by their respective random force standard deviations $\sigma_{F_\text{R}} = \sqrt{\langle \dot{A}_\text{f}^2 \rangle\Gamma(0)}$, and the absolute values are normalized by their maximal value. {\bf\sffamily b}: Random-force distributions for the financial data shown in the main text and calculated using equation~(\ref{eq:fr0}). {\bf\sffamily c and d}: Distributions of the random force corrected for discretization effects using equation~(\ref{eq:fr_corrected}) in Supplementary Information \ref{app:fdt_disc}.}
	\label{fig:random_force_distributions_filtered_trajectories}
\end{figure*}

The random-force generation technique proposed in Supplementary Information \ref{app:SFRF} requires the past random-force trajectory to be Gaussian distributed. However, in Fig.~\ref{fig:random_force_distributions_filtered_trajectories}a and b, the distributions of random-force trajectories computed using equation~(\ref{eq:fr0}) in Supplementary Information \ref{app:fdt_disc} reveal that this is not always the case, especially for the financial data, where the distributions exhibit considerable deviations from Gaussian behavior. 
In fact, the non-Gaussian random-force contributions are removed by our correction method introduced in Supplementary Information \ref{app:fdt_disc}. Adding Gaussian terms to the discrete velocities and accelerations, the corrected random forces of both weather and financial data become Gaussian, as we demonstrate in Fig.~\ref{fig:random_force_distributions_filtered_trajectories}c and d. The only exception is the weekly JPY/USD exchange rate, presumably, due to the short trajectory length.

\section{Performance for Different Numbers of Random-Force Trajectories}
\label{app:performance_avg_size}

\begin{figure*}%[hbt!]
	\centering
	\includegraphics[width=0.5\linewidth]{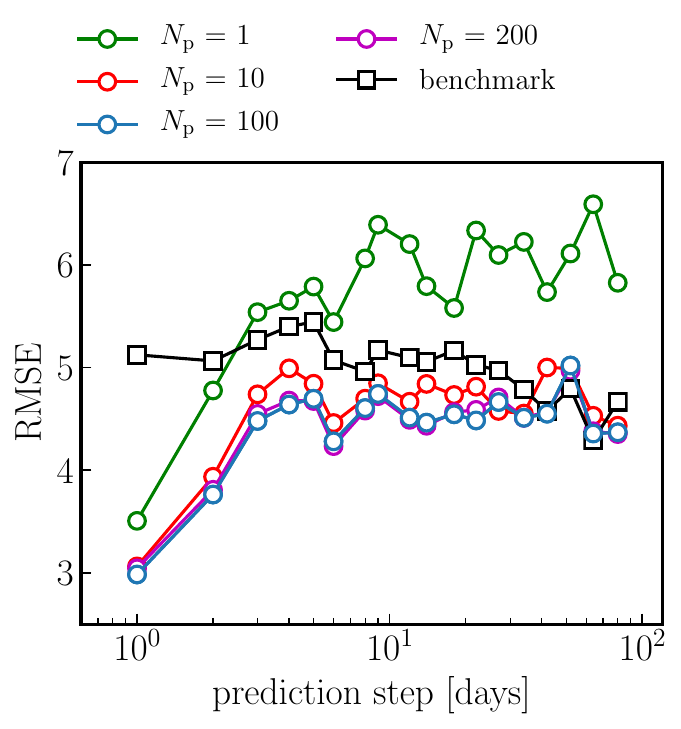}
	\caption{RMSE for the maximal temperature in Berlin (Fig.~\ref{fig:maxTemp2}e in the main text) using the GLE in equation~(\ref{eq:ODE_discret}) with different values of $N_\text{p}$, being the number of random-force trajectories.}
	\label{fig:pred_avg_size}
\end{figure*}

In Fig.~\ref{fig:pred_avg_size}, we show the RMSE for the maximal temperature in Berlin (similar to Fig.~\ref{fig:maxTemp2}e in the main text), using the GLE in equation~(\ref{eq:ODE_discret}) with different values of $N_\text{p}$, the number of random-force trajectories used for a prediction. We find that the higher $N_\text{p}$, the better the performance, saturating above $N_\text{p} = 100$. Consequently, we choose $N_\text{p} = 100$ for the predictions shown in the main text.

\section{Accuracy of the RK4 Method for Simulations of the GLE}
\label{app:test_RK4}

\begin{figure*}
	\centering
	\includegraphics[width=1.0\linewidth]{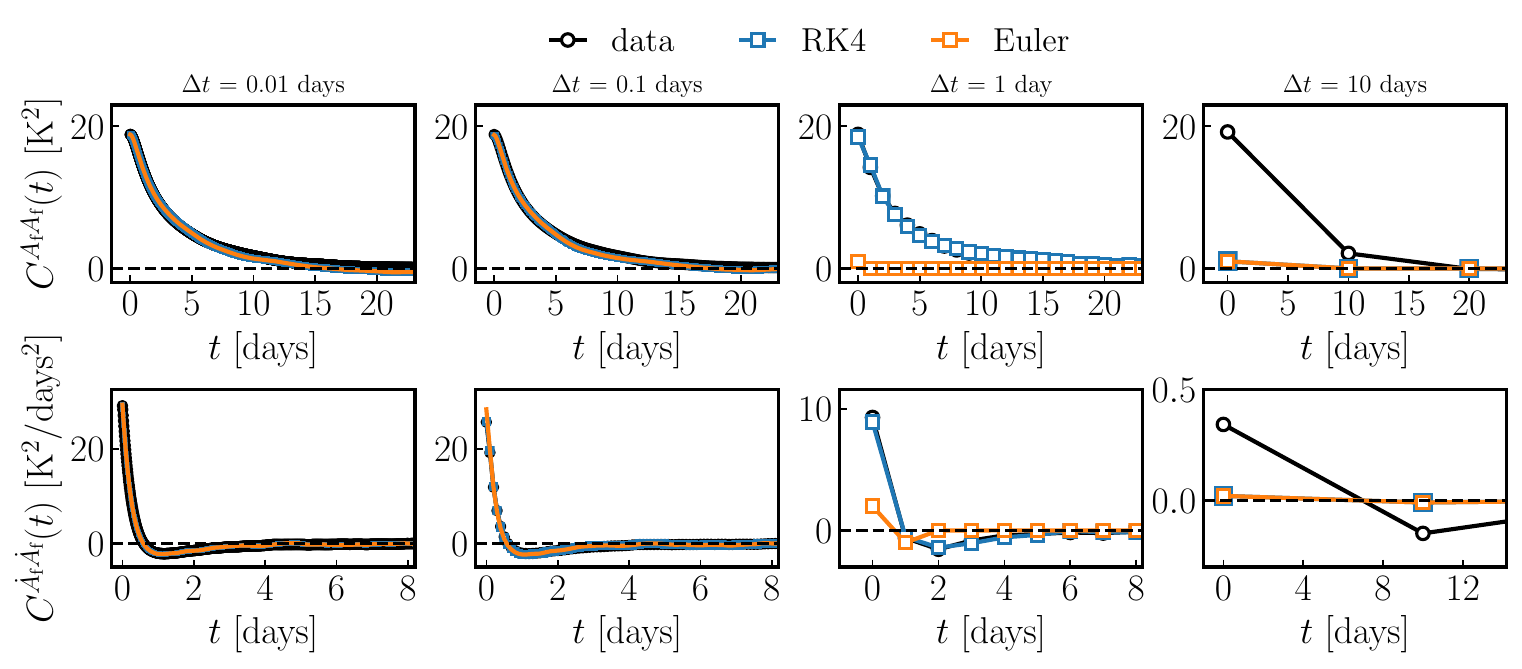}
	\caption{Test of discretization effects on the numerical solution of the GLE with equation~(\ref{eq:ODE_discret}) in the main text, using the RK4 and the Euler integration method with different time steps $\Delta t$. The numerical solution is obtained with the memory kernel $\Gamma(t)$, $k$ and $B$ from the discrete estimation method (Supplementary Information \ref{app:mitterwallner}) for the maximal temperature in Berlin, listed in Table~\ref{tab:fits_kernel} in Supplementary Information \ref{app:fit_kernels_params}. We compare the discrete position autocorrelation function $C^{A_\text{f}A_\text{f}}(t)$  and velocity autocorrelation function $C^{\dot{A}_\text{f}\dot{A}_\text{f}}(t)$ from the numerical solution of equation~(\ref{eq:ODE_discret} using the RK4 (blue) and Euler (orange) methods with the results from quasi-continuous data obtained by a simulation of equation~(\ref{eq:ODE_system0}) with a time step of $\delta t = 0.001$ days (for details see text). The RK4 method with $\Delta t = $ 1 day reproduces the discrete correlation functions of the quasi-continuous data very well, which demonstrates that the RK4 integration technique is appropriate for time-series prediction with $\Delta t = $ 1 day.}
	\label{fig:predRK4Euler}
\end{figure*}

Simulation of the discretized GLE in equation~(\ref{eq:ODE_discret}) in the main text is subject to discretization errors if the discretization time step $\Delta t$ becomes of the order of the time scales $\tau$, $\tau_\text{rel}$, or $\tau_\text{per}$. To investigate discretization effects, we simulate a quasi-continuous trajectory using equation~(\ref{eq:ODE_system0}), which is equivalent to the Mori-GLE in equation~(\ref{eq:mori_GLE}) with the memory kernel given in equation~(\ref{eq:meteo_kernel}) (Supplementary Information \ref{app:model_sys}), by using a discretization time step $\delta t = 0.001$ days that is much smaller than all time scales of the system. For the parameters, we use the obtained values for the maximal temperature in Berlin, listed in Table~\ref{tab:fits_kernel} in Supplementary Information \ref{app:fit_kernels_params}, and produce a trajectory with a length of 5$\cdot 10^7$ steps.
\\ \indent By comparison with the quasi-continuous trajectory, we test the numerical solution of equation~(\ref{eq:ODE_discret}), with the RK4 method and with the Euler integration method with much larger discretization $\Delta t > \delta t$. We use the memory kernel $\Gamma$, $k$ and $b$ of the quasi-continuous simulation data, and use random-force data generated from the method described in Supplementary Information \ref{app:SFRF}. Note that the simulation parameters are the same as for the prediction in Methods \ref{sec:GLE}, except that we use initial conditions $\dot{A}_\text{f}^0 = 0$ and $F_\text{R}^0 = 0$, thus, the simulation does not depend on the past trajectory. All discrete trajectories have the same temporal length as the quasi-continuous trajectory we compare with. \\
\indent In Fig.~\ref{fig:predRK4Euler}, we compare the discrete position autocorrelation function $C^{A_\text{f}A_\text{f}}(t)$  and the velocity autocorrelation function $C^{\dot{A}_\text{f}\dot{A}_\text{f}}(t)$ from the quasi-continuous simulation data (black) with results obtained by direct numerical simulation of equation~(\ref{eq:ODE_discret}) at variable discretization time step $\Delta t$. The RK4 method (blue) reproduces the quasi-continuous simulation data up to a time step of 1 day, which is the time step we use in the main text. The Euler method (orange) only works for time steps up to 0.1 days, which is due to the higher integration error of the Euler method. We find that a simulation of the GLE using equation~(\ref{eq:ODE_discret}) with the RK4 method describes the maximal temperature data we use in the main text very accurate and is, therefore, a proper method for time-series prediction with $\Delta t = 1$ day. The results for time steps $\Delta t = 10$ days could in principle be further improved with higher-order Runge-Kutta methods.

%\newpage
\section{Self-Consistency of the Extraction and Prediction Methods}
\label{app:POC}

\begin{figure*}[h!]
	\centering
	\includegraphics[width=1.0\linewidth]{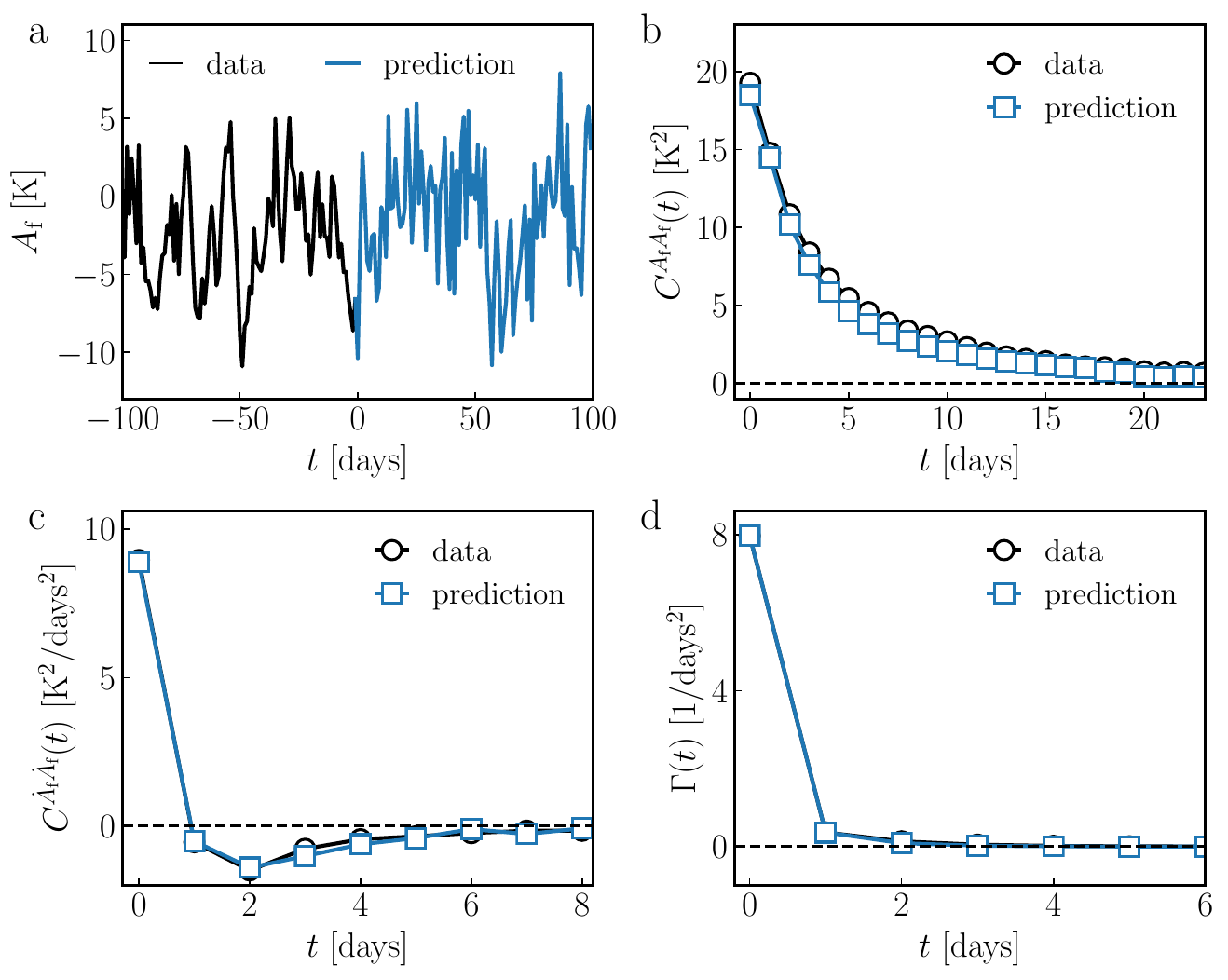}
	\caption{Test of the self-consistency of the GLE prediction method presented in Methods \ref{sec:GLE}, using the data for the daily maximal temperature in Berlin in Fig.~\ref{fig:maxTemp1} in the main text. {\bf\sffamily a}: Section of a predicted trajectory $A_\text{f}$ (blue), appended to the end of the past data (black). Note that the time $t$ is shifted to an arbitrarily chosen initial point, which is the starting point of the prediction. Here, we use the data until November 21, 1998. {\bf\sffamily b and c}: Discrete position autocorrelation function $C^{A_\text{f}A_\text{f}}(t)$ ({\bf\sffamily b}) and velocity autocorrelation function $C^{\dot{A}_\text{f}\dot{A}_\text{f}}(t)$ ({\bf\sffamily c}) from the prediction (blue, $t > 0$) and the past data (black, $t \leq 0$) in ({\bf\sffamily a}).
		{\bf\sffamily d}: Extracted memory kernel from the prediction (blue, $t > 0$) in ({\bf\sffamily a}) compared with the memory kernel extracted from the past data (black, $t \leq 0$), both computed using the discrete estimation method described in Supplementary Information \ref{app:mitterwallner}.}
	\label{fig:predPOC}
\end{figure*}

We test the self-consistency of the GLE prediction presented in Methods \ref{sec:GLE} using the data of the daily maximal temperature in Berlin in Fig.~\ref{fig:maxTemp1} in the main text. For this purpose, we use the memory kernel $\Gamma(t)$, which is extracted from a trajectory up to a randomly chosen time point (here November 21, 1998) by the discrete estimation technique explained in Supplementary Information \ref{app:mitterwallner} to generate a random-force trajectory based on the past random-force trajectory, using the method introduced in Supplementary Information \ref{app:SFRF}, with a length of 10$^5$ steps and with discretization time $\Delta t$ = 1 day. Next, we use this sample and the past time-series data to generate a trajectory of $A_\text{f}$ using equation~(\ref{eq:ODE_discret}) in the main text. In Fig.~\ref{fig:predPOC}a, we show a short section of the predicted trajectory, alongside with a section of the past data up to the time where the prediction starts. From the discrete velocitiy autocorrelation function of the predicted trajectory, we compute the memory kernel $\Gamma$ using the discrete estimation method described in Supplementary Information \ref{app:mitterwallner}. 
\\ \indent We demonstrate in Fig.~\ref{fig:predPOC}b-d that the discrete position autocorrelation function $C^{A_\text{f}A_\text{f}}(t)$, the velocity autocorrelation function $C^{\dot{A}_\text{f}\dot{A}_\text{f}}(t)$, computed from the discrete velocities $\dot{A}_\text{f}((i+1/2)\Delta t) = (A_\text{f}((i+1)\Delta t) - A_\text{f}(i\Delta t))/\Delta t$, and the memory kernel from the prediction match very well the ones from the past data, illustrating the self-consistency of the prediction and discrete estimation methods. 

\section{Hyperparameter Choice for LSTM Predictions}
\label{app:perfornance_lstm}

\begin{figure*}
	\centering
	\includegraphics[width=1\linewidth]{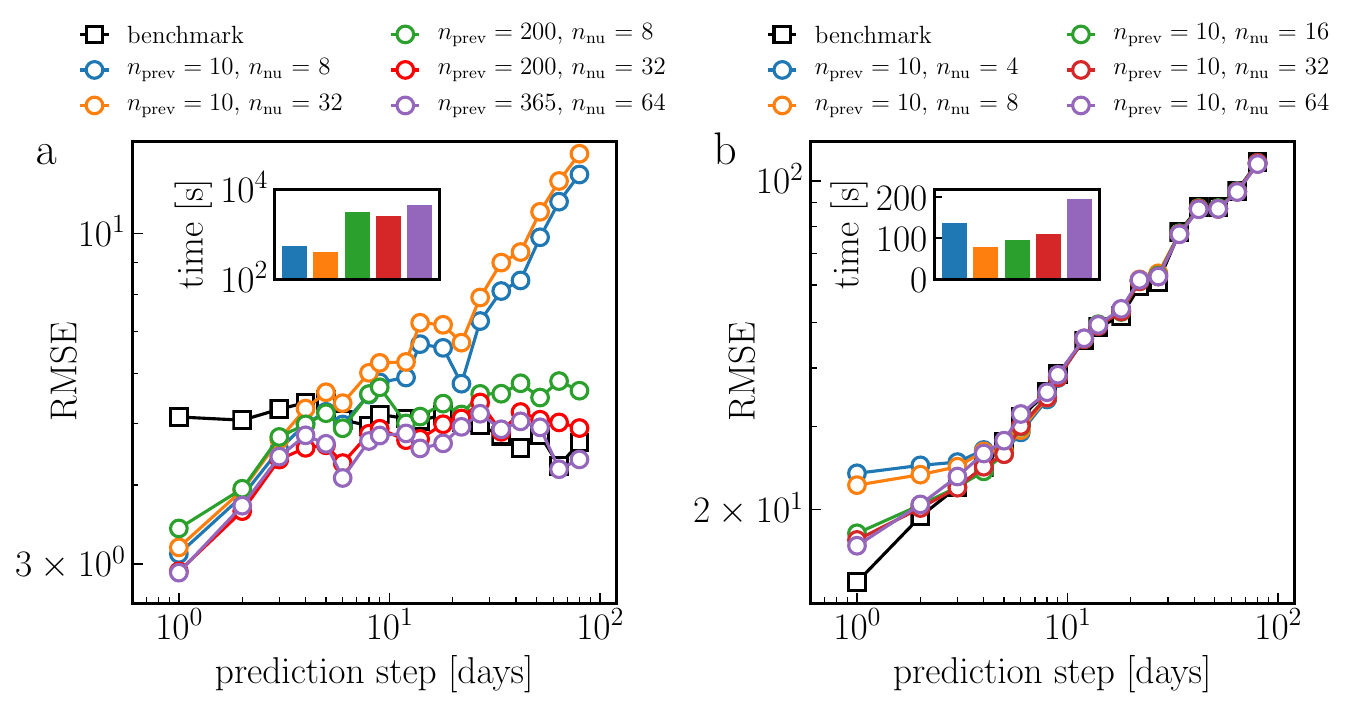}
	\caption{Hyperparameter optimization for LSTM neural network weather and financial data predictions.  {\bf\sffamily a}: RMSE (given in equation~(\ref{eq:RMSE}) in the main text) of LSTM predictions of the maximal temperature in Berlin (see Fig.~\ref{fig:maxTemp2}e in the main text for details) for different input lengths $n_{\text{prev}}$ and number of neural units in the hidden layer $n_\text{nu}$ (defined in Methods \ref{sec:LSTM}). The inset shows the computation time for a prediction over 360 days. {\bf\sffamily b}: Results for the S$\&$P 500 stock market index (displayed in Fig.~\ref{fig:finance2}b in the main text).}
	\label{fig:performance_lstm_gridsearch}
\end{figure*}

The training and prediction using LSTM neural networks (see Methods \ref{sec:LSTM}) depend on the choice of hyperparameters, in particular, the length of the input sequence $n_{\text{prev}}$ and the number of neural units $n_{\text{nu}}$ in the hidden layer. In Fig.~\ref{fig:performance_lstm_gridsearch}, we investigate the prediction error of LSTM neural networks after training with different values for $n_{\text{prev}}$ and $n_{\text{nu}}$ in the single input and hidden layers we use. For the maximal temperature in Berlin (a), an input length of $n_{\text{prev}} = 10$ is not sufficient to approach the benchmark for a high number of prediction steps. This is only possible with a higher input length of $n_{\text{prev}}$ = 200, and $n_{\text{nu}}$ = 32 units in the hidden layer. The performance can be slightly improved by increasing both hyperparameters, accompanied by a higher computation time (see inset). We, therefore, consider the data for $n_{\text{prev}}$ = 200 and $n_{\text{nu}}$ = 32 (red) to be a fair comparison with the GLE prediction (Fig.~\ref{fig:maxTemp2}e in the main text). 
\\ \indent Since the memory time for the financial data is below the time resolution (shown in Supplementary Information \ref{app:fit_kernels_params}), for the S$\&$P 500 stock market index in Fig.~\ref{fig:performance_lstm_gridsearch}b, we only investigate the effect of $n_{\text{nu}}$ on the prediction and keep $n_{\text{prev}} = 10$ fixed. The RMSE improves with increasing $n_{\text{nu}}$. For $n_{\text{nu}}$  = 32 units, the RMSE saturates, which is the reason for considering data for $n_{\text{prev}}$ = 10 and $n_{\text{nu}}$ = 32 (red) to constitute a fair comparison with the GLE (Fig.~\ref{fig:finance2}b in the main text). Note that the prediction time in the insets partly appears to decrease with increasing $n_{\text{nu}}$, which is due to a shorter training time.

\section{Derivation of equation~(\ref{eq:trafo})}
\label{app:proofs_sampling}
Given a random data sequence $\mathbf{G}$  coming from the distribution $\mathcal{N}(\mathbf{0},\hat{1})$, we show that the random sequence $\mathbf{F}_\text{R}=\hat{L}\mathbf{G}+\boldsymbol\mu$ follows the distribution $\mathcal{N}(\boldsymbol{\mu},\hat{L}\hat{L}^\text{T})$. For this, we compute the probability density function of $\mathbf{F}_\text{R}$ via
\begin{eqnarray}
	P_{F_\text{R}}(\mathbf{F}_\text{R}) = \frac{1}{\sqrt{\left(2\pi |\hat{1}| \right)}}\int\mathrm{d}^M G\,\delta(\mathbf{F}_\text{R} - \hat{L}\cdot\mathbf{G}-\boldsymbol\mu)\,e^{-\frac{1}{2}\mathbf{G}^\text{T}\hat{1}\mathbf{G}}.
\end{eqnarray}
Using the substitution $\mathbf{F}_\text{R} = \hat{L}\cdot\mathbf{G}+\boldsymbol\mu$ or equivalently $\mathbf{G}=\hat{L}^{-1}(\mathbf{F}_\text{R} - \boldsymbol\mu)$ and the corresponding transformation of the differential $\mathrm{d}^M F_\text{R} = \sqrt{| \hat{L}\hat{L}^\text{T}|} d^M G$, it immediately follows that
\begin{eqnarray}
	P_{F_\text{R}}(\mathbf{F}_\text{R}) &= \frac{1}{\sqrt{\left(2\pi |\hat{L}\hat{L}^\text{T}|\right)}}e^{-\frac{1}{2}\left( \mathbf{F}_\text{R} - \boldsymbol\mu \right)^\text{T} \left(\hat{L}\hat{L}^\text{T}\right)^{-1}\left(\mathbf{F}_\text{R} - \boldsymbol\mu \right)},
\end{eqnarray}
i.e. $\mathbf{F}_\text{R}$ follows the distribution $\mathcal{N}(\boldsymbol\mu,\hat{L}\hat{L}^\text{T})$.

\section{Derivation of equation~(\ref{eq:conditional})}
\label{app:proofs_sampling2}

For the proof of equation~(\ref{eq:conditional}), we assume that all inverse matrices exist. We prove the validity of equation~(\ref{eq:conditional}) by factorizing the multivariate Gaussian probability distribution given by
\begin{eqnarray}
	\label{eq:pr1}
	\mathcal{N}(\mathbf{0}, \hat{C}) = \frac{1}{\sqrt{\left(2\pi |\hat{C}|\right)}} e^{-\frac{1}{2}\mathbf{F}_\text{R}^\text{T}\:\cdot \:\hat{C}^{-1}\:\cdot\:\mathbf{F}_\text{R}},
\end{eqnarray}
into the marginal and conditional probability density functions using the Bayesian theorem, i.e. $ P_{F_\text{R}^\text{f}}\left(\mathbf{F}_\text{R}^\text{f} , \mathbf{F}_\text{R}^\text{p}\right) = P_{F_\text{R}^\text{f}}\left(\mathbf{F}_\text{R}^\text{p}\right)	P_{F_\text{R}^\text{f}}\left(\mathbf{F}_\text{R}^\text{f} | \mathbf{F}_\text{R}^\text{p}\right)$. This is accomplished by partitioning the exponent $\mathbf{F}_\text{R}^\text{T}\cdot \hat{C}^{-1}\cdot\mathbf{F}_\text{R}$ in equation~(\ref{eq:pr1}) using $ \hat{C}$ from equation~(\ref{eq:part_cov}) according to
\begin{subequations}
	\begin{eqnarray}
		\mathbf{F}_\text{R}^\text{T}\cdot \hat{C}^{-1}\cdot\mathbf{F}_\text{R} = \mathbf{F}_\text{R}^\text{T}\cdot\hat{S}\cdot\mathbf{F}_\text{R} = \left( \begin{array}[t]{c c}
			(\mathbf{F}_\text{R}^\text{p})^\text{T} & (\mathbf{F}_\text{R}^\text{f})^\text{T}
		\end{array}\right)\left(\begin{array}[c]{c c}
			\hat{S}_\text{pp} & \hat{S}_\text{pf}\\
			\hat{S}_\text{pf}^\text{T} & \hat{S}_\text{ff}
		\end{array}\right)\left(\begin{array}{c}
			\mathbf{F}_\text{R}^\text{p}\\
			\mathbf{F}_\text{R}^\text{f}
		\end{array}\right)\label{eq:pr2},\\
		= (\mathbf{F}_\text{R}^\text{p})^\text{T}\hat{S}_\text{pp}\mathbf{F}_\text{R}^\text{p} + (\mathbf{F}_\text{R}^\text{p})^\text{T}\hat{S}_\text{pf}\mathbf{F}_\text{R}^\text{f} + (\mathbf{F}_\text{R}^\text{f})^\text{T}\hat{S}_\text{pf}^\text{T}\mathbf{F}_\text{R}^\text{p} + (\mathbf{F}_\text{R}^\text{f})^\text{T}\hat{S}_\text{ff}\mathbf{F}_\text{R}^\text{f} ,\label{eq:pr3}
	\end{eqnarray}
\end{subequations}
where we define $\hat{S}\equiv \hat{C}^{-1}$ using the block matrices $\hat{S}_\text{pp} \in \mathbb{R}^{N_\text{past}\times N_\text{past}}$, $\hat{S}_\text{pf} \in \mathbb{R}^{N_\text{past}\times N_\text{fut}}$ and $\hat{S}_\text{ff} \in \mathbb{R}^{N_\text{fut}\times N_\text{fut}}$. To relate the block matrices of $\hat{S}$ to the block matrices of the covariance matrix $\hat{C}$ in equation~(\ref{eq:part_cov}), we consider the inverse of $\hat{S}$ \cite{dennis_s_bernstein_author_matrix_2009}, using Schur complements,
\begin{eqnarray}
	\label{eq:pr4}
	&\hat{C} = \hat{S}^{-1} = \nonumber\\
	& \left(\begin{array}{c c}
		\left( \hat{S}_\text{pp} - \hat{S}_\text{pf}\hat{S}_\text{ff}^{-1}\hat{S}_\text{pf}^\text{T} \right)^{-1} & -\left( \hat{S}_\text{pp} - \hat{S}_\text{pf}\hat{S}_\text{ff}^{-1}\hat{S}_\text{pf}^\text{T} \right)^{-1}\hat{S}_\text{pf}\hat{S}_\text{ff}^{-1}\\
		-\hat{S}_\text{ff}^{-1}\hat{S}_\text{pf}^\text{T}\left( \hat{S}_\text{pp} - \hat{S}_\text{pf}\hat{S}_\text{ff}^{-1}\hat{S}_\text{pf}^\text{T} \right)^{-1} & \left(\hat{S}_\text{ff} - \hat{S}_\text{pf}^\text{T}\hat{S}_\text{pp}^{-1}\hat{S}_\text{pf}\right)^{-1}
	\end{array}\right).
\end{eqnarray}
Here we used that $\hat{S} \hat{S}^{-1} = \hat{S}^{-1}  \hat{S}= \hat{1}$ (see Supplementary Information \ref{app:proof_invert_cov}). Comparing equation~(\ref{eq:pr4}) with equation~(\ref{eq:part_cov}) gives
\begin{subequations}
	\label{eq:pr5}
	\begin{align}
		\hat{C}_\text{pp} &= \left( \hat{S}_\text{pp} - \hat{S}_\text{pf}\hat{S}_\text{ff}^{-1}\hat{S}_\text{pf}^\text{T} \right)^{-1},\label{eq:pr51}\\
		\hat{C}_\text{ff} &= \left( \hat{S}_\text{ff} - \hat{S}_\text{pf}^\text{T}\hat{S}_\text{pp}^{-1}\hat{S}_\text{pf} \right)^{-1},\label{eq:pr52}\\
		-\hat{S}_\text{ff}^{-1}\hat{S}_\text{pf}^\text{T} \hat{C}_\text{pp}& = \hat{C}_\text{pf}^\text{T} \,\Rightarrow\, \hat{S}_\text{ff}^{-1}\hat{S}_\text{pf}^\text{T} = -\hat{C}_\text{pf}^\text{T}\hat{C}_\text{pp}^{-1}.\label{eq:pr53}
	\end{align}
\end{subequations}
To make use of the relations in equations~(\ref{eq:pr5}),  we add and subtract $(\mathbf{F}_\text{R}^\text{p})^\text{T}\hat{S}_\text{pf}\hat{S}_\text{ff}^{-1}\hat{S}_\text{pf}^\text{T}\mathbf{F}_\text{R}^\text{p}$ from equation~(\ref{eq:pr3}). Subsequently, equation~(\ref{eq:pr3}) can be written as
\begin{align}
	\mathbf{F}_\text{R}^\text{T} \cdot \hat{C}^{-1}\cdot\mathbf{F}_\text{R}  = &  \\ 
	=  (\mathbf{F}_\text{R}^\text{p})^\text{T}\hat{C}_\text{pp}^\text{-1}\mathbf{F}_\text{R}^\text{p}   + & (\mathbf{F}_\text{R}^\text{p})^\text{T}  \hat{S}_\text{pf}\mathbf{F}_\text{R}^\text{f}  + (\mathbf{F}_\text{R}^\text{f})^\text{T}\hat{S}_\text{pf}^\text{T}\mathbf{F}_\text{R}^\text{p} + (\mathbf{F}_\text{R}^\text{f})^\text{T}\hat{S}_\text{ff}\mathbf{F}_\text{R}^\text{f} + (\mathbf{F}_\text{R}^\text{p})^\text{T}\hat{S}_\text{pf}\hat{S}_\text{ff}^{-1}\hat{S}_\text{pf}^\text{T}\mathbf{F}_\text{R}^\text{p},\\
	=  (\mathbf{F}_\text{R}^\text{p})^\text{T}\hat{C}_\text{pp}^\text{-1}\mathbf{F}_\text{R}^\text{p}  + & (\mathbf{F}_\text{R}^\text{p})^\text{T}  \hat{S}_\text{pf}\mathbf{F}_\text{R}^\text{f} + (\mathbf{F}_\text{R}^\text{f})^\text{T}\hat{S}_\text{pf}^\text{T}\mathbf{F}_\text{R}^\text{p} + (\mathbf{F}_\text{R}^\text{f})^\text{T}\left(\hat{C}_\text{ff}^{-1} + \hat{S}_\text{pf}^\text{T}\hat{S}_\text{pp}^{-1}\hat{S}_\text{pf}\right)\mathbf{F}_\text{R}^\text{f} \nonumber \\  -   (\mathbf{F}_\text{R}^\text{p})^\text{T}   \hat{S}_\text{pf}\hat{C}_\text{pf}^\text{T} & \hat{C}_\text{pp}^{-1} \mathbf{F}_\text{R}^\text{p}, \\
	=  (\mathbf{F}_\text{R}^\text{p})^\text{T}\hat{C}_\text{pp}^\text{-1}\mathbf{F}_\text{R}^\text{p}   - &
	\left(  \hat{C}_\text{pf}^\text{T} \hat{C}_\text{pp}^{-1}\right)^\text{T}(\mathbf{F}_\text{R}^\text{p})^\text{T}\left(\hat{C}_\text{ff}-\hat{C}_\text{pf}^\text{T}\hat{C}_\text{pp}^{-1}\hat{C}_\text{pf}\right)^{-1} \mathbf{F}_\text{R}^\text{f} \nonumber \\
	-  (\mathbf{F}_\text{R}^\text{f})^\text{T} & \left(\hat{C}_\text{ff}-\hat{C}_\text{pf}^\text{T}\hat{C}_\text{pp}^{-1}\hat{C}_\text{pf}\right)^{-1}  \hat{C}_\text{pf}^\text{-T}\hat{C}_\text{pp}^\text{-1}\mathbf{F}_\text{R}^\text{p} + (\mathbf{F}_\text{R}^\text{f})^\text{T}\left(\hat{C}_\text{ff}-\hat{C}_\text{pf}^\text{T}\hat{C}_\text{pp}^{-1}\hat{C}_\text{pf}\right)^{-1} \mathbf{F}_\text{R}^\text{f} \nonumber \\ +   \left(  \hat{C}_\text{pf}^\text{T} \hat{C}_\text{pp}^{-1}\right)^\text{T}&(\mathbf{F}_\text{R}^\text{p})^\text{T} \left(\hat{C}_\text{ff}-\hat{C}_\text{pf}^\text{T}\hat{C}_\text{pp}^{-1}\hat{C}_\text{pf}\right)^{-1} \hat{C}_\text{pf}^\text{T}\hat{C}_\text{pp}^{-1} \mathbf{F}_\text{R}^\text{p} ,
	\\=  (\mathbf{F}_\text{R}^\text{p})^\text{T}\hat{C}_\text{pp}^\text{-1}\mathbf{F}_\text{R}^\text{p}   + & \left(\mathbf{F}_\text{R}^\text{f} - \hat{C}_\text{pf}^\text{T}\hat{C}_\text{pp}^{-1}\mathbf{F}_\text{R}^\text{p}\right)^\text{T}  \left(\hat{C}_\text{ff}-\hat{C}_\text{pf}^\text{T}\hat{C}_\text{pp}^{-1}\hat{C}_\text{pf}\right)^{-1}  \left(\mathbf{F}_\text{R}^\text{f} - \hat{C}_\text{pf}^\text{T}\hat{C}_\text{pp}^{-1}\mathbf{F}_\text{R}^\text{p}\right). 	\label{eq:pr6}
\end{align}
Finally, we make use of the identity for the determinant of block matrices \cite{dennis_s_bernstein_author_matrix_2009}
\begin{eqnarray}
	|\hat{C}| &= \Bigl|\left(\begin{array}{c c}
		\hat{C}_\text{pp} & \hat{C}_\text{pf}\\
		\hat{C}_\text{pf}^\text{T} & \hat{C}_\text{ff}
	\end{array}\right)\Bigr| =\Bigl|\left(\hat{C}_\text{pp}\right)\Bigr|\cdot \Bigl|\left( \hat{C}_\text{ff}-\hat{C}_\text{pf}^\text{T}\hat{C}_\text{pp}^{-1}\hat{C}_\text{pf}\right)\Bigr|.\label{eq:pr7}
\end{eqnarray}
Inserting equation~(\ref{eq:pr6}) and equation~(\ref{eq:pr7}) into equation~(\ref{eq:pr1}) leads to the factorization
\begin{eqnarray}
	\mathcal{N}(\mathbf{0}, \hat{C}) = \mathcal{N}(\mathbf{0}, \hat{C}_\text{pp})\,\, \mathcal{N}(\hat{C}_\text{pf}^\text{T}\hat{C}_\text{pp}^{-1}\mathbf{F}_\text{R}^\text{p}, \hat{C}_\text{ff}-\hat{C}_\text{pf}^\text{T} \hat{C}_\text{pp}^{-1}\hat{C}_\text{pf}),
\end{eqnarray}
where $\mathcal{N}(\hat{C}_\text{pf}^\text{T}\hat{C}_\text{pp}^{-1}\mathbf{F}_\text{R}^\text{p}, \hat{C}_\text{ff}-\hat{C}_\text{pf}^\text{T} \hat{C}_\text{pp}^{-1}\hat{C}_\text{pf})$ is the conditional probability density function $P_{F_\text{R}}(\mathbf{F}_\text{R}^\text{f}|\mathbf{F}_\text{R}^\text{p})$ in equation~(\ref{eq:conditional}).

\section{Proof of \textbf{${S} \hat{S}^{-1} = \hat{S}^{-1}  \hat{S}= \hat{1}$}}
\label{app:proof_invert_cov}
For the proof of $\hat{S}^{-1}  \hat{S}= \hat{1}$, we make use of the fact that $\hat{S}^{-1} = \hat{C}$ is positive definite, i.e. for all non-zero eigenvectors $\mathbf{x}$: $\mathbf{x}^\text{T}  \hat{C} \mathbf{x} > 0$. Since $ \hat{C}$ is symmetric, it can be diagonalized, i.e. there exists an orthogonal matrix $\hat{Q}$ and a diagonal matrix $\hat{D}$ such that
\begin{equation}
	\hat{C} = \hat{Q} \hat{D} \hat{Q}^\text{T},
\end{equation}  
where $\hat{Q}$ contains the eigenvalues $c_1,c_2,...,c_\text{n}$ of $\hat{C}$ on its diagonal. For any non-zero vector $\mathbf{x}$
\begin{equation}
	\mathbf{x}^\text{T} 	\hat{C} \mathbf{x} = \mathbf{x}^\text{T}  \hat{Q} \hat{D} \hat{Q}^\text{T} \mathbf{x}.
\end{equation}
Let $\mathbf{y} =  \hat{Q}^\text{T} \mathbf{x}$. Note that $\mathbf{y} \neq 0$ since $\hat{Q}$ is orthogonal (and hence invertible). Thus,
\begin{equation}
	\mathbf{x}^\text{T}  \hat{C} \mathbf{x}  = \mathbf{y}^\text{T}  \hat{D} \mathbf{y} = \sum_{i=1}^n c_\text{i} y_\text{i}^2.
\end{equation}
Since $\hat{C}$ is positive definite, $\sum_{i=1}^n c_\text{i} y_\text{i}^2 > 0$ for all non-zero $\mathbf{y}$. This implies that each eigenvalue satisfies $c_\text{i}> 0$, because if any $c_\text{i} \leq 0$, there would exist some non-zero $\mathbf{y}$ making the sum non-positive, contradicting the positive definiteness of $\hat{C}$. Since a matrix is invertible if and only if all its eigenvalues are non-zero, it follows that $\hat{S}^{-1}  \hat{S}= \hat{1}$. The same is true for $\hat{S}\hat{S}^{-1}  = \hat{1}$ since the inverse of a positive definite matrix is also positive definite.

\end{document}